\documentclass[10pt]{article}
\usepackage{float}
\usepackage{graphicx}
\usepackage[margin=1.5cm]{geometry}
\graphicspath{{dir_arxiv_fig/}}
\usepackage{epstopdf}
\usepackage[CaptionAfterwards]{fltpage}
\usepackage{subfig}
\usepackage{amsopn}
\usepackage{amsmath}
\usepackage{amssymb}
\usepackage{amsfonts}
\usepackage{ulem}
\usepackage{xcolor}

\usepackage{amsmath}
\usepackage{amsfonts}
\usepackage{amssymb}
\usepackage{bbm}
\usepackage{graphicx}

\usepackage{xcolor}
\definecolor{LightGray}{rgb}{0.80,0.80,0.85}

\newcommand{\angstrom}{\AA}

\newcommand{\imunit}{\mathrm{i}}
\newcommand{\euler}{\mathrm{e}}

\newcommand{\Real}{\mathbb{R}}
\newcommand{\Complex}{\mathbb{C}}
\newcommand{\Integer}{\mathbb{Z}}
\newcommand{\Leb}{L}

\newcommand{\hypothesis}{\text{\tt Hyp}}
\newcommand\givenori[1][]{\:#1\vert\:}
\newcommand\givenmed{\givenori[]}
\newcommand\givenbig{\givenori[\big]}
\newcommand\semicolonori[1][]{\:#1;\:}
\newcommand\semicolonmed{\semicolonori[]}

\newcommand\commaori[1][]{\:#1;\:}
\newcommand\commamed{\commaori[]}

\newcommand{\vect}[1]{\boldsymbol{#1}}
\newcommand{\fourier}[1]{\hat{#1}}
\newcommand{\bessel}[1]{\check{#1}}

\newcommand{\wfourier}[1]{\widehat{#1}}
\newcommand{\rotation}{{\cal R}}

\newcommand{\ftranslation}{\widehat{\cal T}}

\newcommand{\inverse}[1]{{#1}^{-1}}
\newcommand{\transpose}[1]{{#1}^{\intercal}}

\newcommand{\ctranspose}[1]{{#1}^{\dagger}}
\newcommand{\ictranspose}[1]{{#1}^{-\dagger}}
\newcommand{\vzero}{{\vect{0}}}

\newcommand{\vx}{{\vect{x}}}
\newcommand{\vchi}{{\vect{\chi}}}

\newcommand{\vk}{{\vect{k}}}
\newcommand{\vkappa}{{\vect{\kappa}}}
\newcommand{\vd}{{\vect{\delta}}}

\newcommand{\vA}{{\vect{A}}}

\newcommand{\cX}{{\cal X}}

\newcommand{\cR}{{\cal R}}

\newcommand{\cJ}{{\cal J}}

\newcommand{\cL}{{\cal L}}
\newcommand{\cN}{{\cal N}}

\newcommand{\cl}{{\ell}}

\newcommand{\cD}{{\cal D}}
\newcommand{\cE}{{\cal E}}
\newcommand{\xA}{          {A} }
\newcommand{\fA}{{\wfourier{A}}}

\newcommand{\fB}{{\wfourier{B}}}
\newcommand{\xF}{          {F} }
\newcommand{\fF}{{\wfourier{F}}}

\newcommand{\fG}{{\wfourier{G}}}

\newcommand{\fsigma}{{\fourier{\sigma}}}
\newcommand{\strikefsigma}{\text{\sout{\ensuremath{\fsigma}}}}
\newcommand{\strikemu}{\text{\sout{\ensuremath{\mu}}}}
\newcommand{\xlambda}{{         \lambda }}

\newcommand{\xS}{          {S} }
\newcommand{\fS}{{\wfourier{S}}}

\newcommand{\fSlice}{\widehat{\text{\small Sl}}}

\newcommand{\bA}{{\bessel{A}}}
\newcommand{\bB}{{\bessel{B}}}

\newcommand{\xmax}{{X}}
\newcommand{\kmax}{{K}}

\newcommand{\lmax}{{L}}

\newcommand{\dx}{{\Delta x}}     
\newcommand{\dk}{{\Delta k}}     

\newcommand{\XA}{[{\tt X0}]}
\newcommand{\XB}{[{\tt X1}]}
\newcommand{\YA}{[{\tt Y0}]}
\newcommand{\YB}{[{\tt Y1}]}
\newcommand{\YC}{[{\tt Y2}]}
\newcommand{\Zone}{[\Delta{\tt Z1}]}
\newcommand{\Ztwo}{[\Delta^{2}{\tt Z2}]}

\newcommand{\mutauemp}{\mu_{\tau}^{\emp}}
\newcommand{\mutauequa}{\mu_{\tau}^{\equa}}
\newcommand{\mutaupole}{\mu_{\tau}^{\pole}}

\newcommand{\Ptauprior}{P(\tau)}
\newcommand{\argmin}{\operatorname*{arg\,min}}
\newcommand{\argmax}{\operatorname*{arg\,max}}
\newcommand{\bigO}{\mathcal{O}}

\newcommand{\nimage}{J}
\newcommand{\tnimage}{\tilde{\nimage}}

\newcommand{\nvolume}{I}
\newcommand{\npixel}{N}

\newcommand{\signal}{{\text{\tiny signal}}}
\newcommand{\noise}{{\text{\tiny noise}}}

\newcommand{\true}{{\text{\tiny true}}}

\newcommand{\critical}{{\text{\tiny crit}}}
\newcommand{\synthetic}{{\text{\tiny syn}}}
\newcommand{\lowtemp}{{\text{\tiny zero}}}

\newcommand{\opt}{{\text{\tiny opt}}}

\newcommand{\est}{{\text{\tiny est}}}
\newcommand{\loc}{{\text{\tiny loc}}}
\newcommand{\emp}{{\text{\tiny emp}}}

\newcommand{\pre}{{\text{\tiny old}}}

\newcommand{\pos}{{\text{\tiny new}}}

\newcommand{\dof}{{\text{dof}}}

\newcommand{\hk}{\hat{\boldsymbol{k}}}
\newcommand{\hkappa}{\widehat{\boldsymbol{\kappa}}}

\newcommand{\kpolara}{\theta}
\newcommand{\kazimub}{\phi}

\newcommand{\epolara}{\beta}
\newcommand{\eazimub}{\alpha}
\newcommand{\egammaz}{\gamma}

\newcommand{\bp}{{\text{\tiny bp}}}
\newcommand{\var}{{\text{\tiny var}}}
\newcommand{\lsq}{{\text{\tiny lsq}}}

\newcommand{\equa}{{\text{\tiny equa}}}
\newcommand{\pole}{{\text{\tiny pole}}}
\newcommand{\north}{{\text{\tiny north}}}
\newcommand{\south}{{\text{\tiny south}}}
\newcommand{\dipole}{{\text{\tiny dipole}}}
\newcommand{\period}{\text{.}}
\newcommand{\comma}{\text{,}}

\begin{document}

\begin{flushleft}
  {\Large
    \textbf\newline{Appendix for: Sensitivity Analysis of Cryo-Electron Microscopy Reconstructions via the Log-Likelihood Hessian}
  }
  \newline
  \\
Aaditya V. Rangan\textsuperscript{1,2*},
Wai-Shing Tang\textsuperscript{2,3},
Kexin Zhang\textsuperscript{4},
Pilar Cossio\textsuperscript{2,3},
\\
\bigskip
\textbf{1} Courant Institute of Mathematical Sciences, New York University, New York, NY, USA
\\
\textbf{2} Center for Computational Mathematics, Flatiron Institute, New York, USA
\\
\textbf{3} Center for Computational Biology, Flatiron Institute, New York, USA
\\
\textbf{4} Molecular Biophysics \& Biochemistry, Yale University, New Haven, USA
\bigskip
* avr209@nyu.edu

\end{flushleft}

\begin{abstract}

Cryo-electron microscopy (cryo-EM) is a technique in structural biology which is used to produce high-resolution 3D reconstructions of molecules.
One common step in the cryo-EM pipeline involves \textit{ab-initio} single-particle reconstruction (SPR).
This is a type of inverse-problem which takes as input a collection of 2D-images -- each a projection of a molecule from an unknown viewing-angle -- and attempts to reconstruct the 3D-volume representing the underlying molecular density.

Most methods for solving SPR search for a solution which optimizes a posterior likelihood of generating the observed image-data, given the reconstructed volume.
Within this framework, it is natural to study the Hessian of the log-likelihood: the eigenvectors and eigenvalues of the Hessian determine how the likelihood changes with respect to perturbations in the solution, which can give insight into the sensitivity of the solution to aspects of the input, such as the orientation distribution of image viewing-angles.

In this paper, we describe a simple strategy for estimating the smallest eigenvalues and eigenvectors (i.e., the `softest modes') of the Hessian of the log-likelihood for the \textit{ab-initio} single-particle reconstruction problem.
This strategy involves rewriting the log-likelihood as a 3D-integral.

Once we have estimated the softest modes, we can use them to determine which parts of the reconstructed volume are trustworthy, and which are unreliable, and how this unreliability might depend on the data-set and the imaging parameters.
We believe that this kind of analysis can be used alongside more traditional strategies for sensitivity analysis, as well as in other applications, such as free-energy estimation.

\end{abstract}

\section{Introduction}\label{sec_Introduction}

Cryo-electron microscopy (cryo-EM) is a popular technique in structural biology which is used to produce high-resolution 3D reconstructions of macromolecular structures and biomolecules, such as proteins \cite{Goncharov1987,vanHeel1987,Goncharov1988,Jonic2005,Cheng2015,Nogales2015,Elmlund2015,Murata2017,Sigworth2016,Bhella2019}.

A typical cryo-EM analysis pipeline involves many steps (see Appendix section 1 for a brief overview), one of which is called the `single-particle reconstruction', or `SPR' problem.
This inverse-problem involves thousands of experimental images that are averaged together to generate a 3D volumetric map of a biomolecule, under the assumption that all projections arise from a single, uniform molecular conformation \cite{PR16,TWSMR16,Sigworth2016,HEIMOWITZ2018215,MCSWEENEY2020719}.
Because this is a non-convex inverse problem, validating the result is non-trivial , and it can be challenging given the low signal-to-noise ratios in most experiments \cite{Sigworth1998,Grigorieff2007,EMAN2,Yang2008,Singer2009,Singer2011,Scheres2012,Shkolnisky2012,Lyumkis2013,Wang2013,Grigorieff2016,Punjani2017,PunjaniBrubaker2017}.
Thus, it is natural to ask whether or not SPR produces reliable results.

The validity and nominal resolution of SPR is typically assessed using methods involving split-set validation in the reconstruction \cite{henderson2012outcome,afonine2018new,lander2024single}, such as the Gold-Standard Fourier Shell Correlation (FSC) \cite{rosenthal2003optimal}.
This approach splits the dataset into two independent halves and measures the correlation between the two reconstructions\cite{Scheres2008,Scheres2009}.
Other strategies, such as ResMap and MonoRes, compare the signal-strength of features within the reconstructed volumes to estimates of the noise within the data, providing estimates of local error for those volumes \cite{Kucukelbiretal_ResMap_2014,Vilasetal_MonoRes_2018}.

While these methods offer a form of statistical validation for SPR, they are rather agnostic regarding how the reliability of the solution is related to the structure of the SPR inverse-problem itself.
For example, the FSC is typically used as a validation of a particular algorithm used to solve an SPR inverse-problem.
Similarly, ResMap and MonoRes are usually used to estimate the signal-to-noise ratio of regions of a particular volumetric solution to an SPR inverse-problem.
These methods aren't typically used to determine if there might be other `nearby' solutions to the SPR problem that are almost as good.
Similarly, these methods typically do not comment on whether or not certain regions with low signal-to-noise are unavoidable, or how those regions might change if the parameters of the SPR problem were to be modified.
It is these latter questions which motivate the present study.

In this paper, we propose a complementary approach to the kinds of statistical validation mentioned above.
We present a method for analyzing how sensitive solutions of the inverse-problem might be to small perturbations in the problem parameters, and whether or not that solution remains a faithful representation of the true biomolecule.

Our approach begins with the observation that, given how SPR is typically defined, a solution to an SPR problem is a reconstructed volume that optimizes a posterior likelihood of generating the observed picked-particle images, given the microscope parameters used to image those particles (see \cite{Wang2013,Singer2009,Shkolnisky2012,Goncharov1987,Goncharov1988,vanHeel1987,Jonic2005,Yang2008} and software from \cite{Punjani2017,Scheres2012,Grigorieff2016,Lyumkis2013,Grigorieff2007,EMAN2}).
That is, any particular solution to an SPR inverse-problem lies at a critical point, typically an optima, of a likelihood-landscape determined by the underlying dataset and a few additional problem parameters \cite{Sigworth1998,Sigworth2010}.
By looking around this local optima, we can check to see if there are other `nearby' solutions that might be almost as good as the optimal one.
By analyzing those nearby solutions, and how they depend on the problem parameters, we can gain insight into what kinds of errors are easy to make when solving the SPR problem, and under what conditions these errors might manifest.

\section{Problem Setup}

As an example of this perspective, let's first consider a $4.5\angstrom$ resolution reconstruction of the trpv1 capsaicin receptor (the transient receptor potential cation channel subfamily V member 1 referenced in {\tt EMD-5778}), which we'll refer to as $\xF^{\opt}$.
A level-surface of $\xF^{\opt}$ is shown in Fig \ref{fig_landscape_FIGAG0_0}A.
For this example, $\xF^{\opt}$ was obtained by applying an SPR reconstruction algorithm to a stack of synthetic images corresponding to various projections of $\xF^{\opt}$, each labeled as $\xA_{j}$ for an image-specific projection-direction $\tau_{j}^{\true}$.
The $\tau_{j}^{\true}$ themselves were drawn from the empirical distribution of viewing-angles for this molecule, as referenced in {\tt EMPIAR-10005}, and the contrast-transfer function (CTF) was taken to be unity.
This is an idealized setting, allowing us to take advantage of an `image-aligned' low-noise limit in our calculations.
These simplifications are for illustrative purposes only, and our methodology can still be applied even with real noisy images and real CTFs in the `high degree-of-freedom limit', as we'll discuss later on.

Now the reconstruction algorithm which produced $\xF^{\opt}$ did so by searching for a volume which maximizes the `likelihood' $P(\{\xA_{j}\}\givenmed \xF)$ of producing the data (in this case the collection of synthetic images $\{\xA_{j}\}$), given a putative volume $\xF$ and a standard forward model of image-generation (described in Appendix sections 4,5).
The $\xF^{\opt}$ obtained will be a local minimum of the negative-log-likelihood-function $\mathcal{L}(\{\xA_{j}\}\givenmed \xF)\propto-\log\left(P(\{\xA_{j}\}\givenmed \xF)\right)$.
We remark that, because we are operating in an image-aligned regime, $\mathcal{L}$ will also depend on the implicitly-determined estimated alignments $\tau_{j}^{\est}$ of each the images (each depending on $\xF$ as well).

\begin{FPfigure}
\centering
\includegraphics[width=1.0\textwidth]{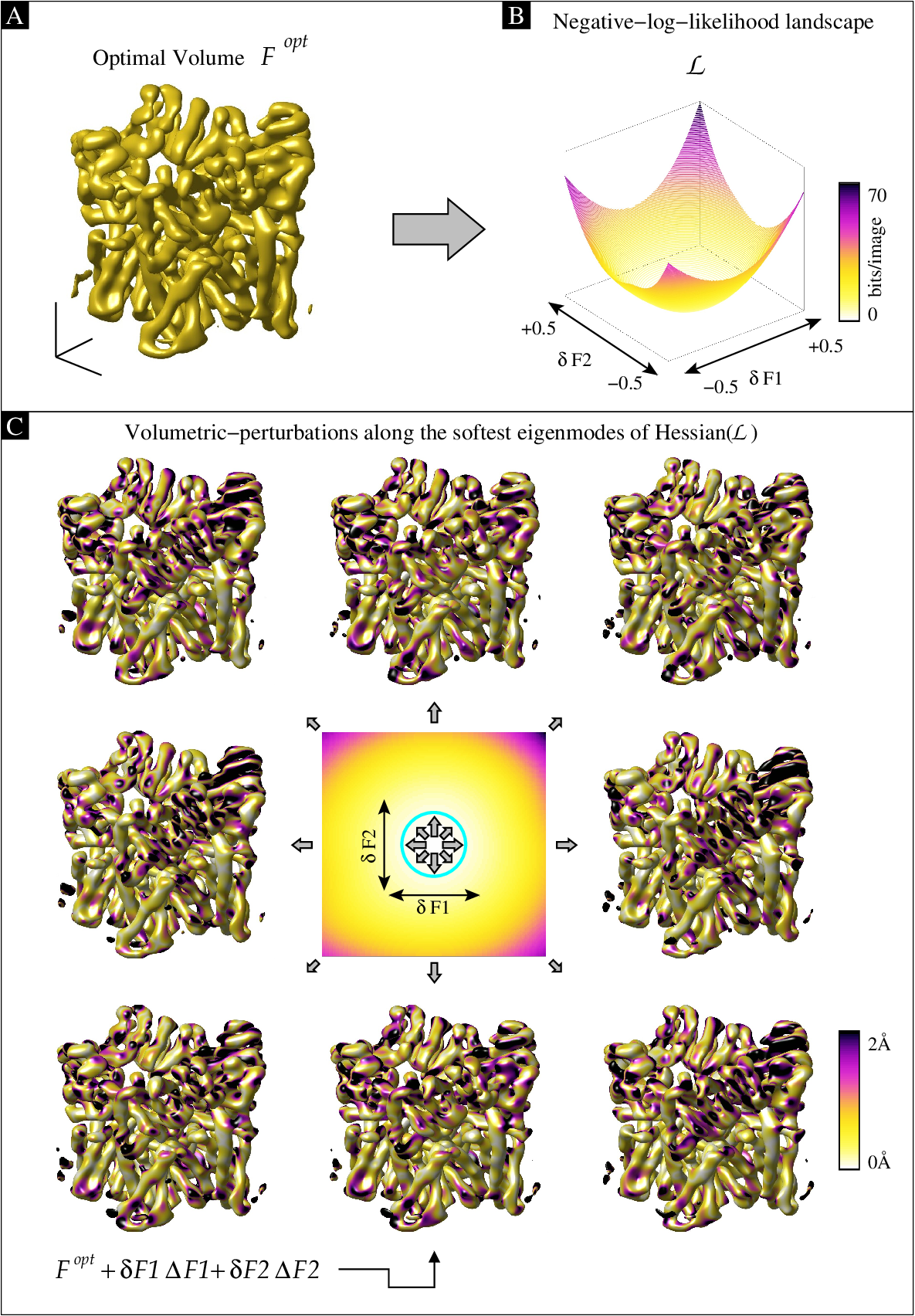}
\caption{
A likelihood landscape.
In panel-A we show a level-surface from a $4.5\angstrom$ reconstruction $\xF^{\opt}$ of the TRPV1 molecule from {\tt EMD-5778}.
In panel-B we show a subset of the graph of the negative-log-likelihood function $\mathcal{L}(\{\xA_{j}\}\givenmed \xF)$.
The two `flat' axes correspond to two independent directions in volume-space (with a unit scale equal to the l2-norm of $\xF^{\opt}$), while the height corresponds to the value of $\mathcal{L}$; lower values correspond to more likely volumes, with the minimum $\mathcal{L}$ corresponding to the most likely $\xF^{\opt}$, situated in the center of the graph.
Note that, close to the optimal $\xF^{\opt}$, the graph of $\mathcal{L}$ looks like a quadratic bowl.
In the center of panel-C we represent the same graph of $\mathcal{L}(\{\xA_{j}\}\givenmed \xF)$, this time as a heat-map, with white to purple indicating low to high values of $\mathcal{L}$, matching the graph from panel-B.
For each of the directions we illustrate a level-set of the perturbed volume $\xF^{\opt}+\Delta\xF$, with $\Delta\xF$ lying on the cyan circle (in the direction indicated).
Each of these level-sets is colored in accordance with the magnitude of its local perturbation from the original level-set shown in Fig \ref{fig_landscape_FIGAG0_0}A; yellow regions indicate little to no shift, while dark purple regions indicate a significant shift of $\sim 2\angstrom$in the direction of the surface-normal (see colorbar).
Note that, depending on the direction of $\Delta\xF$, different regions of the volume $\xF^{\opt}+\Delta\xF$ can be affected.
The deformations shown here correspond to the periphery of the $95\%$-credible-interval associated with the solution $\xF^{\opt}$ to the SPR-problem, given $J\gtrsim 2000$ images; i.e., these deformed volumes can't easily be ruled-out in favor of $\xF^{\opt}$ unless the number of images available for the reconstruction is $\gtrsim 2000$.
This $95\%$-credible-interal is nearly circular in this case, with a border shown by the cyan-circle in the center of panel-C.
}
\label{fig_landscape_FIGAG0_0}
\end{FPfigure}

As is typically the case, the graph of the negative-log-likelihood $\mathcal{L}(\{\xA_{j}\}\givenmed \xF)$ is high-dimensional; the family of input-volumes $\xF$ has many degrees of freedom.
In Fig \ref{fig_landscape_FIGAG0_0}B we show a 2-dimensional cross-section of the graph of $\mathcal{L}(\{\xA_{j}\}\givenmed \xF)$, with the xy-plane corresponding to two (orthonormal) directions in volume-space.
A heat-map of this same graph is shown in the center of Fig \ref{fig_landscape_FIGAG0_0}C, and the axes have been centered so that $\xF^{\opt}$ lies at the origin (i.e., in the center of the diagram).

As described above, the minimum of $\mathcal{L}(\{\xA_{j}\}\givenmed \xF)$ occurs at the center of the diagram: $\xF^{\opt}$ is a local maximum of $P(\{\xA_{j}\}\givenmed \xF)$.
Consequently, we expect $\mathcal{L}(\{\xA_{j}\}\givenmed \xF)$ to increase quadratically as we move away from $\xF^{\opt}$, at least for small perturbations $\delta\cdot\Delta \xF$ (i.e., if we consider nearby volumes $\xF^{\opt}+\delta\cdot\Delta \xF$).
Put another way, a cross-section of the graph of $\mathcal{L}(\{\xA_{j}\}\givenmed \xF)$ along a particular perturbation-direction $\Delta \xF$ will look like a parabola when we are close to $\xF^{\opt}$.
If the concavity of this parabola is high (i.e., the parabola is steep), then that perturbation-direction $\Delta \xF$ corresponds to perturbations that are very unlikely to have produced the images $\{\xA_{j}\}$.
These `hard' modes correspond to directions in volume-space that point to nearby volumes that can easily be `ruled-out' due to their mismatch to the data.
By contrast, if $\Delta \xF$ is chosen so that the concavity of the parabola is low (i.e., the parabola is shallow), then $\Delta \xF$ points to nearby volumes that can be reasonably well-matched to the data (i.e., are relatively likely to produce the images $\{\xA_{j}\}$).
These `soft' modes can't be ruled-out as easily.

The 2-dimensional cross-section of the graph of $\mathcal{L}(\{\xA_{j}\}\givenmed \xF)$ shown in Fig \ref{fig_landscape_FIGAG0_0}B is specifically chosen to illustrate the two `softest' modes for this SPR problem.
That is, we have used the local structure of $\mathcal{L}(\{\xA_{j}\}\givenmed \xF)$ to calculate the two (orthogonal) directions in volume space for which the corresponding volumetric perturbations are as difficult to rule-out as possible; the term `soft-mode' is borrowed from the molecular-dynamics literature \cite{LGSL2013}.
The softest-mode (associated with the smallest possible concavity) corresponds to the horizontal axis in Fig \ref{fig_landscape_FIGAG0_0}C, while the second-softest-mode corresponds to the vertical.

In Fig \ref{fig_landscape_FIGAG0_0}C we also illustrate the volumetric-perturbations associated with $\xF:=\xF^{\opt}+\delta\cdot\Delta \xF$ for $\Delta \xF$ along the eight directions in the plane shown.
These perturbations lie on the cyan circle around the origin of the likelihood-heatmap.
The images shown are level-surfaces of the perturbed volume $\xF$, with the surface color ranging from yellow to purple based on how much it moved from the original surface of $\xF^{\opt}$.

As we'll describe in the following sections, the features of the softest-modes, as well as their associated alignment-perturbations, can help quantify which regions of the reconstructed molecule are least/most trustworty (given the data available), and yield other insights into the reconstruction process.

{\bf Our approach:\ }
With the above illustrative example in mind, any candidate solution to an SPR problem takes the form of a `model' $\xF$ that optimizes the likelihood $P(\text{\small data}\givenmed\text{\small model})$.
Because the solution lies at a critical-point of $\mathcal{L}:=-\log P$, any perturbations to the solution will result in a shift to the overall likelihood which depends on the Hessian of $\mathcal{L}$.

While it is natural to consider the Hessian of $\mathcal{L}$, it may not be immediately clear how to calculate the spectrum of this object.
For example, the posterior likelihood $P(\text{\small data}\givenmed\text{\small model})$ is usually defined using a standard noise-model (e.g., independent gaussian noise applied to each pixel in each picked-particle image) \cite{Sigworth1998,Sigworth2010}.
In this scenario the likelihood $P(\text{\small data}\givenmed\text{\small model})$ is typically written as a product of the individual image-likelihoods $P(\xA_{j}\givenmed\text{\small model})$, each of which involves an accumulation of many exponential terms.
Taking a second-derivative of the overall negative-log-likelihood with respect to the model-parameters can sometimes be difficult, especially because the model can involve many degrees of freedom (e.g., the number of voxels in the volume $\xF(\vx)$ and the number of possible viewing-angles can both be large).

In this paper we make a simple observation.
Namely, the negative-log-likelihood $\mathcal{L}(\text{\small data}\givenmed\text{\small model})$ can be approximated as a single three-dimensional integral.
This simplification allows for the efficient calculation of $\mathcal{L}$, as well as its gradient and Hessian.
Importantly, this approximation is valid in two commonly encountered regimes: (i) the `low-noise' limit used when the noise-level is known and small, and (ii) the `high degree-of-freedom' limit used when the noise-level is unknown and possibly large.
The low-noise regime is often used in theoretical treatments of the SPR problem, while the high-degree-of-freedom limit is often seen when SPR is applied to real data-sets, as we demonstrate in Appendix section 20.

Using the simple observation above, we can approximate the spectrum of the Hessian of the negative-log-likelihood.
In particular, we can estimate the extremal eigenvalues and eigenvectors of the Hessian.
The eigenvectors corresponding to the smallest eigenvalues will then correspond to the `softest-modes' for the SPR problem.
I.e., the softest-mode indicates the volumetric- and viewing-angle-perturbations which are most easily confused for the current solution.
The softest-eigenvalue is a measure of the fisher-information restricted to the direction of this particular perturbation.

Calculation of the softest-modes allows for various kinds of sensitivity analysis.
For example, we can use the soft-subspace to determine which parts of the reconstructed volume are trustworthy, and which are unreliable.
We can also use this approach to determine the benefit of adding more images to the image-pool; often the information gained will depend on how those new images are aligned to the volume, as well as other image properties such as the defocus-level.
Additionally, when using cryo-EM data to estimate free-energy landscapes, we can use the spectrum of the hessian to help guide our choice of collective-variables.

To summarize, we believe that this kind of analysis can be used alongside more traditional strategies for sensitivity analysis, such as bootstrapping and crossvalidation.
For example, the structure of the soft-subspace can be a useful complement to the fourier-shell-correlation, which is often used to characterize the robustness of an ab-initio reconstruction \cite{Henderson2012,Penczek2014,Heymann2015,Rosenthal2015}.

This paper is structured as follows.
We first briefly review the mathematical details underlying our approach.
We then provide several examples illustrating different applications of this kind of analyis.
We conclude by discussing some extensions of this work.

\section{Methods Summary}
\label{sec_Methods_Summary}

In this section we briefly summarize our methodology, with the details deferred to the Appendix.

As described above, our methodology predominantly applies to the single-volume reconstruction problem.
To set the stage, let's assume that we are given a collection of $\nimage$ picked-particle images $\{\xA_{j}\}$.
To ease the presentation for now, we'll ignore the image-specific displacements $\{\vd_{j}\}$ and contrast-transfer-functions $\{CTF_{j}\}$.

For any particular image $\xA$ and volume $\xF$, we can define the likelihood: $P(\xA \givenmed \xF,\Ptauprior)$, which is the probability of observing the image $\xA$, given that the true signal is produced by the volume $\xF$.
In this expression the term $\Ptauprior$ is the prior-distribution of viewing-angles $\tau\in SO3$ associated with the data, and we assume that the image-generation model involves an image-noise $\sigma$ associated with each of the $n_{\dof}$ image degrees-of-freedom.

The setting we consider assumes that a single-particle reconstruction algorithm has already been run, and has produced a volume $\xF$ which optimizes the posterior likelihood of observing the collection of images $\{\xA_{j}\}$:
\begin{eqnarray}
P\left(\{\xA_{j}\}\givenbig \xF,\Ptauprior\right) = \prod_{j=1}^{\nimage} P\left(\xA_{j}\givenbig \xF,\Ptauprior\right) \period
\label{eq_main_P_AA_given_F}
\end{eqnarray}
For our purposes we'll perform most of our calculations in fourier-space, referring to $\fA_{j}$ and $\fF$ as the 2d- and 3d-fourier-transforms of $\xA_{j}$ and $\xF$, respectively, and using $\fsigma$ and $n_{\dof}$ to denote the image-noise and number of image degrees-of-freedom in fourier-space.

Under certain cirumstances, we can approximate the likelihood in Eq. \ref{eq_main_P_AA_given_F} with a so-called `image-aligned' limit:
\[ P\left(\{\fA_{j}\}\givenbig \fF,\{\tau^{\opt}_{j}\}\right) \propto \exp\left(-\zeta^{2}\cdot \cL\left( \{\fA_{j}\}\givenbig \fF,\{\tau^{\opt}_{j}\} \right)\right) \comma \]
where $\zeta^{2}$ is a regime-dependent prefactor and the negative-log-likelihood $\cL$ takes the form of a sum of $2$-dimensional integrals:
\begin{eqnarray}
\cL\left( \{\fA_{j}\}\givenbig \fF,\{\tau^{\opt}_{j}\} \right) \ \sim \ \frac{1}{2}\sum_{j=1}^{\nimage} \iint_{\Real^{2}} \left|\fA_{j}(\vk)-\fS\left(\vk;\tau^{\opt}_{j};\fF\right)\right|^{2} d\vk \period
\label{eq_main_P_AA_given_F_maximum_likelihood}
\end{eqnarray}
In this last expression we use $\fS(\vk;\tau;\fF)$ to denote the fourier-transform of a $2$-dimensional projection of any given volume $\fF$ along viewing-angle $\tau$, and denote by $\tau^{\opt}_{j}$ the `optimum' viewing-angle for image-$j$.
As we'll discuss below, the viewing-angle $\tau^{\opt}_{j}$ implicitly depends on $\fF$, and is often referred to as the `optimal-alignment' of $\fA_{j}$ to $\fF$.

This image-aligned limit holds in two closely related regimes.
The first is the `low-noise' limit, where the image-noise $\fsigma$ is known and small.
This is a regime that is commonly studied when analyzing the SPR-problem, and in this regime the prefactor $\zeta^{2} \sim 1/\fsigma^{2}$.
The image-aligned limit also often holds in a `high-degree-of-freedom' limit, where the image-noise $\fsigma$ is not known precisely, but is marginalized over when calculating the likelihood \cite{TSRC2024}.
In this more practical scenario, the prefactor $\zeta^{2} \sim \frac{1}{2}n_{\dof}\cdot p_{1}$, where $p_{1}$ is a regression-coefficient dependent on the data (as described in Appendix section 20).

The main advantage of Eq. \ref{eq_main_P_AA_given_F_maximum_likelihood} is that the image-aligned limit allows for the negative-log-likelihood to be rewritten as a single $3$-dimensional integral:
\begin{equation}
\cL\left( \{\fA_{j}\}\givenbig \fF,\{\tau^{\opt}_{j}\} \right) \ \sim \
\frac{1}{2}\iiint_{\Real^{3}} \left(M_{20}(\vkappa)\left|\fF(\vkappa)\right|^{2} - 2\Re\left(M_{11}^{\dag}(\vkappa)\fF(\vkappa)\right) + M_{02}(\vkappa)\right) \frac{d\vkappa}{\left|\vkappa\right|} \comma
\label{eq_main_ssnll_q3d_pre}
\end{equation}
where $M_{20}$, $M_{11}$,  $M_{02}$ are each $3$-dimensional functions corresponding to the various moments of the data, and the integral is performed over frequency-space $\vkappa\in\Real^{3}$.

Considering the image-pool $\{\fA_{j}\}$ to be fixed, we can consider the negative-log-likelihood $\cL$ to be a function of the volume $\fF$ and optimal-alignments $\{\tau^{\opt}_{j}\}$.
By using both Eq. \ref{eq_main_P_AA_given_F_maximum_likelihood} and \ref{eq_main_ssnll_q3d_pre}, we can take various derivatives of this log-likelihood.
For our purposes we will assume that, with the image-pool fixed, the volume and the optimal-alignments $\transpose{[\fF,\{\tau_{j}\}]}$ are situated at a local minimum of $\cL\left( \{\fA_{j}\}\givenbig \fF,\{\tau^{\opt}_{j}\} \right)$.
Consequently, we expect that perturbations to these quantities will produce second-order perturbations to $\cL$ of the form:
\begin{multline}
\cL\left(\{\fA_{j}\}\givenbig \fF + \Delta\fF,\{\tau^{\opt}_{j} + \Delta\tau_{j} \} \right)
= \\
\cL\left(\{\fA_{j}\}\givenbig \fF,\{\tau^{\opt}_{j}\} \right)
+
\frac{1}{2}
\cdot
\left[
\begin{array}{c}
\Delta\fF \\
\{\Delta\tau_{j}\} \\
\end{array}
\right]^{\dagger}
\cdot
\left[
\begin{array}{cc}
H_{FF} & H_{F\tau} \\
H_{\tau F} & H_{\tau\tau}
\end{array}
\right]
\cdot
\left[
\begin{array}{c}
\Delta\fF \\
\{\Delta\tau_{j}\} \\
\end{array}
\right]
\period
\label{eq_main_ddssnll_block}
\end{multline}
In this formula we use `$\dagger$' to denote the adjoint (i.e., conjugate-transpose).
We also use the terms $H_{FF}$, $H_{F\tau}$, $H_{\tau F}$ and $H_{\tau\tau}$ to refer to blocks of the second-derivative (i.e., Hessian) of $\cL$.
The diagonal-blocks $H_{FF}$ and $H_{\tau\tau}$ are trivial to calculate: the former is diagonal in fourier-space while the latter decomposes into small block-diagonal elements for each image.
The off-diagonal-blocks $H_{F\tau}$ and $H_{\tau F}$ require slightly more care.
In our calculation we use the volumetric-representation in Eq. \ref{eq_main_ssnll_q3d_pre} to calculate $H_{F\tau}$ for any particular choice of $\{\Delta\tau_{j}\}$, and we use the 2-dimensional representation in Eq. \ref{eq_main_P_AA_given_F_maximum_likelihood} to calculate $H_{\tau F}$ for any particular choice of $\Delta\fF$.

In most practical scenarios of interest the volume $\fF$ and optimal-alignments $\{\tau^{\opt}_{j}\}$ are implicitly related: the optimal-alignment $\tau^{\opt}_{j}$ for any particular image $\fA_{j}$ is the viewing-angle for which that image aligns best to $\fF$.
Thus, we can consider the $\tau^{\opt}_{j}$ each to be functions of $\fF$, and correspondingly interpret $\cL$ as a function of $\fF$ alone.

With this perspective the negative-log-likelihood takes the form:
\begin{eqnarray}
\cL\left(\{\fA_{j}\}\givenbig \fF + \Delta\fF \right)
=
\cL\left(\{\fA_{j}\}\givenbig \fF \right)
+
\frac{1}{2}
\cdot
\Delta\fF^{\dagger}
\cdot
\left[ H \right]
\cdot
\Delta\fF
\comma
\label{eq_main_ddssnll_block_implicit}
\end{eqnarray}
with $[H]$ representing the Hessian of $\cL\left(\{\fA_{j}\}\givenbig \fF \right)$ with respect to volumetric-perturbations of $\fF$.

With this formalism in place, we can immediately see that the influence of any perturbation $\Delta\fF$ on the likelihood is determined by the rayleigh-quotient.
That is to say, the log-likelihood-ratio 
\begin{eqnarray}
\Delta\cL := \cL\left(\{\fA_{j}\}\givenbig \fF + \Delta\fF \right) - \cL\left(\{\fA_{j}\}\givenbig \fF \right) = 
\frac{1}{2}
\cdot
\Delta\fF^{\dagger}
\cdot
\left[ H \right]
\cdot
\Delta\fF
\comma
\label{eq_main_nllr}
\end{eqnarray}
contrasting the original volume $\fF$ with the perturbed volume $\fF + \Delta\fF$ is given by the rayleigh-quotient $\Delta\fF^{\dagger}\cdot [H]\cdot \Delta\fF$.
Thus, given any volumetric-perturbation $\Delta\fF$, the rayleigh-quotient associated with that $\Delta\fF$ can be thought of as a measure of `information' in that direction.

More generally, the critical-points of the rayleigh-quotient correspond to the eigenvectors of $[H]$.
Eigenvectors corresponding to the largest eigenvalues correspond to volumetric-perturbations $\Delta\fF$ for which the nearby volumes $\fF\pm\Delta\fF$ are most easily distinguishable from the original volume $\fF$.
I.e., these `stiff-modes' correspond to volumetric-perturbations which produce a significant mismatch between the projections of the perturbed volume and the image-pool.
On the other hand, the eigenvectors corresponding to the smallest eigenvalues indicate volumetric-perturbations for which the nearby volumes are \textit{least} distinguishable from the original $\fF$.
I.e., these `soft-modes' correspond to volumetric-perturbations which produce only subtle shifts to the relevant projections of the perturbed volume; shifts that might still allow for plausible alignment to the images.

With this in mind, we are motivated to estimate the spectrum of $[H]$ from Eq. \ref{eq_main_ddssnll_block_implicit}, specifically the lower-tail of the spectrum, comprising the smallest eigenvalues as well as their corresponding eigenvectors (i.e., the softest-modes).
In terms of computational details, the Hessian $[H]$ involves an accumulation of terms combining the various blocks $H_{FF}$, $H_{F\tau}$, $H_{\tau F}$ and $H_{\tau\tau}$ from Eq. \ref{eq_main_ddssnll_block} with the volumetric-perturbations $\Delta\fF$ and the (implicitly-determined) alignment-perturbations $\{\Delta\tau_{j}\}$.
As a result, the $[H]$ from Eq. \ref{eq_main_ddssnll_block_implicit} is not quite as easy to diagonalize as the $H_{FF}$ from Eq. \ref{eq_main_ddssnll_block}.
However, we can still compute quantities such as $[H]\cdot\Delta\fF$ relatively easily, allowing us to use iterative strategies such as lanczos-iteration to estimate the extreme eigenvalues and eigenvectors of $[H]$.
During this calculation we make sure to focus only on true deformations, projecting away any combinations of volumetric- and alignment-perturbations $[\Delta\fF,\{\Delta\tau_{j}\}]$ which correspond to trivial rigid rotations.

As mentioned above, the details of this calculation are provided in the Appendix.
Importantly, for our calculation, the soft-modes (as well as the rest of the spectrum of the Hessian) depend on the volume $\fF$ and the image-pool $\{\fA_{j}\}$, as well as on the optimal-alignments for the images (which are determined implicitly).
More generally, the soft-modes also depend on the image-specific CTF-functions $\{CTF_{j}\}$ (see applications below) as well as other image-parameters, such as the image-specific displacements $\{\vd_{j}\}$ (mentioned in the discussion).

\section{Applications}
\label{sec_Applications}

As a straightforward application, one can simply inspect the softest-modes of a given single-particle reconstruction (SPR) problem.
The softest-modes can then be used to directly estimate how robust that reconstruction might be.

\subsection{Real data: sensitivity of reconstructed maps}

Two illustrations of this approach are shown in Fig \ref{fig_p_reco_empi_FIGA_0}.
The first example in Fig \ref{fig_p_reco_empi_FIGA_0}A,C  involves a $10\angstrom$ resolution reconstruction of trpv1.
This reconstruction was carried out using using several thousand images from {\tt EMPIAR-10003}, as well as their corresponding contrast-transfer-functions (CTFs).
The optimal reconstruction $\xF^{\opt}$, shown in Fig \ref{fig_p_reco_empi_FIGA_0}A, is reasonably robust, but there are certain portions of the reconstruction that are more trustworty than others.
To quantify this effect, we calculate the softest-mode of the Hessian of the negative-log-likelihood for this SPR problem.
In terms of technical details, we carry out this calculation after marginalizing over the noise and appealing to the high-degree-of-freedom limit (see Appendix section 20).

This softest-mode, labeled $\Delta\xF_{1}$, is illustrated in Fig \ref{fig_p_reco_empi_FIGA_0}C.
On the left of Fig \ref{fig_p_reco_empi_FIGA_0}C we show two level-sets for the volumetric-representation of $\Delta\xF_{1}(\vx)$.
As mentioned above, the 3-dimensional function $\Delta\xF_{1}(\vx)$ isn't the entire story, and the softest-mode carries with it an implicitly-defined alignment-perturbation, which we denote as $\Delta\tau(\Delta\xF_{1})$, shown in the middle of Fig \ref{fig_p_reco_empi_FIGA_0}C.
Together, the volumetric- and alignment-perturbations associated with $\Delta\xF_{1}$ point towards `nearby' solutions to the SPR-problem which are almost as likely as the optimal $\xF^{\opt}$.
Exactly how likely or unlikely these nearby solutions might be depends on the magnitude of the `softest-eigenvalue' associated with this softest-mode.

For this particular SPR-problem, the softest-eigenvalue is of moderate size, corresponding to roughly $1500$ bits/image.
To give a sense of what this means, we present a volumetric-perturbation of the form $\xF^{\opt}+\frac{1}{8}\Delta\xF_{1}$ on the right of Fig \ref{fig_p_reco_empi_FIGA_0}C.
Given the magnitude of the softest-eigenvalue for this SPR-problem, the choice of $\delta\xF_{1}=1/8$ places the perturbed volume $\xF^{\opt}+\delta\xF_{1}\Delta\xF_{1}$ onto the boundary of the $95\%$-credible-interval associated with the trpv1-reconstruction $\xF^{\opt}$ shown in Fig \ref{fig_p_reco_empi_FIGA_0}A, given $J\gtrsim 1000$ images; i.e., this deformed volume can't easily be ruled-out in favor of $\xF^{\opt}$ unless the number of images available for the reconstruction is $\gtrsim 1000$.

As a point of contrast, we can consider a similar calculation, carried out on a $15\angstrom$ resolution reconstruction of rib80s (the Plasmodium falciparum 80S ribosome bound to emetin) created using several thousand images with their associated contrast-transfer-functions from {\tt EMPIAR-10028} (see Fig \Ref{fig_p_reco_empi_FIGA_0}B and Fig \ref{fig_p_reco_empi_FIGA_0}D).
The magnitude of the softest-eigenvalue for the rib80s reconstruction is roughly $18500$ bits/image, which is quite a bit larger than the softest-eigenvalue for the trpv1 reconstruction in Fig \Ref{fig_p_reco_empi_FIGA_0}A,C.
This implies that the SPR-problem for rib80s is better posed, and that the results are more reliable.
Quantitatively, this means that the perturbed volume $\xF^{\opt}+\delta\xF_{1}\Delta\xF_{1}$ shown on the right of Fig \ref{fig_p_reco_empi_FIGA_0}D, which again has $\delta\xF_{1}=\frac{1}{8}$, places the perturbed volume onto the boundary of the $95\%$-credible-interval associated with the rib80s-reconstruction shown in Fig \ref{fig_p_reco_empi_FIGA_0}B, given only $J\gtrsim 100$ images.
In other words, a deformation of equivalent relative-magnitude to that shown in Fig \ref{fig_p_reco_empi_FIGA_0}C can already be ruled-out when the number of images available for the rib80s reconstruction is only in the hundreds.

Note that, in both cases, the structure of the softest-mode can indicate which portions of the reconstructed volume and which image-alignments are more trustworty (yellow) than others (purple).

\begin{FPfigure}
\centering
\includegraphics[width=1.0\textwidth,trim=000 000 000 000,clip]{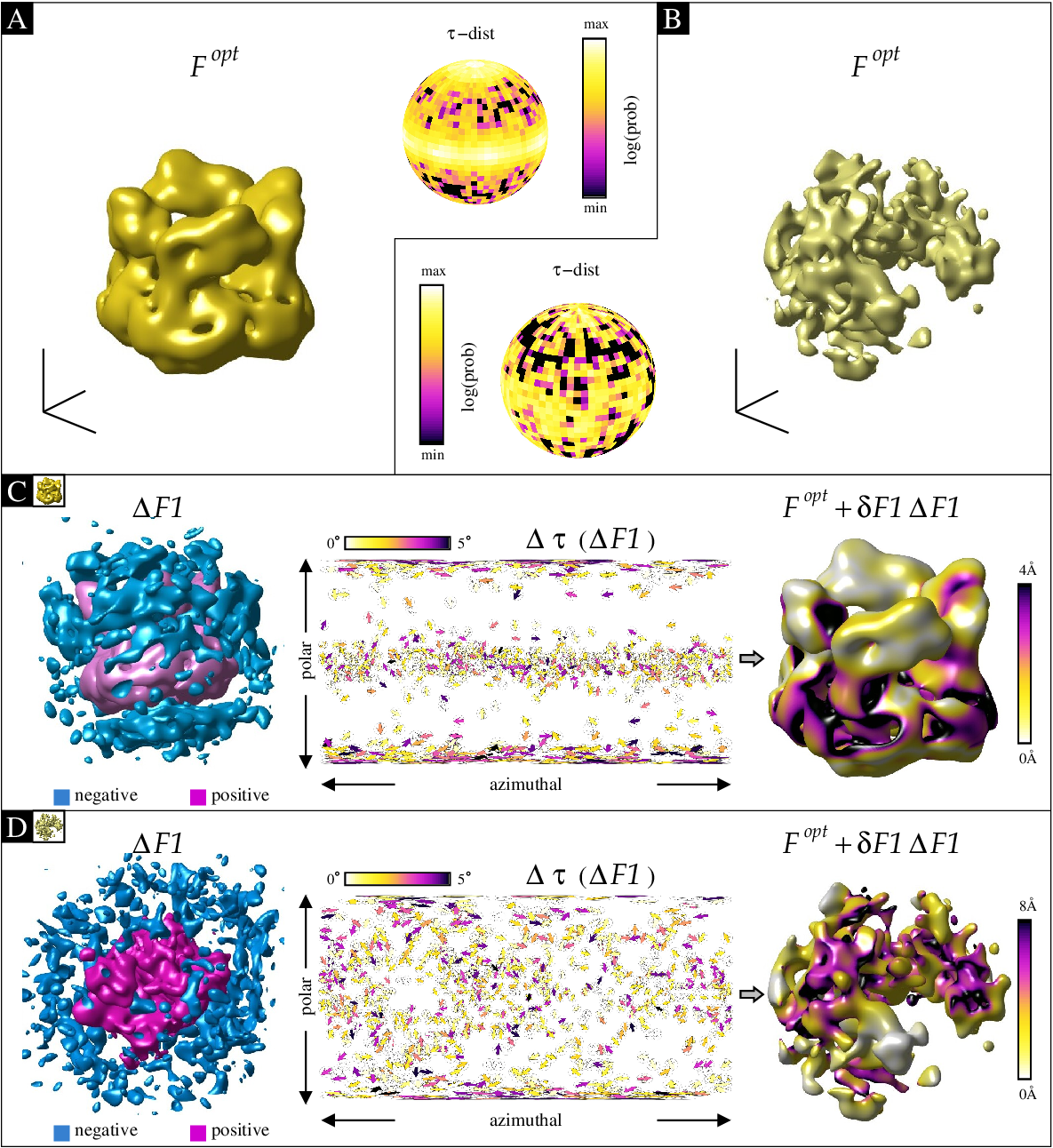}
\caption{
Illustrations of the softest-modes for two single-particle-reconstructions.
In panel-A we show a level-set from a $10\angstrom$ resolution reconstruction of trpv1 created using several thousand images from {\tt EMPIAR-10003}, as well as the corresponding viewing-angle distribution for the images used.
In panel-C we illustrate features of the softest-mode $\Delta\xF_{1}$ of the Hessian of the negative-log-likelihood for this reconstruction.
On the far left we show two level-sets for the volumetric-representation of this softest-mode $\Delta\xF_{1}(\vx)$, with negative- and positive-zones colored blue and pink, respectively.
On the far right we show a level-set of a perturbed volume $\xF^{\opt} + \delta\xF_{1}\Delta\xF_{1}$, with perturbation-magnitude $\delta\xF_{1}=\frac{1}{8}$.
The level-set of the perturbed-volume has its surface colored in accordance with the magnitude of its local perturbation from the original level-set shown in panel-A; yellow regions indicate little to no shift, while dark purple regions indicate a significant shift of $\sim 4\angstrom$in the direction of the surface-normal (see colorbar).
In the center we show the aligment-perturbation $\Delta\tau$ corresponding to $\Delta\xF_{1}$ (i.e., implicitly determined by the volumetric-features of $\Delta\xF_{1}$).
We represent the perturbations for the image-alignments as follows: each image is depicted as an arrow in the plane.
The location of each arrow indicates the optimal projection-direction of that image as aligned to $\xF^{\opt}$, with polar- and azimuthal-angle corresponding to the vertical and horizontal axes, respectively.
Meanwhile, each arrow's direction indicates the perturbation to that alignment (as re-aligned to $\xF^{\opt}+\delta\xF_{1}\Delta\xF_{1}$).
The color of each arrow corresponds to the magnitude of the perturbation; yellow arrows indicate little to no shift, while dark purple arrows indicate a significant shift to that image's alignment as a result of the volumetric-perturbation.
Regions with no arrows correspond to projection-directions that either had no corresponding images in the data-set, or for which the alignment-perturbation had a relative magnitude $<10^{-4}$.
Note that the structure of this softest-mode can indicate which portions of the reconstructed volume and which image-alignments are more trustworty (yellow) than others (purple).
Panel-B and panel-D illustrate a similar calculation, carried out on a $15\angstrom$ resolution reconstruction of rib80s created using several thousand images from {\tt EMPIAR-10028}.
}
\label{fig_p_reco_empi_FIGA_0}
\end{FPfigure}

\subsection{Sensitivity of reconstructed maps to the viewing-angle distribution}
\label{sec_Applications_trpv1}

As indicated in Fig \ref{fig_landscape_FIGAG0_0} and Fig \ref{fig_p_reco_empi_FIGA_0}, the structure of the likelihood-landscape for different SPR-problems will be a function of the SPR problem-parameters.
Typically, the softest-modes -- or more generally the `softest-subspace' -- will depend strongly on the viewing-angle distribution of the images used for the reconstruction.

To illustrate this, we conduct a simple case-study again involving the trpv1-molecule, shown in Fig \ref{fig_trpv1_x0_zoom_ssnll_k80_taudist_FIGA_0}A.
To focus on the effects of viewing-angle-distribution, we consider an idealized scenario involving synthetic-images, each with unit CTF.
With this idealization, we consider three different SPR-problems, corresponding to three different distributions of viewing-angle (referred to as the $\tau$-distribution in Fig \ref{fig_trpv1_x0_zoom_ssnll_k80_taudist_FIGA_0}B).
The first of these distributions (top) is concentrated towards the north- and south-poles; the middle distribution is concentrated towards the equator; the final distribution (bottom) is uniform.

For each of these viewing-angle distributions we can calculate the lower-tail of the spectrum of the Hessian of the associated negative-log-likelihood, as shown in the bar-graph of Fig \ref{fig_trpv1_x0_zoom_ssnll_k80_taudist_FIGA_0}B.
We can immediately see that, for this case-study, the polar-distributed SPR-problem is much more poorly conditioned than the equatorially- and uniformly-distributed SPR-problems.
In addition to having a lower value for its softest-eigenvalue, the polar-distributed SPR-problem actually has many soft-modes (corresponding to a soft-subspace of dimension $3-6$); by contrast, the equatorially- and uniformly-distributed SPR-problems are far more robust.

The two softest-modes for the polar-distributed SPR-problem are shown in Fig \ref{fig_trpv1_x0_zoom_ssnll_k80_taudist_FIGA_0}C, while the softest-mode for the equatorially- and uniformly-distributed SPR-problems are shown in Fig \ref{fig_trpv1_x0_zoom_ssnll_k80_taudist_FIGA_0} D,E.
In each case the deformation shown involves a perturbation with magnitude $\delta\xF_{1}$ or $\delta\xF_{2}$ equal to $\frac{1}{8}$.

As we've done above, we can use the magnitude of the soft-eigenvalues to quantify how sensitive these SPR-problems are.
The softest-deformation shown in Fig \ref{fig_trpv1_x0_zoom_ssnll_k80_taudist_FIGA_0}C lies on the $95\%$-credible-interval for the polar-distributed SPR-problem given $J\gtrsim 500K$ images, while the second-softest deformation lies on the $95\%$-credible-interval given $J\gtrsim 250K$ images.
This soft-subspace reveals deformations that are quite difficult to rule-out unless many images are used in the reconstruction.
By contrast, the softest-deformations shown in Fig \ref{fig_trpv1_x0_zoom_ssnll_k80_taudist_FIGA_0}D,E lie on the $95\%$-credible-interval for the equatorially- and uniformly-distributed SPR-problems, given only $J\gtrsim 5K$ images.
That is to say, these softest-deformations (applied with equivalent relative-magnitude to those shown in \ref{fig_trpv1_x0_zoom_ssnll_k80_taudist_FIGA_0}C) can be easily ruled-out with only five thousand images.

A similar calculation involving the ribosome is shown in Supplemental Figure \ref{fig_rib80s_x0_zoom_ssnll_k80_taudist_FIGA_0}.

\begin{FPfigure}
\centering
\includegraphics[width=0.8\textwidth,trim=000 000 000 000,clip]{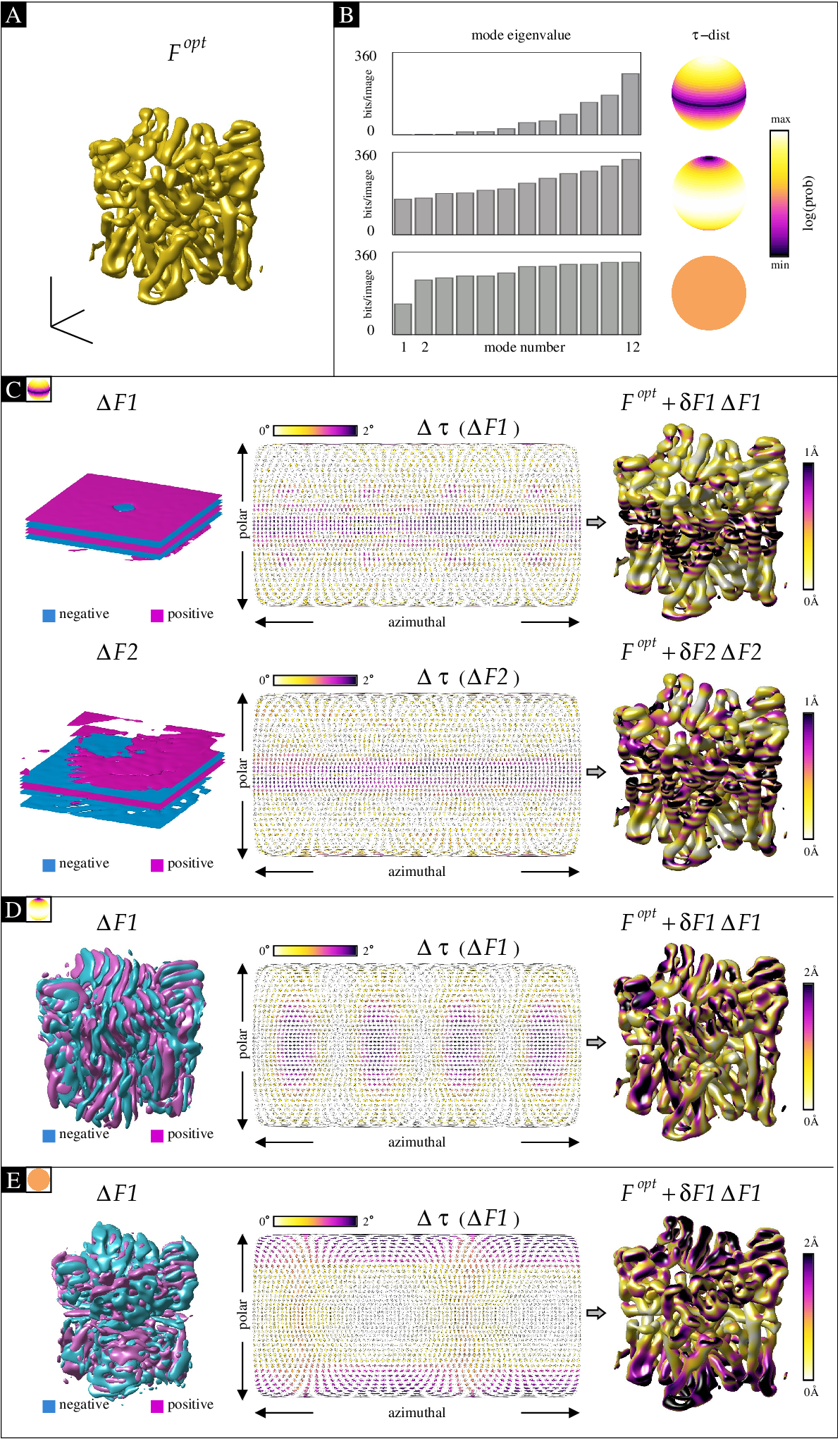}
\caption{
Illustrations of the softest-subspace for three SPR-problems with different viewing-angle distributions.
In panel-A we show a level-set from a $4.5\angstrom$ resolution reconstruction of trpv1 corresponding to the {\tt EMD-5778} reconstruction taken from {\tt EMPIAR-10003}.
We consider three idealized viewing-angle distributions, shown on the left in panel-B.
These distributions are, respectively, (i) concentrated towards the poles, (ii) concentrated towards the equator, and (iii) uniform.
For each of these viewing-angle-distributions we calculate the smallest-eigenvalues of the Hessian of the negative-log-likelihood for the associated SPR-problem, assuming synthetic-images and a unit CTF.
The lower-tail of the resulting spectrum is shown in panel-B.
Note that the polar-distribution yields many small eigenvalues (i.e., a softest-subspace of dimension $\sim 3-6$), indicating a poorly-conditioned SPR-problem.
By contrast, the equatorial- and uniform-distributions have much larger eigenvalues, indicating a much more robust SPR-problem.
Panel-C illustrates the two softest-modes $\Delta\xF_{1}$ and $\Delta\xF_{2}$ for the polar-distributed SPR-problem, using the same format as Fig \ref{fig_p_reco_empi_FIGA_0}C,D.
Panels D,E show the softest-mode for the equatorially- and uniformly-distributed SPR-problems.
In each case the magnitude of $\delta\xF_{1}$,$\delta\xF_{2}$ is taken to be $\frac{1}{8}$.
Note that, as discussed in the main text, the different viewing-angle distributions lead to different sensitivity-maps.
}
\label{fig_trpv1_x0_zoom_ssnll_k80_taudist_FIGA_0}
\end{FPfigure}

\begin{FPfigure}
\centering
\includegraphics[width=0.8\textwidth,trim=000 000 000 000,clip]{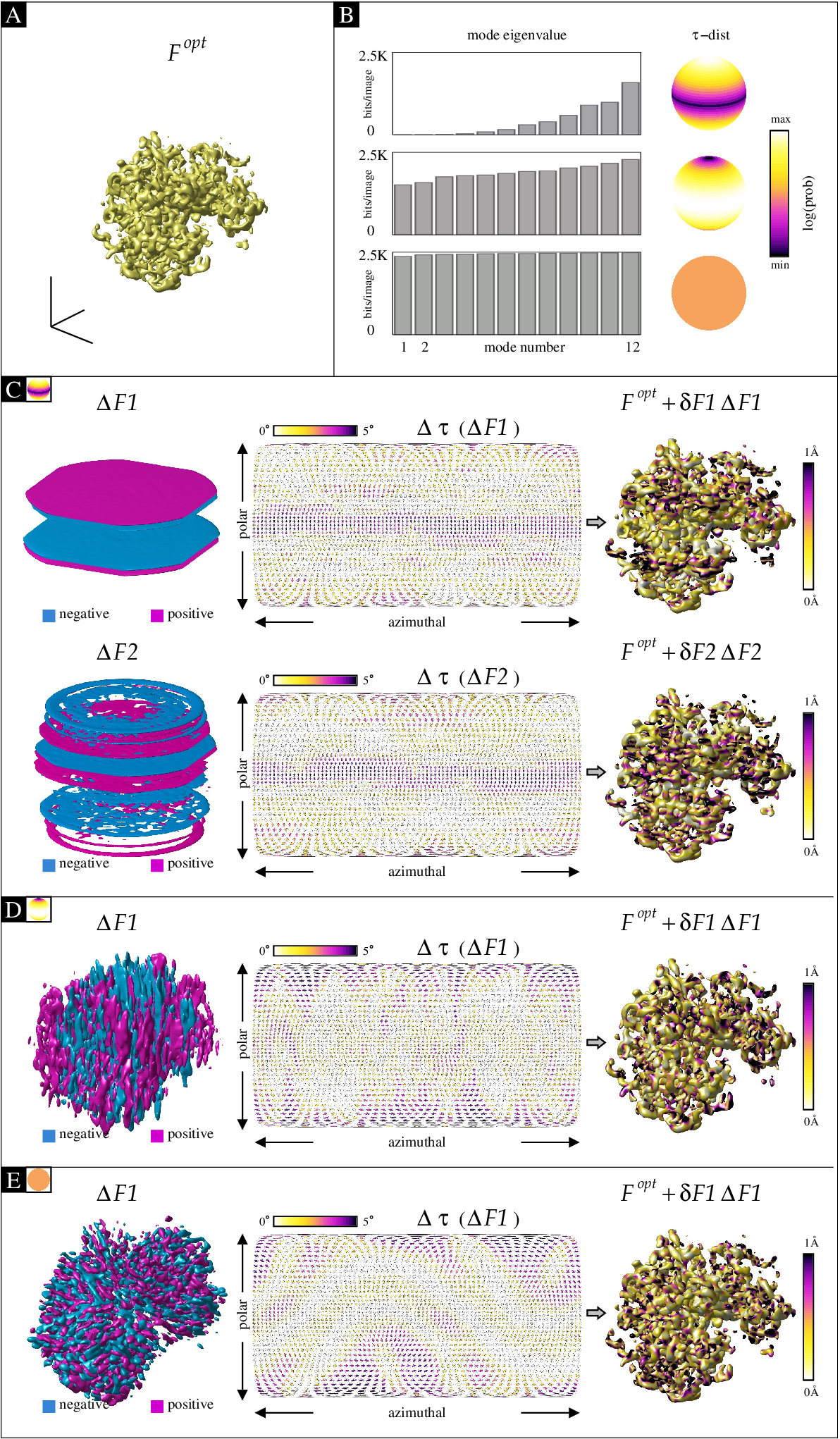}
\caption{
Supplementary Figure 1: Illustrations of the softest-subspace for three SPR-problems involving the ribosome with different viewing-angle distributions.
In panel-A we show a level-set from an $8\angstrom$ resolution reconstruction of rib80s corresponding to the {\tt EMD-2660} reconstruction taken from {\tt EMPIAR-10028}.
The format of this figure is similar to Fig \ref{fig_trpv1_x0_zoom_ssnll_k80_taudist_FIGA_0}.
}
\label{fig_rib80s_x0_zoom_ssnll_k80_taudist_FIGA_0}
\end{FPfigure}

\subsection{Sensitivity of reconstructed maps to the defocus}
\label{sec_Applications_effect_of_defocus}

In the previous section we used the softest-modes to illustrate how changes in the viewing-angle distribution might impact the sensitivity of single-particle-reconstruction.
A similar phenomenon also applies to changes in defocus.
To illustrate this, we use an $8.5\angstrom$ resolution reconstruction of TMEM16F (a Calcium-activated lipid scramblase and ion channel in digitonin with calcium bound, taken from {\tt emd-20244}), shown in Fig \ref{fig_DefV_FIGA_0}A.

Similar to Fig \ref{fig_trpv1_x0_zoom_ssnll_k80_taudist_FIGA_0}, we'll use synthetic-images to focus on the effects of defocus.
However, this time we'll use three of the typical CTF-functions drawn from the data-set at {\tt EMPIAR-10278}.
These CTF-functions involve defocus-values of $11789\angstrom$, $24141\angstrom$ and $25366\angstrom$, respectively, shown in Fig \ref{fig_DefV_FIGA_0}B.

As before, we calculate the tail of the spectrum of the Hessian for these SPR-problems, which are shown in shown in Fig \ref{fig_DefV_FIGA_0}B.
In this case the overall structure of the softest-mode (shown in Fig \ref{fig_DefV_FIGA_0}C,D,E) does not change too dramatically as the defocus increases, but the softest-eigenvalue decreases significantly.
As a consequence, we expect this particular SPR-problem to be more poorly-conditioned for the higher level of defocus.
In terms of quantifying this effect, we can use the softest-eigenvalues to conclude that the deformation shown in Fig \ref{fig_DefV_FIGA_0}E lies on the $95\%$-credible-interval for the high-defocus SPR-problem given $J\gtrsim 50K$ images, while the deformation of equivalent relative-magnitude shown in Fig \ref{fig_DefV_FIGA_0}C lies on the $95\%$-credible-interval for the low-defocus SPR-problm given only $J\gtrsim 20K$ images.
I.e., the deformation in Fig \ref{fig_DefV_FIGA_0}C is easier to rule-out than the equivalently-sized deformation in Fig \ref{fig_DefV_FIGA_0}E.

\begin{FPfigure}
\centering
\includegraphics[width=0.8\textwidth,trim=000 000 000 000,clip]{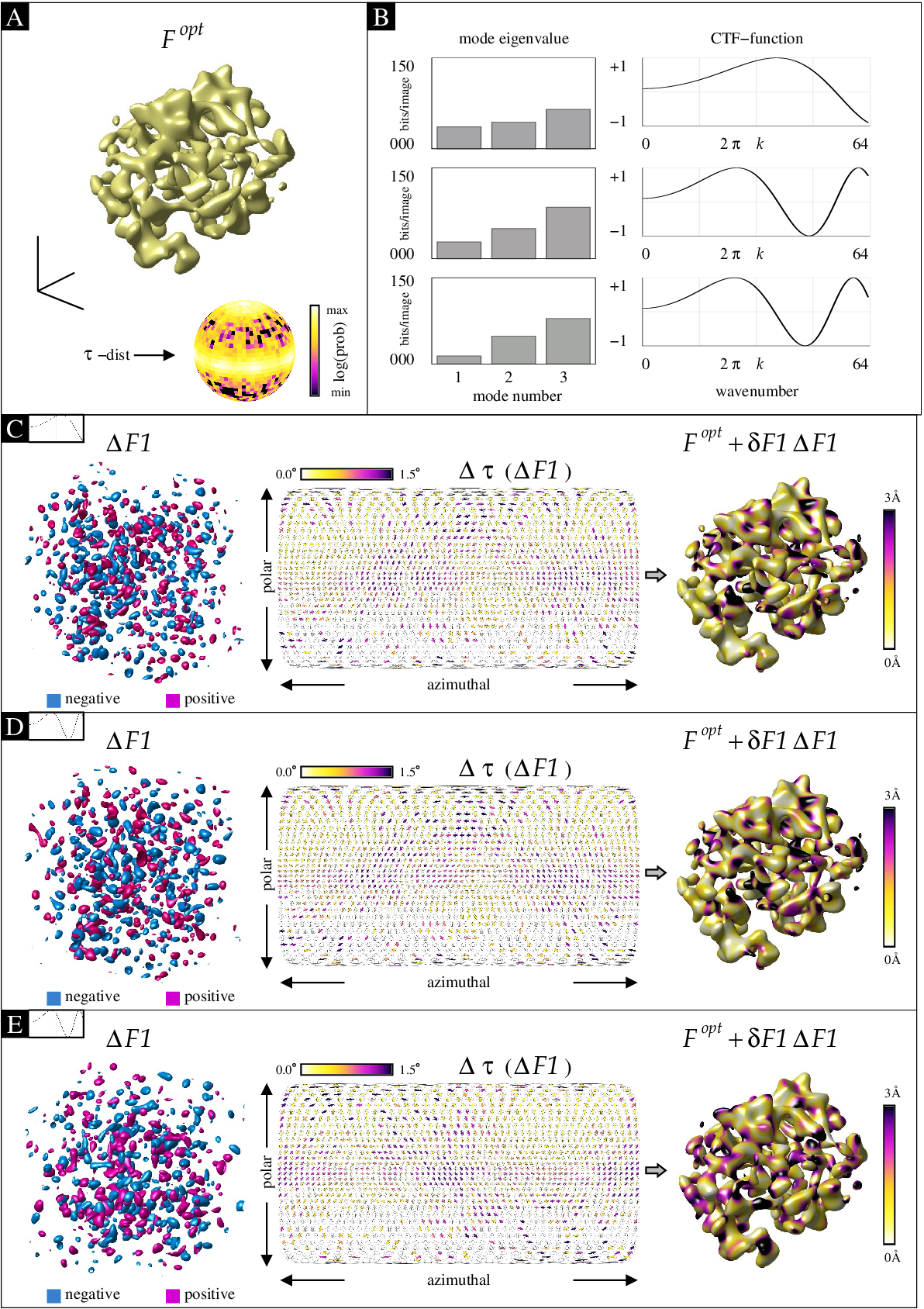}
\caption{
Illustrations of the softest-subspace for three SPR-problems with different levels of defocus.
In panel-A we show a level-set from an $8.5\angstrom$ resolution reconstruction of TMEM16F corresponding to the {\tt EMD-20244} reconstruction taken from {\tt EMPIAR-10278}.
We consider three CTF-functions with defocus-values of $11789\angstrom$, $24141\angstrom$ and $25366\angstrom$, shown on the left in panel-B.
For each of these defocus-values, we calculate the smallest-eigenvalues of the Hessian of the negative-log-likelihood for the associated SPR-problem, assuming synthetic-images drawn from the empirical distribution shown in panel-A.
The lower-tail of the resulting spectrum is shown in panel-B.
Note that the softest-eigenvalue is different for these different defocus-values, indicating changes to the conditioning of the associated SPR-problem.
Panels C,D and E illustrate the softest-modes $\Delta\xF_{1}$ for the different SPR-problems, using the same format as Fig \ref{fig_p_reco_empi_FIGA_0}C,D.
In each case the magnitude of $\delta\xF_{1}$ is taken to be $\frac{1}{8}$.
Note that, as discussed in the main text, the different defocus-values lead to different levels of sensitivity.
}
\label{fig_DefV_FIGA_0}
\end{FPfigure}

\subsection{Sensitivity analysis to select paths in volume-space}
\label{sec_Applications_paths_in_volume_space}

As demonstrated above, the structure of the softest-subspace depends on the viewing-angle distribution, as well as on imaging parameters such as the defocus.
The softest-subspace also obviously depends on the particular volume which is provided as a solution to the single-particle-reconstruction problem.
This dependence can be of interest in certain scenarios, such as free-energy estimation \cite{GBetal2021}.

\begin{figure}[H]
\centering
\includegraphics[width=1.0\textwidth]{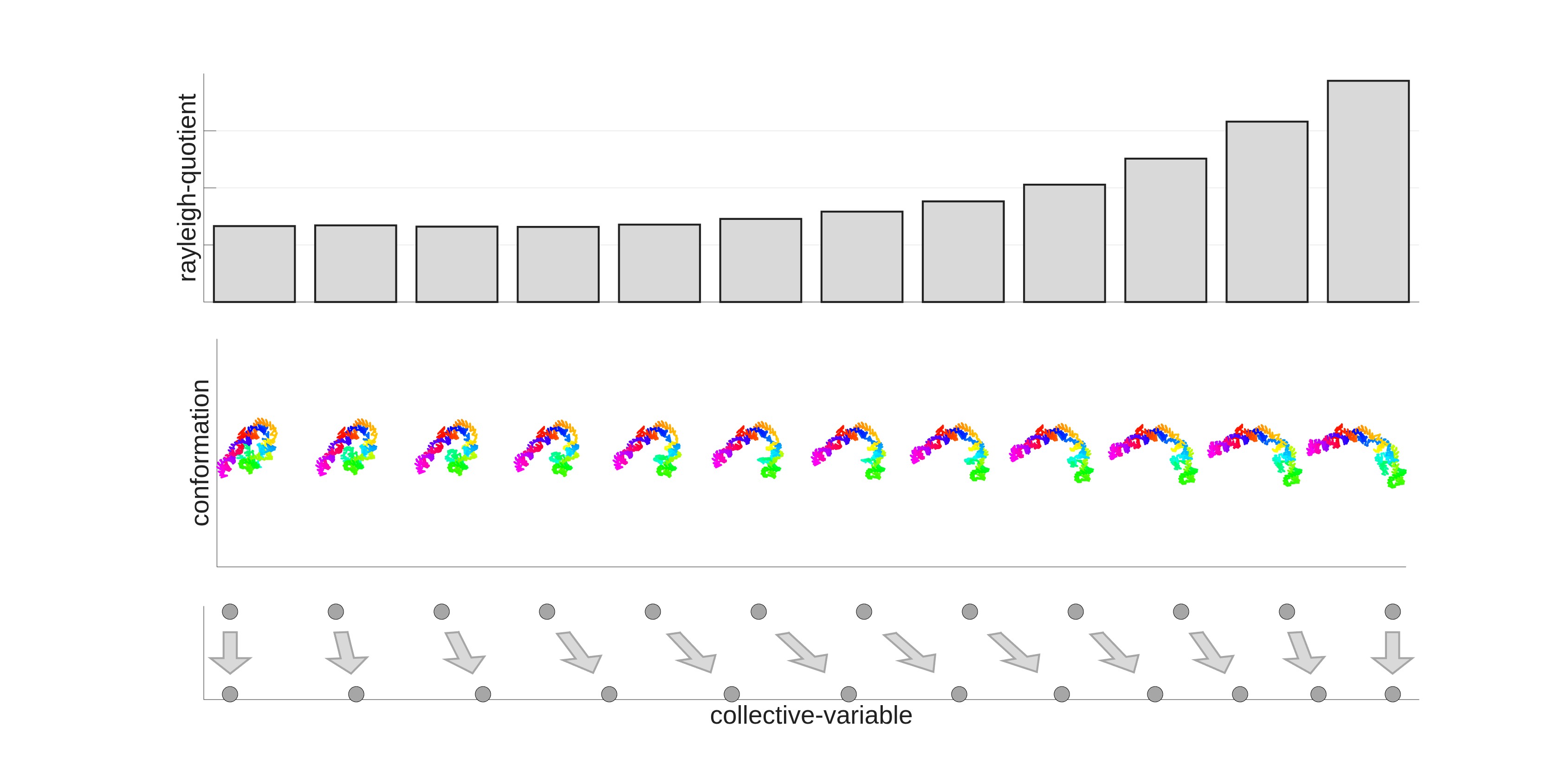}
\caption{
In this figure we illustrate the rayleigh-quotient $\cR:=[\partial_{s}\fF(\cdot;s)]^{\dagger}[H(s)][\partial_{s}\fF(\cdot;s)]$ (top) along the conformational path corresponding to the sequence of volumes shown on the bottom.
The path shown here connects an `open' RNA molecular conformation (left) to a `closed' configuration (right).
The calculation of $\cR$ was performed at $10.7\angstrom$, assuming that the viewing-angle-distribution is uniform (for each conformation), and that the CTF is identically $1$ (although these assumptions can easily be relaxed).
In these calculations we also assume that (i) the values of $\Delta\tau(\tau)$ are determined implicitly, with (ii) $\|\partial_{s}\xF(\vx,s)\|\equiv 1$ at each point along the path.
Portions of the path for which $\cR$ is low correspond to zones where locally resolving the free-energy will be more difficult, due simply to the fact that other nearby conformations produce similar projections.
Conversely, portions of the path for which $\cR$ is high correspond to zones where locally resolving the free-energy will be be more robust, with less opportunity for confusion.
One possible strategy to address these differences in variability might be to adjust the collective-variable to create larger $s$-neighborhoods of each representative $\xF(\cdot;s)$ when the corresponding $\cR$ is smaller, and vice-versa (see bottom row).
}
\label{fig_RNA0_ssnll_k32_ddssnll_lsq_vs_Atom_v_FIGC2_stripped.jpg}
\end{figure}

To illustrate this phenomenon, we consider a one-dimensional family of conformations for an RNA-molecule, shown in Fig \ref{fig_RNA0_ssnll_k32_ddssnll_lsq_vs_Atom_v_FIGC2_stripped.jpg}.
This family of conformations corresponds to a particular path of interest in volume-space, and can be parametrized by a `collective variable' $s$.
As described in \cite{GBetal2021}, it is in principle possible to use cryo-EM data to estimate the free-energy density $G(s)$ along this path.
This kind of free-energy estimate involves calculating, for each volume $\xF(\vx;s)$ along the path, the likelihood that the image-pool might produce that particular volume.
Those values of $s$ for which $\xF(\vx;s)$ is more likely will correspond to lower values of $G(s)$.

When performing this calculation, it is important to consider the interplay between the path-tangent $\partial_{s}\xF(\vx;s)$ (i.e., the direction in volume-space corresponding to perturbations in the path-variable $s$) and the softest-modes of the Hessian $H(s)$ of the likelihood at each location along the path.
Locations $s$ where the path-tangent $\partial_{s}\xF(\vx;s)$ lies along a stiff-mode of the Hessian $H(s)$ correspond, locally, to locations along the path that are highly discriminable from their neighbors (given the image-pool).
In other words, these are locations where projections of $\xF(\vx;s)$ are not easily confused for projections of nearby volumes $\xF(\vx;s\pm\Delta s)$.
Conversely, locations where the path-tangent $\partial_{s}\xF(\vx;s)$ lies parallel to a soft-mode of the Hessian $H(s)$ correspond, locally, to locations along the path that are easily confused with their neighbors.
In other words, these are locations where projections of $\xF(\vx;s)$ are quite similar to projections of nearby volumes $\xF(\vx;s\pm\Delta s)$.

Thus, the reliability of any free-energy estimate $G(s)$ will be influenced by the rayleigh-quotient:
\[ \cR(s) : = [\partial_{s}\fF(\cdot;s)]^{\dagger}[H(s)][\partial_{s}\fF(\cdot;s)] \period \]
Locations along the path for which $\cR(s)$ is small will correspond to locations where the estimate of $G(s)$ is inherently unreliable, simply due to the poor conditioning of the likelihood function.
Estimates of $\cR(s)$ are shown in Fig \ref{fig_RNA0_ssnll_k32_ddssnll_lsq_vs_Atom_v_FIGC2_stripped.jpg}.

For the collective-variable shown in Fig \ref{fig_RNA0_ssnll_k32_ddssnll_lsq_vs_Atom_v_FIGC2_stripped.jpg}, the values of $\cR(s)$ are smallest when the RNA-molecule is `closed'; i.e., there are many projections along which neighboring closed configurations seem quite similar to one another.
By contrast, the values of $\cR(s)$ are largest when the RNA-molecule is `open'; i.e., there are comparatively fewer projections along which neighboring open configurations seem similar.
These differences in $\cR(s)$ can be important when interpreting free-energy estimates.
Moreover, it may be worth considering $\cR(s)$ when designing a collective variable.
By ensuring that the path-tangent travels along stiff-modes of the Hessian of the likelihood-function for each point in volume-space, it may be possible to mitigate one source of uncertainty in the final free-energy estimate.
In scenarios where the collective-variable is predetermined, one can still attempt to address the differences in uncertainty by reparametrizing the collective-variable path.
One such possibility is hinted at towards the bottom of Fig \ref{fig_RNA0_ssnll_k32_ddssnll_lsq_vs_Atom_v_FIGC2_stripped.jpg}; the goal of this reparametrization is to allow for more images in the image-pool to be used when estimating the free-energy at points along the path where $\cR$ is small.
One such possiblity could be to bunch up the representative points $s$ along the path where $\cR$ is large, while spreading out the representative points where $\cR$ is small.

\section{Discussion}
\label{sec_Discussion}

In cryo-EM, single-particle reconstruction algorithms typically search for volumes which maximize a likelihood function.
Consequently, the effect of any local perturbation depends on the Hessian of the log-likelihood in Eq. \ref{eq_main_P_AA_given_F}.
In the context of single-particle-reconstruction (SPR), this log-likelihood can often be approximated in the image-aligned limit, accommodating both the low-noise limit as well as many scenarios involving noise-marginalization.
This image-aligned limit yields the simple accumulation of 2d-integrals shown in Eq. \ref{eq_main_P_AA_given_F_maximum_likelihood}.
This simple structure allows the log-likelihood to be represented as a volumetric integral akin to Eq. \ref{eq_main_ssnll_q3d_pre}.
Together, these 2d- and 3d-representations of the log-likelihood allow us to recover the extremal eigenvalues and eigenvectors of the Hessian, opening the door to several kinds of sensitivity analyses.
We believe that these analyses can complement more traditional strategies for sensitivity analysis, such as bootstrapping and crossvalidation (e.g., measuring the fourier-shell-correlation between reconstructions built using disjoint subsets of the image-pool) \cite{Henderson2012,Penczek2014,Heymann2015,Rosenthal2015}.

In the applications described above, we have illustrated how the `softest-modes' of the Hessian can be used to quantify how robust a particular single-particle reconstruction might be.
For example, one can examine the softest-subspace to identify which regions of the molecule are most uncertain.
The same can be said for the image-alignments, and the softest-modes can quantify which (implicitly determined) image-alignments are most uncertain.
The softest-subspace of the Hessian can also be used to probe how the uncertainty depends on properties of the image-pool, such as the viewing-angle distribution and range of defocus-values.

We expect this kind of analysis to be useful in scenarios where the viewing-angle distribution is naturally nonuniform, and where experimentalists have a range of defocus values to choose from.
One such example might be the reconstruction of membrane proteins from picked-particle images of cellular lammelae \cite{LG2023,ZCRLG2024}.
There are scenarios where the picked-particles can be biased towards regions where the cell membrane intersects the micrograph transversely.
In such scenarios, the membrane-proteins in any picked-particle images would typically be viewed from a `side' (i.e., a direction tangent to the membrane) rather than from the `top' (i.e., the direction orthogonal to the membrane).
As illustrated above, nonuniformity in the viewing-angle distribution can result in large uncertainties for particular regions of the molecule.
This can be particularly problematic when these large uncertainties occur in regions of interest (e.g., the pores of a membrane channel).
By analyzing how the softest-modes of the Hessian depend on the viewing-angle distribution and the imaging parameters, one can determine which additional viewing-angles (and choice of defocus-values) would help reduce this uncertainty the most.

While most of our examples involve single-particle reconstruction, the structure of the Hessian can provide important information in other applications as well.
For example, when using cryo-EM data to estimate free-energy landscapes one often considers paths in conformation-space.
Each location along a conformational-path can be associated with a conformation-specific Hessian, which in turn produces conformation-specific soft-modes.
In such a scenario, the relative orientation between the conformational-tangent-space and the softest-subspace of the conformation-specific Hessian plays a role in determining the local sensitivity of any free-energy estimate.

The analysis of the Hessian can also shed light on some of the difficulties associated with multi-particle reconstruction.
As described in Appendix section 21, multi-particle reconstruction can be ill-posed and/or yield spurious heterogeneity even in situations where the analogous single-particle reconstruction problem is well-posed.
While we do not have a generic solution to this ill-posedness, we suggest applying the analysis above -- namely calculating and inspecting the softest-modes -- for each of the volumes produced in a multi-particle reconstruction.
These softest-modes can reveal regions of high sensitivity which could indicate ill-posedness.

Finally, while our presentation above largely ignored image-specific translations $\{\vd_{j}\}$, one can easily incorporate these translations into the definitions of the likelihood and the corresponding Hessian.
In future work we will expand our computational strategy to account for such translations, and plan to use strategies from \cite{RSAB20} to accelerate this computation.

{\bf Acknowledgements:}\ 
The authors would like to acknowledge the support and advice of Leslie Greengard, director of the Center of Computational Mathematics at the Flatiron Institute.
We would also like to acknowledge the support and advice of Nikolaus Grigorieff, head of the Grigorieff Lab at the University of Massachusetts Chan Medical School.

\part{Appendix}
\label{sec_Appendix}
In this part we introduce the notation used in the main text and describe the objects we are dealing with in the context of cryo-EM.
We first discuss images, and then turn to volumes in section \ref{sec_Volume_notation}.
Many of the conventions we establish for the former will carry over to the latter.
When possible, we will use notation similar to \cite{RSAB20,Rangan22,RG23}.
\section{Background}\label{sec_Background}

Electron cryomicroscopy (cryo-EM) is a popular technique in structural biology which is used to produce accurate volumetric maps of the electron density of macromolecular structures and biomolecules, such as proteins \cite{Goncharov1987,vanHeel1987,Goncharov1988,Jonic2005,Cheng2015,Nogales2015,Elmlund2015,Murata2017,Sigworth2016,Bhella2019}.
Certain stages of a cryo-EM analysis pipeline involve solving an inverse-problem: e.g., given a collection of picked-particle images, the goal is to find the density map (i.e., volume) which produced them \cite{Grigorieff2007,EMAN2,Yang2008,Singer2009,Singer2011,Scheres2012,Shkolnisky2012,Lyumkis2013,Wang2013,Grigorieff2016,Punjani2017,PunjaniBrubaker2017}.
In this paper we focus on one such inverse-problem, so-called `\textit{ab-initio} single-particle reconstruction'.
Below we'll describe a simple strategy for analysing the robustness of solutions to this problem, and provide numerical methods for quantifying the uncertainty in the solution.

To provide some background, a standard single-particle reconstruction pipeline typically involves several stages:
\begin{enumerate}
\item \label{step_prep} {\em Freezing a suspension containing multiple copies of the molecule:}\ 
  As described in \cite{PR16,TWSMR16}, there are several techniques for preparing a suspension of the molecule of interest, with details that depend on the properties of the molecule and its surrounding buffer.
  The goal for this stage is to produce a thin frozen film which can be penetrated by the transmission electron microscope, while also containing multiple well separated copies of the molecule.
\item \label{step_imag} {\em Imaging the suspension to produce micrographs:}\ 
  The electron-beam interacts with the electron-density in the frozen film, and elastically scattered electrons in the beam are subject to a density-dependent phase-shift \cite{Sigworth2016}.
  This phase-shift can be measured at each point in the image-plane, producing a `micrograph'.
  The values of the phase-shift can be related to the electron density within the sample.
  The details of this transformation depends on imaging-parameters such as the defocus level, which determines the contrast transfer function (CTF).
\item \label{step_pick} {\em Scanning the micrographs for `picked-particles':}\ 
  Once the micrographs are recorded, individual molecules within the micrograph must be `picked out'.
  There are several strategies for this `particle picking' task \cite{HEIMOWITZ2018215,MCSWEENEY2020719,LANGLOIS20141,KUMAR200441}, and it is common practice to cluster the picked-particles, filter and denoise them \cite{Park2010,FRANK199654,MCSWEENEY2020719}.
  The goal for this stage is to produce a collection of picked-particle images, each corresponding to a well-centered projection of a single molecule.
\item \label{step_reco} {\em Analysis of the picked-particle images:}\ 
  At this stage, the viewing-direction (i.e., viewing-angle) of each picked-particle projection is unknown.
  Determining these viewing angles is typically done alongside reconstructing the molecular volume itself (see \cite{Scheres2012} but also \cite{Singer2011,Shkolnisky2012,Wang2013}).
  Prior information regarding the molecule can often be used to aid in the reconstruction of the viewing-angles and volume.
  However, when such prior information is absent or deliberately ignored to avoid bias, the reconstruction process is referred to as \textit{ab-initio} single-particle reconstruction \cite{Sigworth1998,Grigorieff2007,Scheres2009,Scheres2012,Scheres2012b,Bell2016}).
\end{enumerate}

Generally speaking, if we were to know the density $\xF(\vx)$ of a molecular volume, as well as a collection of image-specific viewing-angles $\{\tau_{j}\}$, we could use a forward model for the imaging process (involving the imaging-parameters and noise-level) to estimate what each picked-particle image $\{\xA_{j}\}$ might look like \cite{Sigworth1998,Sigworth2010}.
The \textit{ab-initio} single-particle reconstruction mentioned in the final step \ref{step_reco} above involves the reverse: taking the picked-particle images $\{\xA_{j}\}$ and attempting to infer the volume $\xF(\vx)$ and viewing-angles $\{\tau_{j}\}$.
Thus, \textit{ab-initio} reconstruction is a type of inverse-problem.

In practice, most methods for solving the SPR problem search for a solution which optimizes a posterior likelihood of generating the observed picked-particle images, given the reconstructed volume (see \cite{Wang2013,Singer2009,Shkolnisky2012,Goncharov1987,Goncharov1988,vanHeel1987,Jonic2005,Yang2008} and software from \cite{Punjani2017,Scheres2012,Grigorieff2016,Lyumkis2013,Grigorieff2007,EMAN2}).
More specifically, the solution is typically a `model' that optimizes the log-likelihood $\log P(\text{\small data}\givenmed\text{\small model})$.
Because the solution lies at a critical-point of $\log P(\text{\small data}\givenmed\text{\small model})$, any perturbations to the solution will result in a shift to the overall likelihood which depends on the Hessian of $\log P$.
In the remainder of this Appendix we describe computational strategies for probing the structure of the Hessian.

\section{Image notation}
\label{sec_Image_notation}

We use $\vx,\vk \in\Real^{2}$ to represent spatial position and frequency, respectively.
In polar coordinates these vectors are represented as:
\begin{eqnarray}
\vx &=& (x \cos \theta, x \sin \theta) \\
\vk &=& (k \cos \psi, k \sin \psi) \text{.}
\end{eqnarray}

The Fourier transform of a $2$-dimensional function $A \in \Leb^{2}(\Real^{2})$ is defined as
\begin{equation}
  \fA(\vk) := \iint_{\Real^{2}} A(\vx) \euler^{-\imunit \vk \cdot \vx} \mathop{d\vx}~.
  \label{eq_Ahat}
\end{equation}
We recover $A$ from $\fA$ using the inverse Fourier transform:
\begin{equation}
A(\vx) = \frac{1}{(2\pi)^{2}} \iint_{\Real^{2}} \fA(\vk) \euler^{+\imunit \vk \cdot \vx} \mathop{d\vk} \text{.}
\end{equation}
The inner product between two functions $A, B \in \Leb^{2}(\Real^{2})$ is written as
\begin{equation}
  \langle A, B\rangle = \iint_{\Real^{2}} A(\vx)^{\dagger} B(\vx) \mathop{d\vx}
  ~,
\end{equation}
where $z^{\dagger}$ is the complex conjugate of $z \in \Complex$.
We will also use Plancherel's theorem \cite{bracewell},
\begin{equation}
  \langle A, B\rangle = \frac{1}{(2\pi)^{2}} \langle\fA,\fB\rangle
  ~, \qquad \forall A, B \in \Leb^{2}(\Real^{2})~,
  \label{eq_plancherel}
\end{equation}
ignoring the prefactors of $2\pi$ associated with the Fourier transform when they are not relevant.

We represent any given image as a function $A \in \Leb^{2}(\Real^{2})$, with values corresponding to the image intensity at each location.
Any inner product between $A$ and $B$ can be calculated equally well in either real or frequency space.
Because of recent image alignment tools developed in the context of cryo-EM molecular reconstruction \cite{Barnett2017,RSAB20}, we will typically refer to images in frequency space (i.e., $\fA$, rather than $A$).

Abusing notation, we'll refer to $A(\vk)$ and $\fA(\vk)$ in polar coordinates as:
\begin{eqnarray}
A(x, \theta) &:=& A(\vx) = A(x \cos \theta, x \sin \theta) \\
\fA(k, \psi) &:=& \fA(\vk) = \fA(k \cos \psi, k \sin \psi) \text{.}
\end{eqnarray}

With this notation, each $\fA(k,\psi)$ for fixed $k$ and $\psi\in[0,2\pi)$ corresponds to a `ring' in frequency space with radius $k$.

\section{Rotating images: the Fourier--Bessel basis}
\label{sec_rotating_images}

Using the notation above, a rotation $\rotation_{\gamma}$ by angle $\gamma$ can be represented as:
\begin{equation}
\label{eq_rotation_definition}
\rotation_{\gamma} A(x, \theta) := A(x, \theta-\gamma) \text{.}
\end{equation}
Since rotation commutes with the Fourier transform, we have:
\begin{equation}
\wfourier{\rotation_\gamma A}(k, \psi) = \rotation_\gamma \circ \fA (k, \psi) = \fA(k, \psi-\gamma) \text{.}
\end{equation}
In this manner, a rotation of any image by $+\gamma$ can be represented as an angular shift of each image ring by $\psi\rightarrow\psi-\gamma$.

The effect of a particular rotation can also be applied in a different basis: the Fourier--Bessel basis.
To define the Fourier--Bessel coefficients of an image we recall that, for each fixed $k$, the image ring $\fA(k, \psi)$ is a $2\pi$-periodic function of $\psi$.
Thus, we can represent each image ring $\fA(k, \psi)$ as a Fourier series in $\psi$, obtaining
\begin{equation}
\fA(k, \psi) = \sum_{q=-\infty}^{+\infty} \bA(k,q) \euler^{+\imunit q \psi} \text{,}
\end{equation}
for $q\in\Integer$.
The Fourier--Bessel coefficients $\bA(k,q)$ of the image ring $\fA(k,\psi)$ are given by
\begin{equation}
\bA(k,q) = \frac{1}{2\pi} \int_{0}^{2\pi} \fA(k, \psi) \euler^{-\imunit q \psi} \mathop{d\psi}~.
\label{eq_aqk}
\end{equation}
These coefficients can be represented in a more traditional fashion by recalling that the Bessel function $J_{q}(kx)$ can be written as:
\begin{eqnarray}
J_{q}(kx) & = & \frac{1}{2\pi}\int_{0}^{2\pi} \euler^{\imunit kx\sin(\psi)-\imunit q \psi} \mathop{d\psi}~ \\
 & = & \frac{1}{2\pi}\int_{0}^{2\pi} \euler^{-\imunit kx\cos(\psi+\pi/2)-\imunit q \psi} \mathop{d\psi}~ ,
\end{eqnarray}
which, when combined with the definition of the Fourier transform, immediately implies that:
\begin{eqnarray}
\bA(k,q) & = & \iint A(x, \theta) \frac{1}{2\pi} \int_{0}^{2\pi} \euler^{-\imunit kx\cos(\psi-\theta) - \imunit q \psi} \mathop{d\psi} xdx\mathop{d\theta}~ \\
 & = & \iint A(x, \theta) \euler^{-\imunit q(\theta+\pi/2)} \frac{1}{2\pi} \int_{0}^{2\pi} \euler^{-\imunit kx\cos(\psi+\pi/2) - \imunit q \psi} \mathop{d\psi} xdx\mathop{d\theta}~\\
 & = & \iint A(x, \theta) \euler^{-\imunit q(\theta+\pi/2)}J_{q}(kx) xdx\mathop{d\theta}~,
\end{eqnarray}
which is the inner product between the original image (in real space) and a `Fourier--Bessel' function.

The rotation $\rotation_{\gamma}$ can now be represented as:
\begin{eqnarray}
\rotation_{\gamma}\fA(k, \psi) & = & \fA(k,\psi-\gamma) \\
                      & = & \sum_{q=-\infty}^{+\infty} \bA(k,q) \euler^{+\imunit q (\psi-\gamma)} \\
                      & = & \sum_{q=-\infty}^{+\infty} \bA(k,q)\cdot\euler^{-\imunit q\gamma} \euler^{+\imunit q \psi} \text{,}
\label{eq_Fourier_Bessel_rotation}
\end{eqnarray}
such that the Fourier--Bessel coefficients of the rotated image ring $\rotation_{\gamma}\circ \fA(k,\cdot)$ are given by the original Fourier--Bessel coefficients $\bA(k,q)$, each multiplied by the phase factor $\euler^{-\imunit q\gamma}$.
Note that \eqref{eq_Fourier_Bessel_rotation} naturally allows for rotation by any $\gamma$, even values of $\gamma$ that may not lie on the polar grid used to discretize the images.

In a typical discretization scheme (see below) the images $\fA$ and $\fB$ each require the storage of $\bigO(\npixel^{2})$ values, as do the Fourier--Bessel representations $\bA$ and $\bB$.
The calculation of the array $\wfourier{\cX}(q)$ involves $\bigO(\npixel^{2})$ operations and $\bigO(\npixel)$ storage.
Once $\wfourier{\cX}$ is calculated, the inner products $\cX(\gamma)$ can be recovered on a uniform grid of $\bigO(\npixel)$ angles $\gamma\in[0,2\pi)$ with an additional $\bigO(\npixel\log(\npixel))$ operations using the FFT.
More details are given in \cite{Barnett2017,RSAB20,RG23}.

\section{Image noise model}
\label{sec_Image_noise_model}

We assume that the images considered can be modeled as the sum of a `signal' plus a `noise'.
We assume that the signal corresponds to a 2-dimensional projection of a (smooth) molecular electron density function, while the noise corresponds to detector noise \cite{Sigworth1998,Scheres2009,Sigworth2010,Lyumkis2013}.
We do not consider more complicated sources of noise, such as structural noise associated with image preparation (see \cite{Baxter2009}).

Based on these simple assumptions, we'll model the noise in real space as independent and identically distributed (iid), with a variance of $\sigma^{2}$ on a unit scale in 2-dimensional real space.
Thus, the image $A$ evaluated at pixel $(n_{1},n_{2})$ has a value modeled by:
\begin{equation}
A_{n_{1},n_{2}} = A(\vx_{n_{1},n_{2}}) = A^{\signal}_{n_{1},n_{2}} + A^{\noise}_{n_{1},n_{2}} \text{,} 
\end{equation}
where the signal and noise are represented by the arrays $A^{\signal}_{n_{1},n_{2}}$ and $A^{\noise}_{n_{1},n_{2}}$, respectively.
Because each entry in $A^{\noise}_{n_{1},n_{2}}$ corresponds to an average over the area element $\dx^{2}$ associated with a single pixel, we expect the variance of each $A^{\noise}_{n_{1},n_{2}}$ to scale inversely with $\dx^{2}$.
That is to say, we'll assume that each element of the noise array $A^{\noise}_{n_{1},n_{2}}$ is drawn from the standard normal distribution $\cN\left(0,\frac{\sigma^{2}}{\dx^{2}}\right)$, with a mean of $0$ and a variance of $\sigma^{2}/\dx^{2}$.

Given the assumptions above, we expect that $\fA$ will be modeled by:
\begin{equation}
\fA(\vk) = \fA(\vk) = \fA^{\signal}(\vk) + \fA^{\noise}(\vk) \text{,} 
\label{eq_image_noise_model}
\end{equation}
where $\fA^{\noise}$ is now complex and iid.
Because the noise term $A^{\noise}$ is real, the noise term $\fA^{\noise}$ will be complex, with the conjugacy constraint that $\fA^{\noise}(+\vk) = \fA^{\noise}(-\vk)^{\dagger}$.

If we were to sample $\vk$ on the uniform $\npixel\times\npixel$ cartesian grid associated with the standard 2-dimensional fast Fourier transform, then $\vk_{n_{1},n_{2}}$ will correspond to the frequency $\dk\cdot(n_{1},n_{2})$, with frequency spacing $\dk=\pi$ and indices $(n_{1},n_{2})$ considered periodically in the range $[-\npixel/2,\ldots,+\npixel/2-1]$.
In this case the transformation between $A_{n_{1},n_{2}}$ and $\fA_{n_{1},n_{2}}$ will be unitary.
When $(n_{1},n_{2})$ corresponds to an index pair that is periodically reflected onto itself (i.e., the four index pairs $(0,0)$, $(\npixel/2,0)$, $(0,\npixel/2)$ and $(\npixel/2,\npixel/2)$) then $\fA^{\noise}_{n_{1},n_{2}}$ will be real and drawn from $\cN(0,\frac{\sigma^{2}}{\dx^{2}})$.
When $(n_{1},n_{2})$ corresponds to an index pair that is {\em not} periodically reflected onto itself, the real and imaginary components of $\fA^{\noise}(\vk_{n_{1},n_{2}})$ will each be drawn iid from $\cN(0,\frac{\sigma^{2}}{2\dx^{2}})$ subject to the conjugacy constraint above.

Motivated by this observation, we define $\fsigma^{2}=\frac{\pi^{2}\sigma^{2}}{\dx^{2}}$ as the variance on a unit scale in 2-dimensional frequency space.
We expect that the noise term $\fA^{\noise}(\vk)$ integrated over any area element $\dk^{2}$ in frequency space will have a variance of $\frac{\fsigma^{2}}{\dk^{2}}$.

\section{Image-image similarity: likelihood}
\label{sec_Image_similarity}

Given one `noisy' image $\fA=\fA^{\signal}+\fA^{\noise}$ and another `noiseless' image $\fB=\fB^{\signal}$, we can derive the standard likelihood of observing any value of $\fA$, given that the signals $\fA^{\signal}$ and $\fB^{\signal}$ are the same:
\begin{equation}
P(\fA^{\signal}=\fB \givenbig \fA) \cdot P(\fA) = P(\fA \givenbig \fA^{\signal}=\fB) \cdot P(\fA^{\signal}=\fB) \text{,}
\end{equation}
which can be rearranged into:
\begin{equation}
P(\fA^{\signal}=\fB \givenbig \fA) = \frac{P(\fA \givenbig \fA^{\signal}=\fB)\cdot P(\fA^{\signal}=\fB)}{P(\fA)} \text{.}
\end{equation}
Assuming a uniform prior $P(\fA^{\signal}=\fB)$, the posterior probability $P(\fA^{\signal}=\fB|\fA)$ is proportional to the likelihood $P(\fA \givenbig \fA^{\signal}=\fB)$.
\begin{multline}
P(\fA(\vk) \givenbig \fB(\vk)) := P(\fA(\vk) \givenbig \fA^{\signal}(\vk)=\fB(\vk)) = \\
\frac{1}{(\sqrt{2\pi}\ \fsigma)^{\dof}} \exp\left( - \frac{1}{2\fsigma^{2}} \iint_{\vk\in\Omega_{\kmax}} |\fA(\vk) - \fB(\vk)|^{2} d\vk \right) \comma
\label{eq_P_A_given_B}
\end{multline}
where $\dof$ refers to the number of independent degrees of freedom in the image (e.g., $\dof$ could be the number of pixels in the image $A(\vx)$, or the number of quadrature-nodes used to discretize $\fA(k,\psi)$).

When dealing with two noisy images $\fA=\fA^{\signal}+\fA^{\noise}$ and $\fB=\fB^{\signal}+\fB^{\noise}$, as long as $\fA^{\noise}$ is independent from $\fB^{\noise}$ the same general argument applies. The only difference in this scenario is that the effective variance involves both the variance of $\fA^{\noise}$ and $\fB^{\noise}$.

\section{Volume notation}
\label{sec_Volume_notation}

We use $\vchi,\vkappa \in\Real^{3}$ to represent spatial position and frequency, respectively.
in spherical coordinates the vector $\vkappa$ is represented as:
\begin{eqnarray}
\vkappa &=& k\cdot \hkappa, \text{\ \ with\ \ } \hkappa = (\cos \kazimub \sin \kpolara, \sin \kazimub \sin \kpolara , \cos \kpolara) \text{,}
\label{eq_vkappahkappa}
\end{eqnarray}
with polar angle $\kpolara$ and azimuthal angle $\kazimub$ representing the unit vector $\hkappa$ on the surface of the sphere $S^{2}$.

Using a right handed basis, a rotation about the third axis by angle $\eazimub$ is represented as:
\begin{equation}
\rotation_{\eazimub}^{z} =
\left(
\begin{array}{ccc}
+\cos\eazimub & -\sin\eazimub & 0 \\
+\sin\eazimub & +\cos\eazimub & 0 \\
0            & 0            & 1
\end{array}
\right) \text{,}
\end{equation}
and a rotation about the second axis by angle $\epolara$ is represented as:
\begin{equation}
\rotation_{\epolara}^{y} =
\left(
\begin{array}{ccc}
+\cos\epolara & 0 & +\sin\epolara \\
0            & 1 & 0            \\
-\sin\epolara & 0 & +\cos\epolara 
\end{array}
\right) \text{.}
\end{equation}

A rotation $\rotation_{\tau}$ of a vector $\vkappa\in\Real^{3}$ can be represented by the vector of Euler angles $\tau = (\egammaz,\epolara,\eazimub)$:
\begin{equation}
\rotation_{\tau}\cdot\vkappa = \rotation_{\eazimub}^{z}\circ \rotation_{\epolara}^{y}\circ \rotation_{\egammaz}^{z}\cdot \vkappa \text{.}
\end{equation}

We represent any given volume as a function $F\in\Leb^{2}(\Real^{3})$, with values corresponding to the volume intensity at each location $\vchi\in\Omega_{1}$.
We'll refer to $\fF(\vkappa)$ in spherical coordinates as $\fF(k, \hkappa)$.
With this notation, each $\fF(k,\cdot)$ corresponds to a `shell' in frequency space with radius $k$.
The rotation of any volume $\rotation_{\tau}\fF(k,\hkappa)$ corresponds to the function $\fF(k,\rotation_{\tau}^{-1}\hkappa)$.

\section{Rotation in the spherical-harmonic basis}
\label{sec_Spherical_harmonic_basis}

Using the notation above, we can represent a volume $\fF(k,\hkappa)$ as:
\begin{equation}
\fF(k,\hkappa) = \sum_{l=0}^{+\infty}\sum_{m=-l}^{m=+l} \fF_{l}^{m}(k) Y_{l}^{m}(\hkappa) \text{,}
\end{equation}
where $Y_{l}^{m}(\hkappa)$ represents the spherical-harmonic of degree $l$ and order $m$:
\begin{equation}
Y_{l}^{m}(\kpolara,\kazimub) =  Z_{l}^{m} \euler^{+\imunit m \kazimub}P_{l}^{m}(\cos\kpolara) \text{,}
\end{equation}
with $P_{l}^{m}$ representing the (unnormalized) associated Legendre polynomial, and $Z_{l}^{m}$ the normalization constant:
\begin{equation}
Z_{l}^{m} = \sqrt{\frac{2l+1}{4\pi}\times\frac{\left(l-|m|\right)!}{\left(l+|m|\right)!}} \text{.}
\end{equation}
The coefficients $\fF_{l}^{m}(k)$ define the spherical-harmonic expansion of the $k$-shell $\fF(k,\cdot)$.

Using this spherical-harmonic basis allows us to efficiently apply rotations.
For example, given a rotation by $\tau=(\egammaz,\epolara,\eazimub)$, we can represent the rotated volume $\fG:=\rotation_{\tau}\fF$ as:
\begin{equation}
\fG_{l}^{m_{1}} = \sum_{m_{2}=-l}^{m_{2}=+l} \euler^{-\imunit m_{1} \eazimub}d_{m_{1},m_{2}}^{l}(\epolara) \euler^{-\imunit m_{2} \egammaz} \fF_{l}^{m_{2}} \text{,}
\label{eq_rotate_3d}
\end{equation}
where $d_{m_{1},m_{2}}^{l}(\epolara)$ represents the degree $l$ Wigner d-matrix associated with the interior Euler angle $\epolara$.

\section{Volume projection in the spherical-harmonic basis}
\label{sec_Volume_projection}

We denote the projection of a volume $F(\vchi)$ (with $\vchi\in\Real^{3}$) along a particular viewing-angle $\tau$ by the function $S(\vx;\tau;F)$ (with $\vx\in\Real^{2}$).
We'll refer to this projection $S(\vx;\tau;F)$ as the `template' associated with $F(\vchi)$ and viewing-angle $\tau$.

By the Fourier-slice theorem we can decribe the Fourier-transformed template $\fS(\vk;\tau,\fF)$ via:
\begin{eqnarray}
\fS(\vk;\tau;\fF) & = & \left[\fSlice\circ\rotation(\tau)\circ \fF\right](\vk) \comma
\label{eq_template} 
\end{eqnarray}
where $\fSlice$ denotes taking the equatorial slice in the Fourier domain.

In terms of computation, the $\fSlice$ operator can be described by restricting the unit-vector $\hkappa$ in Eq. \ref{eq_vkappahkappa} to the equatorial-plane, corresponding to setting $\epolara\equiv 0$ and associating $\eazimub=:\psi$.
\begin{eqnarray}
\left[\fSlice\circ \fF\right](k,\psi) & := & \sum_{l=0}^{+\infty}\sum_{m=-l}^{m=+l} \fF_{l}^{m}(k) Z_{l}^{m} \euler^{+\imunit m \psi}P_{l}^{m}(0) \period
\label{eq_fSlice} 
\end{eqnarray}

If we consider the effect of a contrast-transfer-function $CTF(\vk)$, then the template associated with $F(\vchi)$ and a viewing-angle $\tau$ can be written as:
\begin{eqnarray}
\fS(\vk;\tau;CTF;\fF) & = & CTF(\vk) \odot \left[\fSlice\circ\rotation(\tau)\circ \fF\right](\vk) \period
\label{eq_CTF_modulated_template} 
\end{eqnarray}

If we were also to consider the effect of an image-dependent displacement $\vd$, then the template associated with $F(\vchi)$ and a viewing-angle $\tau$ can be written as:
\begin{eqnarray}
\fS(\vk;\tau;\vd;CTF;\fF) & = & CTF(\vk) \odot \ftranslation(\vd;\vk) \odot \left[\fSlice \circ \rotation(\tau) \circ \fF\right](\vk) \comma
\label{eq_ftranslated_CTF_modulated_template} 
\end{eqnarray}
where $\ftranslation(\vd;\vk)$ is the fourier-signature corresponding to a 2-dimensional translation by $\vd$:
\begin{eqnarray}
\ftranslation(\vd;\vk) := \exp\left(-i\vd\cdot\vk\right) \period
\label{eq_ftranslation} 
\end{eqnarray}

\section{Image-volume similarity: likelihood}
\label{sec_Volume_likelihood}

\subsection{single-particle:\ }
\label{sec_Volume_likelihood_single_particle}
Referring back to section \ref{sec_Image_similarity} and Eq. \ref{eq_P_A_given_B}, we can express the likelihood that a particular volume $\fF(\vk)$ might produce a particular image $\fA(\vk)$ with CTF-function $CTF(\vk)$ via projection along viewing-angle $\tau$ and translation by a displacement-vector $\vd$:
\begin{eqnarray}
P\left(\fA \givenbig \tau ; \vd ; \fF \right) & = & P\left(\fA \givenbig \fS(\vk;\tau;\vd;CTF;\fF)\right) \period
\label{eq_P_A_given_tau_vd_F} 
\end{eqnarray}
The likelihood $P\left(\fA \givenbig \tau ; \vd ; \fF \right)$ can be marginalized across $\tau$ and $\vd$ to calculate the likelihood that $\fF$ might produce the image $\fA(\vk)$:
\begin{eqnarray}
P\left(\fA \givenbig \fF \right) & = & \int_{\vd\in \Real^{2}} \int_{\tau\in SO3} P\left(\fA \givenbig \tau ; \vd ; \fF \right)\cdot P\left(\tau ; \vd \givenbig \fF\right) d\tau d\vd \\
  & = & \int_{\vd\in \Real^{2}} \int_{\tau\in SO3} P\left(\fA \givenbig \tau ; \vd ; \fF \right)\cdot P\left(\vd \givenbig \tau ; \fF \right) \cdot P\left(\tau \givenbig \fF \right) d\tau d\vd  \comma
\label{eq_P_A_given_F} 
\end{eqnarray}
where $P\left(\tau \givenbig \fF \right)$ is prior distribution attributed to the viewing-angle $\tau$, $P\left(\vd \givenbig \tau ; \fF \right)$ is the prior distribution attributed to $\vd$ for a particular value of $\tau$, and $P\left(\tau ; \vd \givenbig \fF \right)$ is the joint-prior over the alignments $\tau$ and $\vd$. 
To emphasize the dependence of $P(\fA\givenmed\fF)$ on the prior distributions and noise-level, we will sometimes write:
\[ P\left(\fA \givenbig \fF \right) =: P\left(\fA \givenbig \fF \semicolonmed P(\tau;\vd\givenmed\fF) \semicolonmed \fsigma \right) \period \]

In this manuscript we will typically assume that $\delta\equiv 0$.
This corresponds to taking the limit $\sigma_{\vd}\rightarrow 0$, resulting in $P\left(\vd \givenbig \tau ; \fF \right)$ becoming sharply peaked at the origin.
Under this limit we'll suppress the notation for $\vd$.
More generally, we can consider $P\left(\vd \givenbig \tau ; \fF \right)$ to be an isotropic gaussian centered at the origin:
\[ P\left(\vd \givenbig \tau ; \fF \right) \ = \ P(\vd) \ = \ \frac{1}{2\pi \sigma_{\vd}^{2}} \exp\left(-\frac{|\vd|^{2}}{2\sigma_{\vd}^{2}}\right) \comma \]
where $\sigma_{\vd}$ is a fixed small distance (e.g., one tenth the diameter of each picture).

Given a pool of $\nimage$ images $\{\fA_{j}\}$ and their associated CTF-functions $\{CTF_{j}\}$ (both indexed by $j\in 1,\ldots \nimage$), we can accumulate terms from Eq. \ref{eq_P_A_given_F} to calculate the likelihood that $\fF_{i}$ might produce this collection of images:
\begin{eqnarray}
P\left(\{\fA_{j}\} \givenbig \fF \right) = P\left(\{\fA_{j}\} \givenbig \fF \semicolonmed P(\tau;\vd\givenmed\fF) \semicolonmed \fsigma \right) = \prod_{j=1}^{\nimage} P\left(\fA_{j} \givenbig \fF \semicolonmed P(\tau;\vd\givenmed\fF) \semicolonmed \fsigma \right) \period
\label{eq_P_AA_given_F} 
\end{eqnarray}

\subsection{multi-particle:\ }
\label{sec_Volume_likelihood_multi_particle}
In the context of multi-particle reconstruction, we might consider $\nvolume$ different volumes $\{\fF_{i}\}$ (indexed by $i\in 1,\ldots \nvolume$), each with their own prior distributions.
In this context we can calculate the likelihood that the collection $\{\fF_{i}\}$ might produce the image $\fA(\vk)$:
\begin{eqnarray}
P\left(\fA \givenbig \{\fF_{i}\} \right) & = & \sum_{i=1}^{\nvolume} P\left(\fA \givenbig \fF_{i} \right) \cdot P\left( \fF_{i} \right) \\
 & = & \int_{\vd\in \Real^{2}} \int_{\tau\in SO3} \sum_{i=1}^{\nvolume} P\left(\fA \givenbig \tau ; \vd ; \fF_{i} \right) \cdot P\left( \tau ; \vd \givenbig \fF_{i} \right) \cdot P\left( \fF_{i} \right) d\tau d\vd \\
 & = & \int_{\vd\in \Real^{2}} \int_{\tau\in SO3} \sum_{i=1}^{\nvolume} P\left(\fA \givenbig \tau ; \vd ; \fF_{i} \right) \cdot P\left( \tau , \vd , i \givenbig \{\fF_{i}\} \right) d\tau d\vd \comma
\period
\label{eq_P_A_given_FF} 
\end{eqnarray}
where $P\left( \fF_{i} \right)$ is the prior probability that the image $\fA$ was drawn from the volume $\fF_{i}$, and $P\left( \tau , \vd , i  \givenmed \{\fF_{i}\} \right)$ is the prior probability that the image $\fA$ was drawn from the volume $\fF_{i}$ with viewing-angle $\tau$ and displacement $\vd$.
Once again, to emphasize the dependence of $P(\fA\givenmed\{\fF_{i}\})$ on the prior distributions and noise-level, we will sometimes write:
\[ P\left(\fA \givenbig \{\fF_{i}\} \right) =: P\left(\fA \givenbig \{\fF_{i}\} \semicolonmed P\left( \tau , \vd , i  \givenmed \{\fF_{i}\} \right) \semicolonmed \fsigma \right) \period \]

Just as in Eq. \ref{eq_P_AA_given_F}, we can accumulate terms from Eq. \ref{eq_P_A_given_FF} to calculate the likelihood that $\{\fF_{i}\}$ might produce a collection of images:
\begin{multline}
P\left(\{\fA_{j}\} \givenbig \{\fF_{i}\} \right) = P\left(\{\fA_{j}\} \givenbig \{\fF_{i}\} \semicolonmed P\left( \tau , \vd , i  \givenmed \{\fF_{i}\} \right) \semicolonmed \fsigma \right) = \\
\prod_{j=1}^{\nimage} P\left(\fA_{j} \givenbig \{\fF_{i}\} \semicolonmed P\left( \tau , \vd , i  \givenmed \{\fF_{i}\} \right) \semicolonmed \fsigma \right) \period
\label{eq_P_AA_given_FF} 
\end{multline}

\section{Image-volume similarity in the image-aligned limit}
\label{sec_Volume_likelihood_mle}
\subsection{single-particle:\ }
\label{sec_Volume_likelihood_mle_single_particle}
The likelihood in Eq. \ref{eq_P_AA_given_F} depends via Eqs. \ref{eq_P_A_given_F}, \ref{eq_P_A_given_tau_vd_F} and \ref{eq_P_A_given_B} on the estimated noise-level $\fsigma^{2}$, which is related to the signal-to-noise ratio, or `snr', and is sometimes referred to as the estimated `temperature' used in the bayesian framework.
In certain regimes the integral in Eq. \ref{eq_P_A_given_F} becomes dominated by the contribution associated with the `maximum-likelihood' alignment-angle for the image $\fA$.
This maximum-likelihood alignment occurs for the viewing-angle, displacement and volume which maximize the summand $P(\fA \givenmed \tau ; \vd ; \fF)\cdot P( \tau , \vd  \givenmed \fF ) $ in the integrand of Eq. \ref{eq_P_A_given_F}.
This approximation, which we refer to as the `image-aligned' limit, occurs naturally when $\fsigma^{2}\rightarrow 0$, which is termed the `low-noise' limit.
The image-aligned limit also manifests in many practical scenarios when the noise-level is unknown and the number of degrees-of-freedom is large.
For ease of presentation we'll continue our discussion assuming a low-noise limit, and defer our discussion of the noise-marginalized `high-degree-of-freedom' limit to section \ref{sec_noise_marginalized_limit_0}.

To spell out the details, given a particular volume $\fF$, the maximum-likelihood alignment $\{\tau^{\opt},\vd^{\opt}\}$ for an image $\fA$ occurs at:
\begin{eqnarray}
\{ \ \tau^{\opt} \ , \ \vd^{\opt} \ \} := \argmax_{\{\tau,\vd\}} P\left(\fA \givenbig \tau ; \vd ; \fF\right) \comma
\label{eq_P_AA_given_F_maximum_likelihood_alignment}
\end{eqnarray}
where $\tau^{\opt}$ and $\vd^{\opt}$ are restricted to lie in the support of $P\left(\tau;\vd\givenbig \fF\right)$.
With this notation, each of the $\{\tau^{\opt},\vd^{\opt},i^{\opt}\}$ can be thought of as functions of the image $\fA$, the volume $\fF$ and the prior distribution $P(\tau,\vd\givenmed\fF)$.

In the context of single-particle reconstruction, the negative-log-likelihood in the image-aligned limit is dominated by the following leading-order expansion:
\begin{eqnarray}
-\log\left(P\left(\{\fA_{j}\}\givenbig \fF \right)\right) \ \sim \ 
\sum_{j=1}^{\nimage} \iint_{\vk\in\Omega_{\kmax}} \left|\fA_{j}(\vk)-\fS\left(\vk;\tau^{\opt}_{j};\vd^{\opt}_{j};CTF_{j};\fF\right)\right|^{2} d\vk \period
\label{eq_P_AA_given_F_maximum_likelihood}
\end{eqnarray}
This `template-wise' expression represents the likelihood in terms of a collection of 2-dimensional integrals.
Later on below we'll constrast this with an equivalent `volumetric' expression which represents the likelihood in terms of a collection of 3-dimensional integrals.

\subsection{multi-particle:\ }
\label{sec_Volume_likelihood_mle_multi_particle}
In the context of multi-particle reconstruction, the likelihood in Eq. \ref{eq_P_AA_given_FF} depends on Eq. \ref{eq_P_A_given_FF}.
In the image-aligned limit (e.g., when $\fsigma^{2}\rightarrow 0$), the maximum-likelihood alignment occurs for the viewing-angle, displacement and volume-label which maximize the summand $P(\fA \givenmed \tau ; \vd ; \fF_{i})\cdot P( \tau , \vd , i  \givenmed \{\fF_{i}\} ) $ in the integrand of Eq. \ref{eq_P_A_given_FF}.

Given a collection of volumes $\{\fF_{i}\}$, the maximum-likelihood alignment $\{\tau^{\opt},\vd^{\opt}\}$ and volume-label $i^{\opt}$ for an image $\fA$ occurs at:
\begin{eqnarray}
\{ \ \tau^{\opt} \ , \vd^{\opt} \ , \ i^{\opt} \ \} := \argmax_{\{\tau,\vd,i\}} P\left(\fA \givenbig \tau ; \vd ; \fF_{i}\right) \comma
\label{eq_P_AA_given_FF_maximum_likelihood_alignment}
\end{eqnarray}
where $\tau^{\opt}$, $\vd^{\opt}$ and $i^{\opt}$ are restricted to lie in the support of $P\left(\tau,\vd,i \givenmed \{\fF_{i}\} \right)$.
With this notation, each of the $\{\tau^{\opt},\vd^{\opt},i^{\opt}\}$ can be thought of as functions of the image $\fA$, the volume-set $\{\fF_{i}\}$ and the prior distribution $P\left(\tau,\vd,i \givenmed \{\fF_{i}\} \right)$.

Thus, within the context of multi-particle reconstruction, the negative-log-likelihood in the image-aligned limit takes a form very similar to Eq. \ref{eq_P_AA_given_F_maximum_likelihood}:
\begin{eqnarray}
-\log\left(P\left(\{\fA_{j}\}\givenbig\{\fF_{i}\}\right)\right) \ \sim \ 
\sum_{j=1}^{\nimage} \iint_{\vk\in\Omega_{\kmax}} \left|\fA_{j}(\vk)-\fS\left(\vk;\tau^{\opt}_{j};\vd^{\opt}_{j};CTF_{j};\fF_{i^{\opt}_{j}}\right)\right|^{2} d\vk \period
\label{eq_P_AA_given_FF_maximum_likelihood}
\end{eqnarray}

\section{Volume reconstruction in the image-aligned limit via least-squares}
\label{sec_Volume_reconstruction_lsq}

In the image-aligned limit each image $\vA_{j}$ is assigned to a particular viewing-angle $\tau_{j}$, displacement $\vd_{j}$ and volume-label $i$ via Eq. \ref{eq_P_AA_given_FF_maximum_likelihood_alignment}.
Given these maximum-likelihood assignments, each volume can be reconstructed independently.
For example, in the context of single-particle-reconstruction, the single volume $\fF$ can be reconstructed by solving the following linear-system in a least-squares sense:
\begin{eqnarray}
\left[
\begin{array}{c}
CTF_{1} \odot \ftranslation(+\vd_{1}) \odot \left\{ \fSlice\circ \rotation(\tau_{1})\circ \right\} \\
\vdots \\
CTF_{\nimage} \odot \ftranslation(+\vd_{\nimage}) \odot \left\{ \fSlice\circ \rotation(\tau_{\nimage})\circ \right\} \\
\end{array}
\right]
\cdot
\fF^{\lsq}
\approx
\left[
\begin{array}{c}
\fA_{1} \\
\vdots \\
\fA_{\nimage} \\
\end{array}
\right] \period
\label{eq_0lsq_original}
\end{eqnarray}
In this expression, the $\nimage$ viewing angles $\tau_{j}$ and 
displacements $\vd_{j}$ are assumed to be given as input, as are the 
image- or micrograph-specific CTF functions $CTF_{j}$ and 
images $\fA_{j}$. 
The unknown (to be determined) is the volume $\fF^{\lsq}$. 
By a least-squares solution to Eq. \ref{eq_0lsq_original} we specifically mean the volume $\fF^{\lsq}$ for which the residual is proportional to the negative-log-likelihood described in Eq. \ref{eq_P_AA_given_F_maximum_likelihood}.

If it is more convenient, we can center the images on the right-hand-side and rewrite Eq. \ref{eq_0lsq_original} as:
\begin{equation}
\left[
\begin{array}{c}
CTF_{1} \odot \left\{ \fSlice\circ \rotation(\tau_{1})\circ \right\} \\
\vdots \\
CTF_{\nimage} \odot \left\{ \fSlice\circ \rotation(\tau_{\nimage})\circ \right\} \\
\end{array}
\right]
\cdot
\fF^{\lsq}
\approx
\left[
\begin{array}{c}
\ftranslation(-\vd_{1}) \odot \fA_{1} \\
\vdots \\
\ftranslation(-\vd_{\nimage}) \odot \fA_{\nimage} \\
\end{array}
\right]
\period
\label{eq_0lsq_centered}
\end{equation}

In the context of multi-particle reconstruction the pool of images can be sorted by volume-label, and then the same strategy can be applied for each volume independently.

\section{Volume reconstruction in the image-aligned limit via back-projection}
\label{sec_Volume_qbp}

When the number of images $\nimage$ becomes large, the solution to Eq. \ref{eq_0lsq_centered} can be well approximated via a form of back-projection \cite{RG23}.
To motivate this strategy one can rewrite the likelihood as a volumetric-integral (see section \ref{sec_Volume_likelihood_3d}).
Alternatively, one may observe that, if there were no noise and no CTF-attenuation, then the volume and images would solve:
\[
\left[
\begin{array}{c}
\fSlice\circ \rotation(\tau_{1})\circ  \\
\vdots \\
\fSlice\circ \rotation(\tau_{\nimage})\circ  \\
\end{array}
\right]
\cdot
\fF
=
\left[
\begin{array}{c}
\ftranslation(-\vd_{1},\vk)\odot\fA_{1} \\
\vdots \\
\ftranslation(-\vd_{\nimage},\vk)\odot\fA_{\nimage} \\
\end{array}
\right]
\]
with each template $\fSlice\circ\rotation(\tau_{j})\circ\fF$ corresponding to a cross-section of the volume $\fF(\vkappa)$, with $\vkappa\in\Real^{3}$.
Each of these cross-sections can be viewed as a sample of the overall volume $\fF(\vkappa)$, with values given by the appropriate entry on the right-hand-side of the equation above.
With sufficiently many cross-sectional samples, a functional representation of the volume $\fF(\vkappa)$ for any point $\vkappa\in\Real^{3}$ can be reconstructed simply by collecting and averaging the images that correspond to that point.
That is, for each $\vkappa\in\Real^{3}$ we have the formula:
\begin{eqnarray}
\fF^{\bp}(\vkappa) := \frac{1}{\left|\cJ(\vkappa)\right|} \sum_{j'\in\cJ(\vkappa)} \ftranslation(-\vd_{j'},\vk)\cdot\fA_{j'}\left(\vk\right) \comma\text{\ where:\ } \vk:=\vk(\vkappa;j')=\rotation(\tau_{j'})\circ\vkappa \period
\label{eq_0qbp_ctf1}
\end{eqnarray}
In this formula the set $\cJ(\vkappa)$ includes only those image-indices $j'$ for which $\fA_{j'}$ contributes to the point $\vkappa$.
That is to say, for each $\fA_{j'}$ there must exist a point $\vk\in\Real^{2}$ corresponding to a location in the equatorial-plane in $\Real^{3}$ onto which the point $\vkappa\in\Real^{3}$ is rotated by $\rotation(\tau_{j'})$.
This point $\vk\in\Real^{2}$ will implicitly depend on $\vkappa$ as well as the chosen image-index $j'$.
The normalizing factor $1/|\cJ(\vkappa)|$ ensures that $\fF^{\bp}(\vkappa)$ equals the average of all the image-values associated with $\vkappa$.

The above formula can easily be generalized to deal with image-specific CTF-values.
Indeed, by interpreting each image-value $\fA_{j'}(\vk)$ as a noisy sample of $\fF^{\bp}(\vkappa)$, we can think of Eq. \ref{eq_0qbp_ctf1} as a local maximum-likelihood estimate of $\fF^{\bp}(\vkappa)$.
More specifically, one can imagine sampling the image-values $a_{j}$ associated with a particular $\vkappa\in\Real^{3}$ from the corresponding volume-value $f$:
\[ a_{j} = f + \epsilon_{j} \comma \quad\text{\ with each $\epsilon_{j}$ drawn independently from\ }\cN(0,\fsigma^{2}) \period \]
Given a collection of $J$ such observations, the local-maximum-likelihood estimate for $f$ is:
\[ f^{\opt} = \frac{1}{J}\sum_{j=1}^{J} a_{j} \period \]
If each image-value $a_{j}$ is weighted by an image-specific CTF-value $c_{j}$, then we can adjust our model and local maximum-likelihood estimate accordingly:
\[ a_{j} = f\cdot c_{j} + \epsilon_{j} \quad\implies\quad f^{\opt} = \frac{\sum_{j=1}^{J} c_{j}a_{j}}{\sum_{j=1}^{J} c_{j}^{2}} \period \]
Applying this logic to Eq. \ref{eq_0qbp_ctf1} yields:
\begin{eqnarray}
\tnimage(\vkappa) := \sum_{j'\in\cJ(\vkappa)}\left|CTF_{j'}\left(\vk\right)\right|^{2} \comma\text{\ and}
\label{eq_Jqbp_ctf}
\end{eqnarray}
\begin{eqnarray}
\fF^{\bp}(\vkappa) := \frac{1}{\tnimage(\vkappa)}\sum_{j'\in\cJ(\vkappa)} CTF_{j'}\left(\vk\right)\cdot\ftranslation(-\vd_{j'},\vk)\cdot\fA_{j'}\left(\vk\right) \comma
\label{eq_Fqbp_ctf}
\end{eqnarray}
where once again the $\vk\in\Real^{2}$ is implicitly related to $\vkappa$ and $j'$.
This strategy coincides with the CTF-weighting of Bayesian inference described in \cite{Scheres2012,Scheres2012b}.

The term $\tnimage(\vkappa)$ can be interpreted as the accumulated evidence at each point $\vkappa$ in frequency-space.
Similarly, the term $\tnimage(\vkappa)\cdot\fF^{\bp}(\vkappa)$ can be thought of as the accumulated image-data at $\vkappa$.
The quotient of these two terms (i.e., $\fF^{\bp}(\vkappa)$) can be thought of as a first-moment.
Later on below we'll see a similarly constructed second-moment within the likelihood.

\section{Numerical details for back-projection}
\label{sec_Volume_nusht}

Our algorithm attempts to solve the back-projection problem exactly up to a given spherical-harmonic degree $\lmax$ (which is itself related to the maximum frequency $K$ determining the resolution).

To describe our approach in more detail, let's consider the following summation $\Sigma(M)$, originally represented as a 2-dimensional `template-wise' integral:
\begin{equation}
\Sigma(M) = \sum_{j}\int_{\vk\in\Omega(\kmax)\subset\Real^{2}} M_{j}^{\dag}(\vk)\cdot\left[\fSlice\circ\rotation(\tau_{j})\circ\fF\right](\vk) d\vk \comma
\end{equation}
for some collection of data $\{M_{j}(\vk)\}$.

Let's first consider the case where $\nimage=1$, and we have only a single nonzero data-value of $M=1$ associated with the viewing-angle $\tau$ at a point $\vk$ for which $\vkappa=\inverse{\rotation}(\tau)\circ\vk$ is positioned at the north-pole $\vkappa_{0}=\transpose{[0,0,1]}$.
Instead of calculating the sum $\Sigma(M)$ by integrating $M^{\dag}$ against the template $\fSlice\circ\rotation(\tau)\circ\fF$ in 2-dimensions, we can instead just integrate a delta-function `monopole impulse' concentrated at the north-pole against the original volume $\fF$ in 3-dimensions.
Put another way, our single nonzero data-value can be associated with the back-projected volume $\delta(\vkappa-\vkappa_{0})$, again referring to a delta-function impulse concentrated at the north-pole $\vkappa_{0}$.

When restricted to the spherical-shell of radius $1$ and projected onto the spherical-harmonics of degree $l\leq\lmax$, the ideal impulse $\delta(\vkappa-\vkappa_{0})$ (itself fully concentrated at $\vkappa_{0}$) produces a band-limited impulse which is a spherical-sinc that is now centered around (but no longer fully concentrated at) its source at the north-pole:
\[ \tilde{\delta}(\hkappa) = \sum_{l=0}^{\lmax} \sqrt{4\pi} \cdot \sqrt{1+2l} \cdot P_{l}(\cos(\epolara)) = \sum_{l=0}^{\lmax} 4\pi \cdot Y_{l}^{0}(\hkappa) \comma \]
where $P_{l}$ is a legendre-polynomial of degree $l$ and, by construction, the spherical-harmonic coefficients $\tilde{\delta}_{l}^{m}$ are given by $4\pi$ when $m=0$.

Similarly, a single data-value of $M=1$ positioned at the equatorial-point $\vkappa_{1}=\transpose{[\cos(\psi),\sin(\psi),0]}$ will (given our conventions) produce the back-projected impulse $\left[\rotation([0,\pi/2,\psi])\circ\tilde{\delta}\right](\hkappa)$ (i.e., a spherical-sinc with a source positioned at the appropriate point on the equator).

Due to linearity, we can associate the more general data-pool $\{M_{j}(\vk)\}$ with the back-projected shell (for any radius $k$):
\begin{equation}
\tilde{M}(\vkappa) = \sum_{j} \int_{\psi\in[0,2\pi]} M_{j}(\vk) \cdot \left[\inverse{\rotation}(\tau_{j})\circ\rotation([0,\pi/2,\psi])\circ\tilde{\delta}\right](\hkappa) d\psi \comma
\label{eq_qbp_monopole_a}
\end{equation}
which specifies $\tilde{M}(\vkappa)$ at magnitude $k$.
In this expression we implicitly relate $\vkappa\in\Omega(\kmax)\subset\Real^{3}$ to the amplitude and unit-vector $(k,\hkappa)$. Similarly, we relate $\vk\in\Omega(\kmax)\subset\Real^{2}$ to the amplitude and angle $(k,\psi)$. Finally, we relate $\vk = \rotation(\tau_{j})\circ\vkappa$.
In words, this simply amounts to walking through each piece of data $M_{j}$ and, for each point $\vk\in\Real^{2}$, placing an impulse of the appropriate magnitude on the associated point $\vkappa=\inverse{\rotation}(\tau_{j})\circ\vk$ in 3-dimensional space.

With this reorganization we can now represent the sum $\Sigma(M)$ as a 3-dimensional `volumetric' integral:
\begin{equation}
\Sigma(M) = \int_{\vkappa\in\Omega(\kmax)\subset\Real^{3}} \tilde{M}^{\dag}(\vkappa)\cdot\fF(\vkappa)\cdot\frac{d\vkappa}{|\vkappa|} \period
\label{eq_qbp_monopole_b}
\end{equation}

To compute $\tilde{M}(\vkappa)$ efficiently, we employ the following strategy:
\begin{enumerate}
\item First, we apply a linear-transformation $\tilde{D}$ to mollify the impulse $\tilde{\delta}$ and ensure that the support of $\tilde{D}\circ\tilde{\delta}(\hkappa)$ is not too large on the surface of the shell.
\item Alongside $\tilde{D}$, we construct a quadrature-mesh on the surface of the shell which can both (i) resolve the mollified impulse $\tilde{D}\circ\tilde{\delta}$ and (ii) accurately integrate the spherical-harmonics $Y_{l}^{m}$ up to degree $\lmax$.
\item Now, for each $j$ and $q$, the image-value $\fA_{j}(k,\psi_{q})$ is associated with the compactly-supported mollified impulse $\fA_{j}(k,\psi_{q})\cdot\inverse{\rotation}(\tau_{j})\circ\rotation([0,\pi/2,\psi_{q}])\circ\tilde{\delta}(\hkappa)$, which is simply a spherical-sinc with a source at $\inverse{\rotation}(\tau_{j})\circ\transpose{[\cos(\psi),\sin(\psi),0]}$ which we sample on the quadrature-mesh.
\item After accumulating all such values for all $j$ and $q$ on the quadrature-mesh, we use quadrature to recover the spherical-harmonic coefficients of $\tilde{D}\tilde{M}$.
\item Finally, we `undo' the effect of $\tilde{D}$ and calculate $\tilde{M}:=[\tilde{D}]^{-1}\circ [\tilde{D}\tilde{M}]$.
\end{enumerate}

This recipe provides a template to calculate $\tnimage$ and $\tnimage\cdot\fF^{\bp}$:
\begin{eqnarray}
\tnimage(\vkappa) & = & \sum_{j}\int_{\psi\in[0,2\pi]} |CTF_{j}(\vk)|^{2} \cdot \widetilde{\rotation\hspace{-0.125em}\delta} \comma \\
\left[\tnimage\cdot\fF^{\bp}\right](\vkappa) & = & \sum_{j}\int_{\psi\in[0,2\pi]} CTF_{j}(\vk)\cdot\ftranslation(-\delta_{j},\vk)\cdot\fA_{j}(\vk) \cdot \widetilde{\rotation\hspace{-0.125em}\delta} \comma
\end{eqnarray}
\[ \text{with:}\ \ \widetilde{\rotation\hspace{-0.125em}\delta}:=\left[\inverse{\rotation}(\tau_{j})\circ\rotation([0,\pi/2,\psi])\circ\tilde{\delta}\right](\hkappa) \comma \]
where once again $\vkappa=(k,\hkappa)$,  $\vk=(k,\psi)$ and $\vk = \rotation(\tau_{j})\circ\vkappa$.

\subsection{designing $\tilde{D}$:\ }
In this recipe we have the luxury of designing the mollifier $\tilde{D}$. We choose $\tilde{D}$ so that:
\begin{enumerate}
\item $\tilde{D}$ is diagonalized by the spherical-harmonic basis. That is to say: the spherical-harmonic coefficients $[\tilde{D}\circ\tilde{\delta}]_{l}$ are given by $\tilde{D}_{l} \cdot \tilde{\delta}_{l}$, for a collection of coefficients $\tilde{D}_{l}$. This allows for $\tilde{D}$ and $\tilde{D}^{-1}$ to be quickly applied in the spherical-harmonic basis.
\item The majority of the amplitude of $\tilde{D}\circ\tilde{\delta}$ is close to the north- and south-pole. Put another way, we would like:
\[ I^{2}(\epolara_{\north},\epolara_{\south};\tilde{D}) := \int_{\epolara\in [\epolara_{\north},\epolara_{\south}]} \int_{\eazimub\in [0,2\pi]} \left| \tilde{D}\circ\tilde{\delta}(\epolara,\eazimub) \right|^{2} d\eazimub \sin(\epolara)d\epolara  \leq \epsilon_{\pole}^{2} \comma \]
for parameters including cutoff polar-cap angles $\epolara_{\north}$ and $\epolara_{\south}$ and an error-tolerance of $\epsilon_{\pole}^{2}$, where $\epolara_{\north}$ and $\epolara_{\south}$ are as close to $0$ and $\pi$ as possible, while $\epsilon_{\pole}^{2}$ is as small as possible. This constraint allows $\tilde{D}\circ\tilde{\delta}$ to be compactly supported on each spherical shell.
\item The coefficients $\tilde{D}_{l}$ are not too small for $l\leq\lmax$. That is, we would like:
\[ \min(\tilde{D}_{l}) \geq \tilde{D}_{\min} > 0 \comma \]
for some lower-bound $\tilde{D}_{\min}$. This constraint bounds the operator-norm of $\tilde{D}^{-1}$, allowing for both $\tilde{D}$ and $\tilde{D}^{-1}$ to be well-conditioned.
\end{enumerate}

Note that the latter two criteria for $\tilde{D}$ are in conflict with one another: compressing the support of the mollified impulse $\tilde{D}\circ\tilde{\delta}$ can only be done if some of the $\tilde{D}_{l}$ are small.
Thus, while in principle we could use a standard surface-diffusion (i.e., $\tilde{D}_{l}\sim\exp(l(l+1)t)$ for some fixed diffusion-time $t$), we find that more efficient mollifiers can be constructed by solving a quadratic program which balances the constraints placed on $\tilde{D}$.

For example, given a set of coefficients $\tilde{D}_{l}$, we can write $I^{2}(\epolara_{\north},\epolara_{\south};\tilde{D})$ as the quadratic-form:
\[ I^{2}(\epolara_{\north},\epolara_{\south};\tilde{D}) = \sum_{l=0}^{\lmax}\sum_{l'=0}^{\lmax} \tilde{D}_{l} \cdot \tilde{H}(l,l';\epolara_{\north},\epolara_{\south}) \cdot \tilde{D}_{l'} \comma \]
with a kernel:
\[ \tilde{H}(l,l';\epolara_{\north},\epolara_{\south}) = 8\pi^{2} \int_{\epolara\in [\epolara_{\north},\epolara_{\south}]} \sqrt{(1+2l)(1+2l')} P_{l}(\cos(\epolara)) P_{l'}(\cos(\epolara)) \sin(\epolara)d\epolara \period \]
By fixing $\epolara_{\north},\epolara_{\south}$ and $\tilde{D}_{\min}$ and running a quadratic-program with a constraint $\tilde{D}_{l}\geq\tilde{D}_{\min}$, we can quickly find the optimal values of $\tilde{D}_{l}$ and the associated error $\epsilon_{\pole}^{2}$. This procedure can be run as a precomputation, scanning through values of $\epolara_{\north},\epolara_{\south}$ and $\tilde{D}_{\min}$ to find a mollifier which best satisfies the sparsity and error-tolerance constraints provided by the user.

Note that the quadratic-program described above only controls the amplitude of the mollified sinc away from the poles, but does {\em not} control the amplitude of the first- or second-derivatives of the mollified sinc.
Controlling these derivatives can be easily accomplished by adding to $\tilde{H}(l,l';\epolara_{\north},\epolara_{\south})$ additional kernels that integrate higher-order derivatives of the $P_{l}(\cos(\epolara))$.

Note also that, in the discussion above, we have restricted ourselves to mollifiers $\tilde{D}$ that were diagonal in the spherical-harmonic basis.
This is certainly not necessary, as our main design constraint involved being able to apply $\tilde{D}$ `quickly' in the spherical-harmonic basis.
This constraint implies that $\tilde{D}$ could be block-diagonal (e.g., like the wigner-d matrix), or more generally sparse, or even factorizable.
With these extra degrees of freedom available to $\tilde{D}$ one can design mollifiers that accurately capture $\tilde{\delta}$ and its derivatives, all while having sparse support.

\subsection{accommodating derivatives:\ }

We can extend the notion of back-projection to deal with certain differential quantities.
To illustrate this idea, let's consider the following template-wise summation $\Sigma_{\iota}(M)$:
\begin{equation}
\Sigma_{\iota}(M) = \sum_{j}\int_{\vk\in\Omega(\kmax)\subset\Real^{2}} M_{j}^{\dag}(\vk)\cdot\partial_{\tau_{j,\iota}}\left[\fSlice\circ\rotation(\tau_{j})\circ\fF\right](\vk) d\vk \comma
\end{equation}
where $\partial_{\tau_{j,\iota}}$ refers to the $\iota$-component of $\tau_{j}$ (i.e., one of $\{\egammaz_{j},\epolara_{j},\eazimub_{j}\}$).

Again, due to linearity, we can associate the data-pool $\{M_{j}(\vk)\}$ with the back-projected shell:
\begin{eqnarray}
\tilde{M_{\iota}}(\vkappa) & = & \sum_{j} \int_{\psi\in[0,2\pi]} M_{j}(\vk) \cdot \left[\partial_{\tau_{j,\iota}}\inverse{\rotation}(\tau_{j})\circ\rotation([0,\pi/2,\psi])\circ\tilde{\delta}\right](\hkappa) d\psi \comma \\
\ & = &  \sum_{j} \int_{\psi\in[0,2\pi]} M_{j}(\vk) \cdot \left[\inverse{\rotation}(\tau_{j})\circ\tilde{\delta}^{\dipole}_{j,\iota,\psi}\right](\hkappa) d\psi \comma
\label{eq_qbp_dipole_a}
\end{eqnarray}
where the dipole-impulse function $\tilde{\delta}^{\dipole}_{j,\iota,\psi}$ is given by:
\begin{equation}
\tilde{\delta}^{\dipole}_{j,\iota,\psi} = \rotation(\tau_{j})\circ\partial_{\tau_{j,\iota}}\inverse{\rotation}(\tau_{j})\circ\rotation([0,\pi/2,\psi])\circ\tilde{\delta} \period
\end{equation}
This dipole-impulse corresponds to placing two monopole impulses of opposite signs infinitesimally close to one another near the position $\vkappa=\inverse{\rotation}(\tau_{j})\circ\vk$, with the dipole oriented in a direction (along the surface of the sphere) determined by $\tau_{j}$, $\psi$ and $\iota$.

With this reorganization we can once again represent the sum $\Sigma(M)$ as a volumetric integral:
\begin{equation}
\Sigma_{\iota}(M) = \int_{\vkappa\in\Omega(\kmax)\subset\Real^{3}} \tilde{M_{\iota}}^{\dag}(\vkappa)\cdot\fF(\vkappa)\cdot\frac{d\vkappa}{|\vkappa|} \period
\label{eq_qbp_dipole_b}
\end{equation}

\section{Volumetric representation of the likelihood in the image-aligned limit}
\label{sec_Volume_likelihood_3d}

The image-aligned single-particle negative-log-likelihood shown in Eq. \ref{eq_P_AA_given_F_maximum_likelihood} can be written as a volumetric-integral.
To see this more clearly in the context of the low-noise version of the image-aligned limit, let's write:
\begin{multline}
-\log\left(P\left(\{\fA_{j}\}\givenbig \fF \right)\right) \ \sim \ \\
\frac{1}{2\fsigma^{2}}\sum_{j=1}^{\nimage} \iint_{\vk\in\Omega_{\kmax}\subset\Real^{2}} \left|\fA_{j}(\vk)-CTF_{j}(\vk)\odot\ftranslation(\vd;\vk)\odot\left[\fSlice\circ\rotation(\tau_{j})\circ\fF\right](\vk)\right|^{2} d\vk \period
\label{eq_nll_q2d}
\end{multline}
By interchanging the integral and the sum, we can write:
\begin{equation}
-\log\left(P\left(\{\fA_{j}\}\givenbig \fF \right)\right) \ \sim \
\frac{1}{2\fsigma^{2}}\iiint_{\vkappa\in\Omega_{\kmax}\subset\Real^{3}}\tnimage(\vkappa)\left\{ \left|\fF(\vkappa) - \fF^{\bp}(\vkappa)\right|^{2} + \fF^{\var}(\vkappa) \right\} \frac{d\vkappa}{\left|\vkappa\right|} \comma
\label{eq_nll_qbp}
\end{equation}
with $\tnimage$ and $\fF^{\bp}$ given by Eq. \ref{eq_Jqbp_ctf} and \ref{eq_Fqbp_ctf}, and the second-moment $\fF^{\var}$ given by:
\begin{eqnarray}
\fF^{\var}(\vkappa) + \left|\fF^{\bp}(\vkappa)\right|^{2} = \frac{1}{\tnimage(\vkappa)}\sum_{j'\in\cJ(\vkappa)} \left| \ftranslation(\vd;\vk)\odot\fA_{j'}(\vk) \right|^{2} = \frac{1}{\tnimage(\vkappa)}\sum_{j'\in\cJ(\vkappa)} \left| \fA_{j'}(\vk) \right|^{2} \comma
\label{eq_2qbp_ctf}
\end{eqnarray}
\[  \text{\ with:\ } \vk=\rotation(\tau_{j'})\circ\vkappa \period \]
This `volumetric' expression represents the likelihood as a collection of 3-dimensional integrals; this can be compared with the template-wise expression from Eq. \ref{eq_P_AA_given_F_maximum_likelihood}.
Note also that, as mentioned earlier, in the image-aligned-limit the likelihood is maximized when $\fF$ matches the first-moment $\fF^{\bp}$.
This first-moment $\fF^{\bp}$ itself depends on the maximum-likelihood-alignments $\{\tau^{\opt}_{j}\},\{\vd^{\opt}_{j}\}$, the $\{CTF_{j}\}$ and the images $\{\fA_{j}\}$.

Using a very similar strategy, we can rewrite the likelihood in Eq. \ref{eq_P_AA_given_F} as:
\begin{eqnarray}
P\left(\{\fA_{j}\}\givenbig \fF \right) & = & \idotsint\limits_{\{\tau_{j}\},\{\vd_{j}\}} \nonumber \\
 \ & \ & \hspace{-6em} \times\hspace{1em} \frac{1}{(\sqrt{2\pi}\ \fsigma)^{\dof\cdot\nimage}} \exp\left( - \frac{1}{2\fsigma^{2}} \iiint_{\vkappa\in\Omega_{\kmax}\subset\Real^{3}} \tnimage(\vkappa)\left\{ \left|\fF(\vkappa) - \fF^{\bp}(\vkappa)\right|^{2} + \fF^{\var}(\vkappa) \right\}\frac{d\vkappa}{\left|\vkappa\right|}\right) \nonumber \\
 \ & \ & \hspace{-1em} \times\hspace{1em} P\left(\{\tau_{j}\},\{\vd_{j}\}\givenmed\fF\right) \hspace{1em} \times \hspace{1em} d\{\tau_{j}\} d\{\vd_{j}\} \comma
\end{eqnarray}
where the outer integral is over all possible sets of image-alignments $\{\tau_{j}\},\{\vd_{j}\}$.
The terms $\tnimage$, $\fF^{\bp}$ and $\fF^{\var}$ within the integrand are each functions of the alignments $\{\tau_{j}\},\{\vd_{j}\}$.
The probability in the integrand is:
\begin{equation}
P\left(\{\tau_{j}\},\{\vd_{j}\}\givenmed\fF\right) = \prod_{j} P\left(\tau_{j},\vd_{j}\givenmed\fF\right) \comma
\end{equation}
which can be interpreted as the prior-probability of observing a collection of alignments, given the particle $\fF$.

To make further manipulation easier, we'll rewrite Eq. \ref{eq_nll_qbp} as:
\begin{multline}
\cL\left(\{\fA_{j}\}\givenbig \fF; \{\tau^{\opt}_{j}\}; \{\vd^{\opt}_{j}\} \right) \ \sim \ \\
\frac{1}{2}\iiint_{\vkappa} \left(M_{20}(\vkappa)\left|\fF(\vkappa)\right|^{2} - 2\Re\left(M_{11}^{\dag}(\vkappa)\fF(\vkappa)\right) + M_{02}(\vkappa)\right) \frac{d\vkappa}{\left|\vkappa\right|} \comma
\label{eq_ssnll_q3d_pre}
\end{multline}
where $\cL$ is $\fsigma^{2}$ times the negative-log-likelihood, and $M_{20}$, $M_{11}$ and $M_{02}$ represent accumulations of the CTF- and image-data:
\begin{eqnarray}
\cL\left(\{\fA_{j}\}\givenbig \fF; \{\tau^{\opt}_{j}\}; \{\vd^{\opt}_{j}\} \right) & := & -\fsigma^{2}\log\left(P\left(\{\fA_{j}\}\givenbig \fF \right)\right) \comma \\
M_{20}\left(\vkappa; \{\tau^{\opt}_{j}\}; \{\vd^{\opt}_{j}\}; \{CTF_{j}\}]\right) & := & \sum_{j'\in\cJ(\vkappa)} \left|CTF_{j'}\left(\vk\right)\right|^{2} \comma \\
M_{11}\left(\vkappa; \{\tau^{\opt}_{j}\}; \{\vd^{\opt}_{j}\}; \{CTF_{j}\}; \{\fA_{j}\}]\right) & := & \sum_{j'\in\cJ(\vkappa)} CTF_{j'}\left(\vk\right)\cdot\ftranslation(-\vd_{j'},\vk)\cdot\fA_{j'}\left(\vk\right) \comma \\
M_{02}\left(\vkappa; \{\fA_{j}\}]\right) & := & \sum_{j'\in\cJ(\vkappa)} \left|\fA_{j'}\left(\vk\right)\right|^{2} \period
\label{eq_ssnll_q3d_sub}
\end{eqnarray}

\section{Synthetic-images}
\label{sec_synthetic_images}

\subsection{single-particle:\ }
Given a volume $\fF$, we can create a distribution of synthetic-images by adding the appropriate amount of noise to the templates associated with $\fF$.
To construct this distribution we must choose the prior joint-distribution of alignments and CTFs, denoted by $P\left(\tau,\vd,CTF\givenbig\fF\right)$, as well as the temperature $\fsigma$.
Given these ingredients, we can create a synthetic-image $\fA^{\synthetic}(\vk)$ by randomly drawing $\{\tau,\vd,CTF\}$ from $P\left(\tau,\vd,CTF\givenbig\fF\right)$, then producing the template $\fS(\vk;\tau;\vd;CTF;\fF)$, and then finally adding noise in a manner consistent with Eq. \ref{eq_image_noise_model}.
In the image-aligned limit we do the same, except in this case the synthetic-images are simply the templates $\fS(\vk;\tau;\vd;CTF;\fF)$ with no additional noise added.
With this construction, each synthetic-image $\fA^{\synthetic}_{j}(\vk)$ is associated with a particular viewing-angle $\tau_{j}$, displacement $\vd_{j}$ and $CTF_{j}$, and we can express this dependence by writing $\fA^{\synthetic}_{j}\left(\vk;\tau_{j};\vd_{j};CTF_{j};\fF\right)$.

\subsection{consistency between the likelihood and the synthetic-image distribution:\ } 
Note that the likelihood in Eq. \ref{eq_P_AA_given_F} is consistent with the synthetic-images described above.
To step through the details, let's start by considering the simpler case where we ignore displacements and CTFs (i.e., we assume each $\vd_{j}\equiv 0$ and each $CTF_{j}\equiv 1$).
In this simple case the volume $\fF$ is associated with the family of templates $\{\fS(\vk;\tau;\fF)\}$, indexed by the viewing-angle $\tau$.
Given a specific distribution of viewing-angles $P(\tau\givenmed\fF)$, the family of templates $\{\fS(\vk;\tau;\fF)\}$ produces a distribution $P(\fS\givenmed\fF)$ of templates in image-space (here $\fS$ is a dummy-variable indicating the template's location in image-space).

Assuming isotropic noise with variance $\fsigma^{2}$, the distribution of synthetic-images $P(\fA^{\synthetic}\givenmed\fF)$ associated with $P(\fS\givenmed\fF)$ is simply the convolution of $P(\fS\givenmed\fF)$ with a gaussian:
\[ P(\fA^{\synthetic}\givenmed\fF) = P(\fS\givenmed\fF) \star \cN(0,\fsigma^{2}) \period \]
In this formula $\fA^{\synthetic}$ is a dummy variable indicating location in image-space, and $\cN(0,\fsigma)$ represents the gaussian in image-space associated with the isotropic variance of $\fsigma^{2}$ per unit-area in $\vk$-space.

With this perspective, the likelihood $P\left(\{A_{j}\}\givenbig\fF\right)$ in Eq. \ref{eq_P_AA_given_F} is equal to the probability of drawing the image-set $\{\fA_{j}\}$ from the distribution of synthetic-images $P(\fA^{\synthetic}\givenmed\fF)$, provided that the likelihood is calculated using the same prior distribution $P(\tau\givenmed\fF)$ and temperature $\fsigma$ used to generate the distribution of synthetic-images.

\subsection{consistency between $\fF$ and the synthetic-images:\ }
Following directly from the construction of the synthetic-image distribution $P(\fA^{\synthetic}\givenmed\fF)$ above, we can see that the function $\fF$ along with the prior distribution $P(\tau\givenmed\fF)$ are a global maximum of the likelihood:
\begin{eqnarray}
\left\{\ \fF \ , \ P(\tau\givenmed\fF) \ \right\}  & = & \argmax_{\{\fG,P(\tau\givenmed\fG)\}} \lim_{\nimage\rightarrow\infty} P\left(\{A^{\synthetic}_{j}\}\givenbig\fG \semicolonmed P(\tau\givenmed\fG) \semicolonmed \fsigma \right) \comma
\label{eq_A_synthetic_given_F_consistent}
\end{eqnarray}
where we have emphasized the dependence of the likelihood $P\left(\{A^{\synthetic}_{j}\}\givenbig\fG\right)$ on the prior distribution and temperature.
Note that in Eq. \ref{eq_A_synthetic_given_F_consistent} we have assumed (via $\nimage\rightarrow\infty$) that $\{A^{\synthetic}_{j}\}$ is a pool of synthetic-images which is sufficiently large to sample $P(\fA^{\synthetic}\givenmed\fF)$.

To see why Eq. \ref{eq_A_synthetic_given_F_consistent} might be true, we can consider any other candidate template-distribution $P'(\fS)$.
Given any temperature $\fsigma'$, the associated distribution of synthetic-images $P'(\fA^{\synthetic})$ is:
\[ P'(\fA^{\synthetic}) = P'(\fS) \star \cN(0,[\fsigma']^{2}) \period\]

Now, given an infinite number of synthetic-images $\{\fA^{\synthetic}_{j}\}$ sampled from $P(\fA^{\synthetic}\givenmed\fF)$, we can ask the following questions: (i) what is the probability of drawing this image-set from $P(\fA^{\synthetic}\givenmed\fF)$, and (ii) what is the probability of drawing this same image-set from $P'(\fA^{\synthetic})$.
The logarithm of the ratio of the probabilities from (i) and (ii) is proportional to the Kullback-Leibler divergence:
\[ KL( P || P') := \int_{\fA^{\synthetic}} P(\fA^{\synthetic}\givenmed\fF) \ \log\left(\frac{P(\fA^{\synthetic}\givenmed\fF)}{P'(\fA^{\synthetic})}\right) d\fA^{\synthetic} \comma \]
where the integral is carried out over image-space.

The KL-divergence is always nonnegative, and is only equal to $0$ when $P'\equiv P$.
This directly implies that the log-probability-ratio is highest when $P(\fA^{\synthetic}\givenmed\fF)\equiv P'(\fA^{\synthetic})$.
Put more simply: given any distribution $P$, no other distribution $P'$ can possibly produce a collection of samples that better represents $P$.
Consequently, assuming the approximate temperature $\fsigma'$ matches the true temperature $\fsigma$ used to generate the synthetic-images, no template-distribution $P'(\fS)$ can produce a higher likelihood than the template-distribution $P(\fS\givenmed\fF)$.
Thus, no other pair $\{\fG,P(\tau\givenmed\fG)\}$ can generate a higher likelihood than the true $\{\fF,P(\tau\givenmed\fF)\}$.

Note that it is still possible for other volumes to produce the {\em same} likelihood as the true volume $\fF$ (i.e., $\fF$ is not guaranteed to be the only global maximum of Eq. \ref{eq_A_synthetic_given_F_consistent}).
Specifically, another volume $\fG$ might look different from $\fF$, but produce a family of templates $\{\fS(\vk;\tau;\fG)\}$ that overlaps (in image-space) the family of templates $\{\fS(\vk;\tau;\fF)\}$ produced by $\fF$.
In this scenario there exist viewing-angle distributions $P(\tau\givenmed\fG)$ and $P(\tau\givenmed\fF)$ such that the induced distribution $P'(\fS):=P(\fS\givenmed\fG)$ is the same as $P(\fS\givenmed\fF)$. 
Examples along these lines are discussed below.

\subsection{multi-particle:\ }

We can also use a similar strategy to create a pool of synthetic-images associated with multiple volumes $\{\fF_{i}\}$.
In this case we draw alignments, CTFs and volume-labels from the prior joint-distribution $P\left(\tau,\vd,CTF,i\givenbig\{\fF_{i}\}\right)$.

In this multi-volume scenario a statement similar to Eq. \ref{eq_A_synthetic_given_F_consistent} holds.
Suppressing the notation for $\vd$ and the CTF momentarily, we can write:
\begin{eqnarray}
\left\{ \ \{\fF_{i}\} \ , \ P(\tau,i\givenmed\{\fF_{i}\}) \ \right\} =
 \argmax_{\{\{\fG_{i}\},P(\tau,i\givenmed\{\fG_{i}\})\}} \lim_{\nimage\rightarrow\infty} P\left(\{A^{\synthetic}_{j}\}\givenbig\{\fG_{i}\} \semicolonmed P(\tau,i\givenmed\{\fG_{i}\}) \semicolonmed \fsigma \right) \comma
\label{eq_A_synthetic_given_FF_consistent}
\end{eqnarray}
where again we have emphasized the dependence of the likelihood $P\left(\{A^{\synthetic}_{j}\}\givenbig\{\fG_{i}\}\right)$ on the prior distribution and temperature.

Once again, is is possible for other volume-collections to produce the same likelihood as the true volume-collection $\{\fF_{i}\}$.
This phenomenon is much more pronounced in the multi-volume setting than in the single-volume setting, and it is rather straightforward to construct alternative volume-sets $\{\fG_{i}\}$, each with their own viewing-angle distribution $P(\tau\givenmed\fG_{i})$, which collectively produce the same distribution of synthetic-images as $P(\fA^{\synthetic}\givenmed\{\fF_{i}\})$.
Examples along these lines are discussed below.

\section{Local well-posedness of single-particle reconstruction}
\label{sec_local_well_posedness}

For a fixed data-set $\{\fA_{j}\}$, the likelihood $P(\{\fA_{j}\} \givenbig \fF \semicolonmed P(\tau;\vd\givenmed\fF) \semicolonmed \fsigma )$ from Eq. \ref{eq_P_AA_given_F} is usually a nonconvex function of $\fF$, $P(\tau;\vd\givenmed\fF)$ and $\fsigma$.
Even if we assume that the alignment-distribution $P(\tau ; \vd \givenmed \fF)$ and temperature $\fsigma$ are known, finding the `optimal' $\fF^{\opt}$ which maximizes the likelihood is often not easy \cite{Sigworth1998,Scheres2009,Sigworth2010,Scheres2012}.
In general, the iterative strategies used to search for $\fF^{\opt}$ produce an approximate $\fF^{\loc}$ which has a local (but not global) maximum of the likelihood $P(\{\fA_{j}\} \givenbig \fF^{\loc} )$, which we'll denote by $P^{\loc}$.

One standard strategy for measuring the quality of any particular configuration $\{\fF^{\loc} \commamed P(\tau;\vd\givenmed\fF^{\loc}) \}$ is to search for nearby configurations that might have a similar likelihood.
More specifically, we can consider nontrivial infinitesimal perturbations of the form:
\begin{eqnarray}
\left\{\fF^{\loc} \commamed P(\tau;\vd\givenmed\fF^{\loc}) \right\} \rightarrow \left\{ \ \fF^{\loc}+\epsilon\Delta\fF \ \commamed \ P(\tau;\vd\givenmed\fF^{\loc})+\epsilon\Delta P(\tau;\vd) \ \right\} \period
\label{eq_F_P_perturbation}
\end{eqnarray}
In Eq. \ref{eq_F_P_perturbation} the function $\Delta\fF(\vkappa)$ is the Fourier-transform of a function $\Delta F(\vchi)\in\Leb^{2}(\Real^{3})$ representing a perturbation of the volume $F^{\loc}(\vchi)$.
Similarly, the function $\Delta P(\tau;\vd)$ represents a perturbation of the prior distribution $P(\tau;\vd\givenmed\fF^{\loc})$.
We'll assume that both $\Delta\fF(\vkappa)$ and $\Delta P(\tau;\vd)$ have unit norm, and that $\epsilon\in\Real$ is small.
When considering `nontrival' perturbations, we'll restrict our attention only to infinitesimal deformations of the solution, and implicitly exclude those perturbations that correspond to infinitesimal rigid rotations or translations of the original volume $\fF$ and prior-distribution $P(\tau;\vd\givenmed\fF)$.

Assuming that the configuration $\{\fF^{\loc} \commamed P(\tau;\vd\givenmed\fF^{\loc}) \}$ is indeed a local maximum of the likelihood, then the perturbed negative-log-likelihood should be approximately quadratic to leading-order in $\epsilon$ for a typical choice of $\Delta\fF$ and $\Delta P$:
\begin{multline}
-\log\left(P\left(\{\fA_{j}\} \givenbig \fF^{\loc} + \epsilon\Delta\fF(\vkappa) \semicolonmed P(\tau;\vd\givenmed\fF^{\loc}) + \epsilon\Delta P(\tau;\vd) \semicolonmed \fsigma \right)\right) = \\ 
-\log(P^{\loc}) + \frac{1}{2}\sigma_{H}(\Delta\fF,\Delta P)\epsilon^{2} + \bigO(\epsilon^{3}) \period
\label{eq_P_quad}
\end{multline}
In the expression above, $\sigma_{H}(\Delta\fF,\Delta P)$ is the concavity of the negative-log-likelihood in the direction of $\{\Delta\fF,\Delta P\}$.
The second-order term in Eq. \ref{eq_P_quad} is equivalent to a rayleigh-quotient of the quadratic kernel $H$:
\begin{multline}
\sigma_{H}(\Delta\fF,\Delta P) = \\
\idotsint \Delta P(\tau;\vd)\ \Delta\fF(\vkappa)^{\dagger}\cdot H(\vkappa,\tau,\vd,\vkappa',\tau',\vd') \cdot\Delta\fF(\vkappa')\ \Delta P(\tau',\vd') \ d\vkappa' d\tau' d\vd' d\vkappa d\tau d\vd \comma
\label{eq_P_sigma_H}
\end{multline}
where $H$ is the hessian of the negative-log-likelihood with respect to perturbations of $\fF^{\loc}$ and $P(\tau;\vd\givenmed \fF^{\loc})$:
\begin{multline}
H(\vkappa,\tau,\vd,\vkappa',\tau',\vd') = \\
\partial_{\{ \fF(\vkappa) ; P(\tau,\vd\givenmed\fF) \}}\partial_{\{ \fF(\vkappa') ; P(\tau',\vd'\givenmed\fF) \}} \left\{-\log\left(P\left(\{\fA_{j}\} \givenbig \fF^{\loc} \semicolonmed P(\tau,\vd\givenmed\fF^{\loc}) \semicolonmed \fsigma \right)\right)\right\} \comma
\label{eq_P_H}
\end{multline}
which is directly related to a kind of Fisher information.

If the magnitude $\sigma_{H}(\Delta\fF,\Delta P)$ is large for every nontrivial $\{\Delta\fF,\Delta P\}$, then $P^{\loc}$ indicates a sharp minimum of the negative-log-likelihood-function.
In this scenario we can say that the single-particle reconstruction problem is well-posed within a local neighborhood of $\{\fF^{\loc};P(\tau;\vd\givenmed \fF^{\loc})\}$.
Conversely, if there exist nontrivial directions $\{\Delta\fF,\Delta P\}$ for which $\sigma_{H}(\Delta\fF,\Delta P)$ is small, then these directions correspond to nontrivial perturbations (i.e., deformations) along which the likelihood diminishes only gradually.
Put another way, there will be `nearby' configurations that will have likelihoods similar to the original $P^{\loc}$.
Thus, in this latter scenario the single-particle reconstruction problem is poorly-posed near $\{\fF^{\loc};P(\tau;\vd\givenmed \fF^{\loc})\}$.

One could in principle use Eq. \ref{eq_P_sigma_H} to characterize the well-posedness of the reconstruction near any given $\{\fF^{\loc};P(\tau;\vd\givenmed \fF^{\loc})\}$ by finding the nontrivial direction $\{\Delta\fF^{\critical},\Delta P^{\critical}\}$ which minimizes $\sigma_{H}(\Delta\fF,\Delta P)$:
\begin{eqnarray}
\left\{\Delta\fF^{\critical},\Delta P^{\critical}\right\} = \argmin_{\{\Delta\fF,\Delta P\}} \sigma_{H}(\Delta\fF,\Delta P) \quad \text{\ with\ } \quad \sigma_{H}^{\critical} = \sigma_{H}(\Delta\fF^{\critical},\Delta P^{\critical}) \period
\label{eq_sigma_H_crit}
\end{eqnarray}
Note that $\sigma_{H}^{\critical}$ is the smallest eigenvalue of $H$, and $\{\Delta\fF^{\critical},\Delta P^{\critical}\}$ is the associated eigenvector.

We are not aware of an efficient general strategy for calculating $\sigma_{H}^{\critical}$ and $\{\Delta\fF^{\critical},\Delta P^{\critical}\}$.
The main impasse is that we do not know how to apply the hessian $H$ (or its inverse) in a computationally efficient manner.
However, any given nontrival direction $\{\Delta\fF,\Delta P\}$ can be used to give an upper-bound for $\sigma_{H}^{\critical}$.
Because volume-space is high-dimensional, we use the image-aligned limit (described in the next section) to guide our search.

\section{Estimating the hessian in the image-aligned limit}
\label{sec_local_well_posedness_hessian_0}

As mentioned in the previous section, we are not aware of any efficient strategies for calculating the smallest eigenvalue of the hessian of the negative-log-likelihood from Eq. \ref{eq_P_H}.
However, there are strategies that are useful in the image-aligned limit.

To ease the notation, let's ignore the displacements for now, and let's assume for the moment that the viewing-angles $\{\tau_{j}\}$ for each image are given.
Let's use $\langle\cdot,\cdot\rangle$ to represent the standard inner product over $\vk\in\Omega_{\kmax}\subset\Real^{2}$, and $\int_{\vk}$ to represent the integral over $\vkappa\in\Omega_{\kmax}\subset\Real^{3}$ with respect to the weight $d\vkappa/|\vkappa|$.

With this notation we can use a 2-dimensional representation and define $\cL$ to be the negative-log-likelihood scaled by $\fsigma^{2}$ from Eq. \ref{eq_P_AA_given_F_maximum_likelihood}
\begin{eqnarray}
\cL\left(\{\fA_{j}\}\givenbig \fF; \{\tau_{j}\} \right) & := & -\fsigma^{2}\log\left(P\left(\{\fA_{j}\}\givenbig \fF; \{\tau_{j}\} \right)\right) \comma \\
\ & := & \frac{1}{2}\sum_{j=1}^{\nimage} \iint_{\vk} \left| \fS\left(\vk;\tau_{j};CTF_{j};\fF\right) - \fA_{j}(\vk) \right|^{2} d\vk \comma \\
\ & = & \frac{1}{2}\sum_{j=1}^{\nimage}\langle \fS\left(\tau_{j};CTF_{j};\fF\right) - \fA_{j} , \fS\left(\tau_{j};CTF_{j};\fF\right) - \fA_{j} \rangle \comma
\label{eq_ssnll_q2d}
\end{eqnarray}
where we have suppressed reference to $\vk$ in the integrand, and emphazised the dependence on $\tau_{j}$ and $\fF$.

Equivalently, we can use a 3-dimensional representation and write $\cL$ from Eq. \ref{eq_ssnll_q3d_pre} as:
\begin{multline}
\cL\left(\{\fA_{j}\}\givenbig \fF; \{\tau_{j}\} \right) = \ \\
\frac{1}{2}\int_{\vkappa} \left(M_{20}\left(\{\tau_{j}\};\{CTF_{j}\}\right)\cdot\left|\fF\right|^{2} - 2\Re\left(M_{11}^{\dag}\left(\{\tau_{j}\};\{CTF_{j}\};\{\fA_{j}\}\right)\cdot\fF\right) + M_{02}\left(\{\fA_{j}\}\right)\right) \comma
\label{eq_ssnll_q3d}
\end{multline}
where again we have suppressed reference to $\vkappa$ in the integrand, and emphazised the dependence on $\tau_{j}$ and $\fF$.

The 2-dimensional representation Eq. \ref{eq_ssnll_q2d} makes apparent the following relationships:
\begin{eqnarray}
\partial_{\fF}\cL\left(\{\fA_{j}\}\givenbig \fF; \{\tau_{j}\} \right)\cdot\Delta\fF & = & \sum_{j} \Re\langle \fS\left(\tau_{j};CTF_{j};\Delta\fF\right) , \fS\left(\tau_{j};CTF_{j};\fF\right) - \fA_{j} \rangle \comma 
\label{eq_dssnll_q2d_Gv}
\\
\partial_{\tau_{j'}}\cL\left(\{\fA_{j}\}\givenbig \fF; \{\tau_{j}\} \right) & = & \Re\langle \partial_{\tau_{j'}}\fS\left(\tau_{j'};CTF_{j'};\fF\right) , \fS\left(\tau_{j'};CTF_{j'};\fF\right) - \fA_{j'} \rangle \comma 
\label{eq_dssnll_q2d_Gt}
\\
\partial_{\tau_{j'}}\partial_{\fF}\cL\left(\{\fA_{j}\}\givenbig \fF; \{\tau_{j}\} \right)\cdot\Delta\fF & = & \ \nonumber \\
 & \ & \hspace{-16em} \Re\langle \partial_{\tau_{j'}}\fS\left(\tau_{j'};CTF_{j'};\Delta\fF\right) , \fS\left(\tau_{j'};CTF_{j'};\fF\right) - \fA_{j'} \rangle + \Re\langle \fS\left(\tau_{j'};CTF_{j'};\Delta\fF\right) , \partial_{\tau_{j'}}\fS\left(\tau_{j'};CTF_{j'};\fF\right) \rangle \comma
\label{eq_ddssnll_Htv}
\\
\partial_{\tau_{j'}}\partial_{\tau_{j'}}\cL\left(\{\fA_{j}\}\givenbig \fF; \{\tau_{j}\} \right) & = & \ \nonumber \\
 & \ & \hspace{-16em} \Re\langle \partial_{\tau_{j'}}\partial_{\tau_{j'}}\fS\left(\tau_{j'};CTF_{j'};\fF\right) , \fS\left(\tau_{j'};CTF_{j'};\fF\right) - \fA_{j'} \rangle + \Re\langle \partial_{\tau_{j'}}\fS\left(\tau_{j'};CTF_{j'};\fF\right) , \partial_{\tau_{j'}}\fS\left(\tau_{j'};CTF_{j'};\fF\right) \rangle \period
\label{eq_ddssnll_Htt}
\end{eqnarray}
In these expressions $\tau_{j'}$ involves the three euler-angles $\tau_{j'}=\transpose{[\egammaz_{j'},\epolara_{j'},\eazimub_{j'}]}$, and so $\partial_{\tau_{j'}}\cL$ and $\partial_{\tau_{j'}}\partial_{\fF}\cL$ each have three components, while $\partial_{\tau_{j'}}\partial_{\tau_{j'}}\cL$ is a $3\times 3$ tensor.
Note that Eq. \ref{eq_dssnll_q2d_Gv} provides an expression for the directional-derivative of $\cL$, given a particular volumetric-perturbation $\Delta\fF$, while Eq. \ref{eq_dssnll_q2d_Gt} provides a general expression for the gradient of $\cL$ with respect to arbitrary perturbations in $\tau_{j}$.
Similarly, Eq. \ref{eq_ddssnll_Htv} provides an expression for the hessian applied to a particular volumetric-perturbation $\Delta\fF$, while Eq. \ref{eq_ddssnll_Htt} gives a general expression for the hessian of $\cL$ with respect to arbitrary perturbations in $\tau_{j}$.

The 3-dimensional representation Eq. \ref{eq_ssnll_q3d} demonstrates:
\begin{eqnarray}
\partial_{\fF(\vkappa')}\cL\left(\{\fA_{j}\}\givenbig \fF; \{\tau_{j}\} \right) & = & \ \nonumber \\
 & \ & \hspace{-16em} 
\Re\left( M_{20}\left(\vkappa';\{\tau_{j}\};\{CTF_{j}\}\right)\cdot\fF(\vkappa')\right) - \Re\left(M_{11}^{\dag}\left(\vkappa';\{\tau_{j}\};\{CTF_{j}\};\{\fA_{j}\}\right)\right) \comma 
\label{eq_dssnll_q3d_Gv}
\\
\partial_{\tau_{j'}}\cL\left(\{\fA_{j}\}\givenbig \fF; \{\tau_{j}\} \right)\cdot\Delta\tau_{j'} & = & \ \nonumber \\
 & \ & \hspace{-16em}
\frac{1}{2}\int_{\vkappa} \left(\partial_{\tau_{j'}}M_{20}\left(\{\tau_{j}\};\{CTF_{j}\}\right)\cdot\left|\fF\right|^{2} - 2\Re\left(\partial_{\tau_{j'}}M_{11}^{\dag}\left(\{\tau_{j}\};\{CTF_{j}\};\{\fA_{j}\}\right)\cdot\fF\right) \right)\cdot\Delta\tau_{j'} \comma 
\label{eq_dssnll_q3d_Gt}
\\
\partial_{\fF(\vkappa')}\partial_{\fF(\vkappa')}\cL\left(\{\fA_{j}\}\givenbig \fF; \{\tau_{j}\} \right) & = &
M_{20}\left((\vkappa');\{\tau_{j}\};\{CTF_{j}\}\right) \comma
\label{eq_ddssnll_Hvv}
\\
\partial_{\fF(\vkappa')}\partial_{\tau_{j'}}\cL\left(\{\fA_{j}\}\givenbig \fF; \{\tau_{j}\} \right)\cdot\Delta\tau_{j'} & = & \ \nonumber \\
 & \ & \hspace{-16em}
\left( \Re\left(\partial_{\tau_{j'}}M_{20}\left(\vkappa';\{\tau_{j}\};\{CTF_{j}\}\right)\cdot\fF(\vkappa')\right) - \Re\left(\partial_{\tau_{j'}}M_{11}^{\dag}\left(\vkappa';\{\tau_{j}\};\{CTF_{j}\};\{\fA_{j}\}\right)\right) \right)\cdot\Delta\tau_{j'} \period
\label{eq_ddssnll_Hvt}
\end{eqnarray}
In these expressions $\Delta\tau_{j'}$ also involves three euler-angles $\Delta\tau_{j'}=\transpose{[\Delta\egammaz_{j'},\Delta\epolara_{j'},\Delta\eazimub_{j'}]}$, and so $\partial_{\tau_{j'}}\cL\cdot\Delta\tau_{j'}$ and $\partial_{\fF}\partial_{\tau_{j'}}\cL\cdot\Delta\tau_{j'}$ each involve a sum over three components.
Note that Eq. \ref{eq_dssnll_q3d_Gv} provides a general expression for the gradient of $\cL$ with respect to any perturbation in $\fF$, while Eq. \ref{eq_dssnll_q3d_Gt} provides an expression for the directional-derivative of $\cL$ with respect to a particular perturbation in $\tau_{j}$.
Similarly, Eq. \ref{eq_ddssnll_Hvv} provides a general expression for the hessian with respect to arbitrary perturbations in $\fF$, while Eq. \ref{eq_ddssnll_Hvt} gives an expression for the hessian applied to a particular viewing-angle perturbation $\Delta\tau_{j}$.

Note that we can leverage both the 2- and 3-dimensional representations to write the general gradient:
\begin{equation}
G\cdot\Delta
:= 
\left[
\begin{array}{cc}
G_{F} & G_{\tau} 
\end{array}
\right]
\cdot
\left[
\begin{array}{c}
\Delta\fF \\
\{\Delta\tau_{j}\} \\
\end{array}
\right]
=
\left[
\begin{array}{cc}
\partial_{\fF}\cL\cdot\Delta\fF  + \sum_{j'}\partial_{\tau_{j'}}\cL\cdot\Delta\tau_{j'} 
\end{array}
\right]
\period
\label{eq_dssnll_block}
\end{equation}
Specifically, Eq. \ref{eq_dssnll_q3d_Gv} yields the volumetric-component, while Eq. \ref{eq_dssnll_q2d_Gt} yields the alignment-component.

Similarly, given a particular perturbation $\{ \Delta\fF , \{\Delta\tau_{j}\} \}$, we can apply the hessian.
To describe this in more detail, let's refer to the perturbation $\Delta:=\transpose{[\Delta\fF , \{\Delta\tau_{j}\}]}$ as a block-vector.
The first block of $\Delta$ stores the volumetric-perturbation $\Delta\fF$, while the second block stores the alignment-perturbations $\{\Delta\tau_{j}\}$ for each image in sequence.
With this organization the hessian is also divided into blocks:
\begin{equation}
H\cdot\Delta
:= 
\left[
\begin{array}{cc}
H_{FF} & H_{F\tau} \\
H_{\tau F} & H_{\tau\tau}
\end{array}
\right]
\cdot
\left[
\begin{array}{c}
\Delta\fF \\
\{\Delta\tau_{j}\} \\
\end{array}
\right]
=
\left[
\begin{array}{cc}
\partial_{\fF}\partial_{\fF}\cL\cdot\Delta\fF  + \sum_{j'}\partial_{\fF}\partial_{\tau_{j'}}\cL\cdot\Delta\tau_{j'} \\
\partial_{\tau}\partial_{\fF}\cL\cdot\Delta\fF  + \sum_{j'}\partial_{\tau}\partial_{\tau_{j'}}\cL\cdot\Delta\tau_{j'}
\end{array}
\right]
\period
\label{eq_ddssnll_block}
\end{equation}
The first block $H_{FF}$ refers to the volumetric-component, which is diagonalized in $\vk$-space as shown in Eq. \ref{eq_ddssnll_Hvv}.
The second term $H_{F\tau}$ refers to the volumetric-perturbation induced by any given alignment-perturbation, as shown in Eq. \ref{eq_ddssnll_Hvt}.
The third term $H_{\tau F}$ refers to the alignment-perturbation induced by any given volumetric-perturbation, as shown in Eq. \ref{eq_ddssnll_Htv}.
Finally, the block $H_{\tau\tau}$ refers to the alignment-component, which is diagonalized in template-space as shown in Eq. \ref{eq_ddssnll_Htt}.

\subsection{Finding eigenvalues and eigenvectors of $H$:\ }
\begin{figure}[H]
\centering
\includegraphics[width=1.0\textwidth]{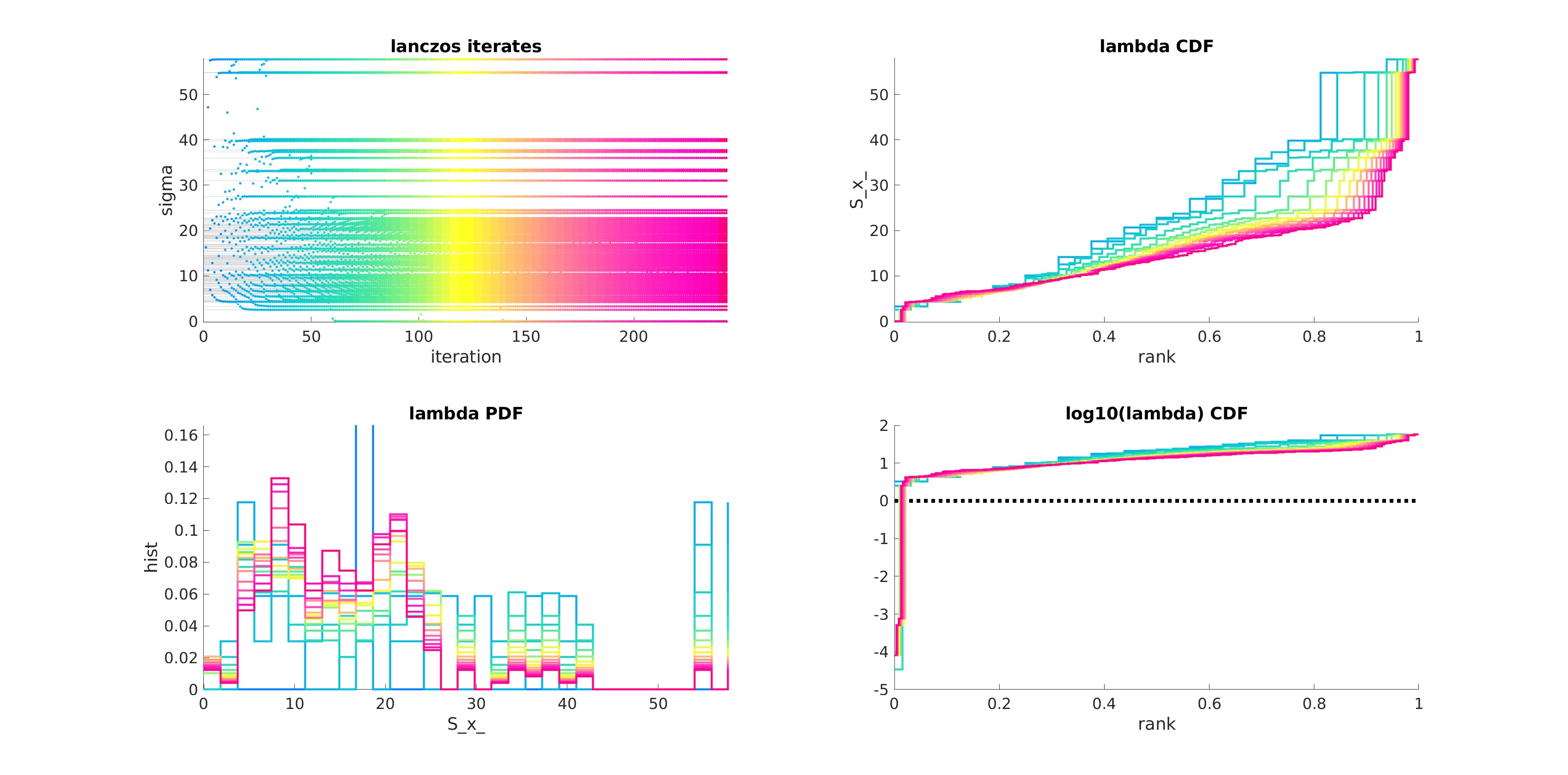}
\caption{
In this figure we illustrate the properties of the hessian $H$, as well as its lanczos-iterates.
For this example we use the TRPV1 molecule at ultra-low resolution: $2\pi\kmax=8$.
This ultra-low resolution allows us to take $\lmax=2\pi\kmax=8$, implying only $(1+\lmax)^{2}=81$ degrees of freedom per frequency-shell.
We also project onto the dominant 3 volumetric-principal-modes (capturing over $99.9\%$ of the alignment-information, as described in \protect\cite{Rangan22}) and determine the alignment-perturbations implicitly (using the first-order scheme described in section \ref{sec_dealing_with_implicitly_defined_tau}).
Thus, the total number of degrees-of-freedom for this simple case-study is $3\times 81=243$.
The first subplot shows the lanczos-iterates along the vertical, with the iteration-number plotted along the horizontal.
The color ranges from cyan (early iterates) to magenta (final iterates).
Note that the outliers in the spectrum are captured quite early on.
The second subplot shows the cumulative-distribution function for the lanczos-iterates, with the different colors corresponding to different iterations (using the same color-scale shown in the top-left).
The third subplot shows the estimated probability-density-function for the lanczos-iterates, while the fourth subplot shows the cumulative-distribution function for the log of the lanczos-iterates.
Note that the log of the determinant of the hessian $\log\det(H)$ is the average of the log of the eigenvalues of $H$, which is proportional to the area under the CDF shown in the bottom-right subplot.
Note also that, once the outliers in the spectrum are well approximated, the $\log\det(H)$ is also reasonably well approximated.
For this example we have $\sim25\%$ relative-error for $\log\det(H)$ by iteration $16$, and $\sim10\%$ relative-error by iteration $64$.
}
\label{fig_trpv1_k8_eig_i1_from_synth_nlt30pm3_p_empirical_summary}
\end{figure}

Given the ability to apply $H$ to any perturbation $\Delta:=\{\Delta\fF,\{\Delta\tau_{j}\}\}$, we can use lanczos-iteration to estimate the extremal eigenvalues and eigenvectors of $H$.
One technical detail that comes into play is that we want to avoid the `trival' eigenvectors of $H$ that correspond to perturbations $\Delta$ which are infinitesimal rigid-rotations of $\{\fF,\{\tau_{j}\}\}$.
Thus, when applying lanczos-iteration we project away the 3-dimensional space associated with these trivial rigid-rotations.
The remaining subspace will include those perturbations which correspond to infinitesimal deformations.
An example illustrating the convergence of lanczos-iteration is shown in Fig \ref{fig_trpv1_k8_eig_i1_from_synth_nlt30pm3_p_empirical_summary}.

Note that, while lanczos-iteration does a good job of focusing on the extremal eigenvalues, it does a poor job of estimating the bulk of the spectrum.
Consequently, while we can expect to use the lanczos-iterates to estimate $\log(\det(H))$ in the image-aligned limit, we don't necessarily expect such an estimate to hold for moderate temperature.
Neverthless, the same strategies used above can still be applied when the temperature is not small after marginalization (see section \ref{sec_noise_marginalized_limit_0}).

\section{Dealing with implicitly defined $\tau_{j}$}
\label{sec_dealing_with_implicitly_defined_tau}

In the construction above we have assumed that the $\{\tau_{j}\}$ are given as input.
Thus, the gradient in Eq. \ref{eq_dssnll_block} and the hessian in Eq. \ref{eq_ddssnll_block} both assume that the overall perturbation $\Delta$ is given as an (independent) combination of a volumetric-perturbation $\Delta\fF$ and an alignment-perturbation $\{\Delta\tau_{j}\}$.
However, in many scenarios it is more appropriate to assume that the volumetric- and alignment-perturbations are implicitly related to preserve the maximum-likelihood alignment (see Eq. \ref{eq_P_AA_given_F_maximum_likelihood_alignment}).
In this section we'll describe how we account for this implicit constraint.

To ease notation, we'll start by considering a single image-index $j$, setting $\fA:=\fA_{j}$ and $\tau:=\tau_{j}$, and suppressing the notation for the $CTF$.
Furthermore, we'll assume that $\tau=\tau^{\opt}$ is determined implicitly by the volume $\fF$.
This implicit constraint minimizes the negative-log-likelihood $\cl$ for that image:
\begin{eqnarray}
\cl(\fA \givenbig \fF,\tau) = \frac{1}{2}\langle \fS(\fF,\tau) - \fA , \fS(\fF,\tau) - \fA \rangle \comma
\end{eqnarray}
where $\fS(\fF,\tau)$ is shorthand for the template $\fS(\vk;\tau;CTF;\fF)$.
Thus, $\tau$ should be a critical-point of $\cl$, implying that:
\begin{eqnarray}
\partial_{\tau_{\iota}}\cl = \frac{1}{2}\langle \partial_{\tau_{\iota}}\fS(\fF,\tau) , \fS(\fF,\tau) - \fA \rangle + \frac{1}{2}\langle \fS(\fF,\tau) - \fA , \partial_{\tau_{\iota}}\fS(\fF,\tau) \rangle = \Re \langle \fS(\fF,\tau) - \fA , \partial_{\tau_{\iota}}\fS(\fF,\tau) \rangle = \vzero \comma
\end{eqnarray}
where $\tau_{\iota}$ runs over each of the three components $\transpose{[\egammaz,\epolara,\eazimub]}$ of $\tau$, and each of the 3 components of the gradient $\partial_{\tau}\cl$ are set $0$.
We expand $\fS(\fF,\tau)$ as follows:
\begin{multline}
\fS(\fF + \Delta\fF,\tau + \Delta\tau) = \fS
 + \partial_{F}\fS\cdot\Delta\fF 
 + \partial_{\tau}\fS\cdot\Delta\tau \\
 + \frac{1}{2}[\Delta\fF]^{\dag}\cdot[\partial_{FF}\fS]\cdot[\Delta\fF] 
 + \frac{1}{2}[\Delta\fF]^{\dag}\cdot[\partial_{F\tau}\fS]\cdot[\Delta\tau] \\
 + \frac{1}{2}\transpose{[\Delta\tau]}\cdot[\partial_{\tau F}\fS]\cdot[\Delta\fF] 
 + \frac{1}{2}\transpose{[\Delta\tau]}\cdot[\partial_{\tau\tau}\fS]\cdot[\Delta\tau]
 \comma
\end{multline}
where each term is evaluated at $(\fF,\tau)$ unless otherwise specified.
Noting that $\fS$ is linear in $\fF$, we can ignore second-order contributions of $\Delta\fF^{2}$ to write:
\begin{eqnarray}
\fS(\fF + \Delta\fF,\tau + \Delta\tau) & = &
   \fS 
 + \partial_{\tau}\fS\cdot\Delta\tau 
 + \frac{1}{2}\transpose{[\Delta\tau]}\cdot[\partial_{\tau\tau}\fS]\cdot[\Delta\tau] \\
 & \ & \ 
 + \Delta\fS 
 + \partial_{\tau}\Delta\fS\cdot\Delta\tau 
 + \frac{1}{2}\transpose{[\Delta\tau]}\cdot[\partial_{\tau\tau}\Delta\fS]\cdot[\Delta\tau] \comma
\end{eqnarray}
where $\Delta \fS:=\fS(\vk;\tau;CTF;\Delta\fF)$ is simply the original template-operator applied to $\Delta \fF$ rather than $\fF$.

Now we can write $\langle \fS - \fA , \partial_{\tau_{\iota}}\fS \rangle$ as: 
\begin{equation}
\begin{split}
\langle \fS - \fA , \partial_{\tau_{\iota}}\fS \rangle =
\\
& \hspace{-4em}
\Bigl<
   \fS - \fA
 + \partial_{\tau}\fS\cdot\Delta\tau 
 + \frac{1}{2}\transpose{[\Delta\tau]}\cdot[\partial_{\tau\tau}\fS]\cdot[\Delta\tau] 
 + \Delta\fS 
 + \partial_{\tau}\Delta\fS\cdot\Delta\tau 
 + \frac{1}{2}\transpose{[\Delta\tau]}\cdot[\partial_{\tau\tau}\Delta\fS]\cdot[\Delta\tau]
\\
& \hspace{-2em}
, \partial_{\tau_{\iota}}\left[
   \fS 
 + \partial_{\tau}\fS\cdot\Delta\tau 
 + \frac{1}{2}\transpose{[\Delta\tau]}\cdot[\partial_{\tau\tau}\fS]\cdot[\Delta\tau] 
 + \Delta\fS 
 + \partial_{\tau}\Delta\fS\cdot\Delta\tau 
 + \frac{1}{2}\transpose{[\Delta\tau]}\cdot[\partial_{\tau\tau}\Delta\fS]\cdot[\Delta\tau] 
\right] \Bigr>
\end{split}
\end{equation}

Removing the lowest-order term and keeping terms up to second-order we have:
\begin{equation}
\begin{split}
\langle \fS - \fA , \partial_{\tau_{\iota}}\fS \rangle =
\\
& \hspace{-6em}
\Bigl<
 + \partial_{\tau}\fS\cdot\Delta\tau 
 + \frac{1}{2}\transpose{[\Delta\tau]}\cdot[\partial_{\tau\tau}\fS]\cdot[\Delta\tau] 
 + \Delta\fS 
 + \partial_{\tau}\Delta\fS\cdot\Delta\tau 
 + \frac{1}{2}\transpose{[\Delta\tau]}\cdot[\partial_{\tau\tau}\Delta\fS]\cdot[\Delta\tau]
\\
& \hspace{-2em}
, \partial_{\tau_{\iota}}\left[
   \fS 
 + \partial_{\tau}\fS\cdot\Delta\tau 
 + \frac{1}{2}\transpose{[\Delta\tau]}\cdot[\partial_{\tau\tau}\fS]\cdot[\Delta\tau] 
 + \Delta\fS 
 + \partial_{\tau}\Delta\fS\cdot\Delta\tau 
 + \frac{1}{2}\transpose{[\Delta\tau]}\cdot[\partial_{\tau\tau}\Delta\fS]\cdot[\Delta\tau] 
\right] \Bigr>
\\
& \hspace{-4em}
+
\Bigl<
   \fS - \fA
, \partial_{\tau_{\iota}}\left[
 + \partial_{\tau}\fS\cdot\Delta\tau 
 + \frac{1}{2}\transpose{[\Delta\tau]}\cdot[\partial_{\tau\tau}\fS]\cdot[\Delta\tau] 
 + \Delta\fS 
 + \partial_{\tau}\Delta\fS\cdot\Delta\tau 
 + \frac{1}{2}\transpose{[\Delta\tau]}\cdot[\partial_{\tau\tau}\Delta\fS]\cdot[\Delta\tau] 
\right] \Bigr>
\end{split}
\end{equation}

or when expanded:
\begin{equation}
\begin{split}
\langle \fS - \fA , \partial_{\tau_{\iota}}\fS \rangle = 
\\
& \hspace{-4em}
 + \langle \partial_{\tau}\fS\cdot\Delta\tau ,
   \partial_{\tau_{\iota}} \fS
 + \partial_{\tau_{\iota}} \partial_{\tau}\fS\cdot\Delta\tau
 + \partial_{\tau_{\iota}} \Delta\fS
\rangle
+ \langle \frac{1}{2}\transpose{[\Delta\tau]}\cdot[\partial_{\tau\tau}\fS]\cdot[\Delta\tau] , 
   \partial_{\tau_{\iota}} \fS
\rangle
\\
& \hspace{+2em}
+ \langle \Delta\fS , 
   \partial_{\tau_{\iota}} \fS
 + \partial_{\tau_{\iota}} \partial_{\tau}\fS\cdot\Delta\tau
 + \partial_{\tau_{\iota}} \Delta\fS
\rangle
+ \langle \partial_{\tau}\Delta\fS\cdot\Delta\tau ,
   \partial_{\tau_{\iota}} \fS
\rangle
\\
& \hspace{+4em}
+ \langle
   \fS - \fA
,
 + \partial_{\tau_{\iota}} \partial_{\tau}\fS\cdot\Delta\tau
 + \partial_{\tau_{\iota}} \frac{1}{2}\transpose{[\Delta\tau]}\cdot[\partial_{\tau\tau}\fS]\cdot[\Delta\tau]
 + \partial_{\tau_{\iota}} \Delta\fS
 + \partial_{\tau_{\iota}} \partial_{\tau}\Delta\fS\cdot\Delta\tau
\rangle
\end{split}
\end{equation}

which contains the following four first-order terms:
\begin{eqnarray}
+ \langle \partial_{\tau}\fS\cdot\Delta\tau , \partial_{\tau_{\iota}}\fS \rangle
+ \langle \Delta\fS , \partial_{\tau_{\iota}}\fS \rangle
+ \langle \fS-\fA , \partial_{\tau_{\iota}}\partial_{\tau}\fS\cdot\Delta\tau \rangle
+ \langle \fS-\fA , \partial_{\tau_{\iota}}\Delta\fS \rangle \period
\end{eqnarray}

as well as the following eight second-order terms:
\begin{eqnarray}
+ \langle \partial_{\tau}\fS\cdot\Delta\tau , \partial_{\tau_{\iota}}\partial_{\tau}\fS\cdot\Delta\tau \rangle
+ \langle \partial_{\tau}\fS\cdot\Delta\tau , \partial_{\tau_{\iota}}\Delta\fS \rangle
+ \langle \frac{1}{2} \Delta\tau^{\intercal}\cdot\partial_{\tau\tau}\fS\cdot\Delta\tau , \partial_{\tau_{\iota}}\fS \rangle
+ \langle \Delta\fS , \partial_{\tau_{\iota}}\partial_{\tau}\fS\cdot\Delta\tau \rangle \\
+ \langle \Delta\fS , \partial_{\tau_{\iota}}\Delta\fS \rangle
+ \langle \partial_{\tau}\Delta\fS\cdot\Delta\tau , \partial_{\tau_{\iota}}\fS \rangle
+ \langle \fS-\fA , \partial_{\tau_{\iota}}\frac{1}{2} \Delta\tau^{\intercal}\cdot\partial_{\tau\tau}\fS\cdot\Delta\tau \rangle
+ \langle \fS-\fA , \partial_{\tau_{\iota}}\partial_{\tau}\Delta\fS\cdot\Delta\tau \rangle \period
\end{eqnarray}

By retaining only first-order terms we see:
\begin{eqnarray}
\sum_{\iota'}\XA_{\iota,\iota'}\cdot \Delta\tau_{\iota'} + \XB_{\iota} = \bigO\left(\Delta\fF^{2}\right) \comma \text{\ where:\ } \\
\XA_{\iota,\iota'} = \Re\left\{ \langle \partial_{\tau_{\iota}}\fS , \partial_{\tau_{\iota'}}\fS \rangle + \langle \fS-\fA , \partial_{\tau_{\iota}\tau_{\iota'}}\fS \rangle \right\} \comma \\
\XB_{\iota} = \Re\left\{ \langle \partial_{\tau_{\iota}}\fS , \Delta\fS \rangle + \langle \partial_{\tau_{\iota}}\Delta\fS , \fS-\fA \rangle \right\} \comma
\end{eqnarray}
where $\XA$ is $\bigO(1)$ and $\XB$ is $\bigO\left(\Delta\fF\right)$, as expected.
This first-order expression implies that:
\begin{eqnarray}
\Delta\tau = -\inverse{\XA}\cdot \XB + \bigO\left(\Delta\fF^{2}\right) \period
\end{eqnarray}

The first-order approximation to $\Delta\tau$ can be improved by using the second-order terms.
Let's first group the second-order terms by their tensor-rank:
\begin{eqnarray}
\YA_{\iota,\iota'',\iota'} & = & \Re\left\{ \langle\partial_{\tau_{\iota''}}\fS,\partial_{\tau_{\iota}}\partial_{\tau_{\iota'}}\fS\rangle + \langle\frac{1}{2}\partial_{\tau_{\iota''}}\partial_{\tau_{\iota'}}\fS,\partial_{\tau_{\iota}}\fS\rangle + \langle\fS-\fA,\frac{1}{2}\partial_{\tau_{\iota}}\partial_{\tau_{\iota''}}\partial_{\tau_{\iota'}}\fS\rangle \right\} \comma \\
\YB_{\iota,\iota'} & = & \Re\left\{ \langle\partial_{\tau_{\iota'}}\fS,\partial_{\tau_{\iota}}\Delta\fS\rangle + \langle\partial_{\tau_{\iota'}}\Delta\fS,\partial_{\tau_{\iota}}\fS\rangle + \langle\Delta\fS,\partial_{\tau_{\iota}}\partial_{\tau_{\iota'}}\fS\rangle + \langle\fS-\fA,\partial_{\tau_{\iota}}\partial_{\tau_{\iota'}}\Delta\fS\rangle \right\} \comma \\
\YC_{\iota} & = & \Re\left\{ \langle\Delta\fS,\partial_{\tau_{\iota}}\Delta\fS\rangle \right\} \comma
\end{eqnarray}
noting that $\YA$ is $\bigO(1)$, $\YB$ is $\bigO\left(\Delta\fF\right)$, and $\YC$ is $\bigO\left(\Delta\fF^{2}\right)$, as expected.

Now, retaining both first- and second-order terms we see:
\begin{multline}
\sum_{\iota'}\XA_{\iota,\iota'}\cdot \Delta\tau_{\iota'} + \XB_{\iota} + \sum_{\iota',\iota''} \Delta\tau_{\iota''}\YA_{\iota,\iota'',\iota'}\Delta\tau_{\iota'} + \\
\sum_{\iota'}\YB_{\iota,\iota'}\Delta\tau_{\iota'} + \YC_{\iota} = \bigO\left(\Delta\fF^{3}\right) \quad \forall \ \tau_{\iota}\in\{\egammaz,\epolara,\eazimub\} \comma
\end{multline}
which, in matrix-vector notation can be written as:
\begin{eqnarray}
\XA_{\iota,:}\cdot\Delta\tau_{:} + \XB + \transpose{\Delta\tau}\cdot\YA_{\iota,:,:}\cdot\Delta\tau + \YB_{\iota,:}\cdot\Delta\tau + \YC_{\iota} = \bigO\left(\Delta\fF^{3}\right) \quad \forall \ \tau_{\iota}\in\{\egammaz,\epolara,\eazimub\} \comma
\end{eqnarray}
where $\XA_{\iota,:}$ and $\YB_{\iota,:}$ are column-vectors produced by slicing $\XA$ and $\YB$ respectively, while $\YA_{\iota,:,:}$ is a matrix produced by slicing $\YA$.
This allows us to solve for $\Delta\tau$ up to second-order:
\begin{eqnarray}
\Delta\tau_{\iota} = -\inverse{\XA}_{\iota,:}\cdot\XB - \inverse{\XA}_{\iota:}\left\{ \transpose{\XB}\cdot\ictranspose{\XA}\cdot\YA_{\iota,:,:}\cdot\inverse{\XA}\cdot\XB - \YB_{\iota,:}\cdot\inverse{\XA}\cdot\XB + \YC_{\iota} \right\} + \bigO\left(\Delta\fF^{3}\right) \period
\end{eqnarray}
Note that, in this expression, we can pick out the first-order solution:
\[ \Zone_{\iota} := -\inverse{\XA}_{\iota,:}\cdot\XB = \bigO(\Delta\fF) \comma \]
as well as the second-order correction:
\[ \Ztwo_{\iota} := - \inverse{\XA}_{\iota:}\left\{ \transpose{\XB}\cdot\ictranspose{\XA}\cdot\YA_{\iota,:,:}\cdot\inverse{\XA}\cdot\XB - \YB_{\iota,:}\cdot\inverse{\XA}\cdot\XB + \YC_{\iota} \right\}  = \bigO(\Delta\fF^{2}) \period \]
Let's refer to the three components of the first- and second-order terms as the vectors $\Zone$ and $\Ztwo$, respectively.
Furthermore, let's re-introduce the image-index $j$, so that:
\[ \Delta\tau_{j} = \Zone_{j} + \Ztwo_{j} + \bigO\left(\Delta\fF^{3}\right) \comma \]
where now we associate $\Delta\tau_{j}$ to the alignment of image-$j$, and note that each term in this expression has three components (corresponding to the index $\iota$, which is no longer written).

With this notation we can now use our expression for the gradient in Eq. \ref{eq_dssnll_block} and Eq. \ref{eq_ddssnll_block} to write:
\begin{equation}
\begin{split}
\cL\left(\{\fA_{j}\}\givenbig \fF+\Delta\fF; \{\tau_{j}+\Delta\tau_{j}\} \right) & = \cL + G\cdot\Delta + \ctranspose{\Delta}\cdot H\cdot\Delta
\\
 & \hspace{-8em}
= \cL
+ \partial_{\fF}\cL\cdot\Delta\fF
+ \sum_{j}\partial_{\tau_{j}}\cL\cdot\Delta\tau_{j}
\\
 & \hspace{-4em}
+ \ctranspose{\Delta\fF}\cdot\partial_{\fF}\partial_{\fF}\cL\cdot\Delta\fF
+ \sum_{j}\ctranspose{\Delta\fF}\partial_{\fF}\partial_{\tau_{j}}\cL\cdot\Delta\tau_{j}
\\
 & \hspace{-4em}
+ \sum_{j}\transpose{\Delta\tau_{j}}\cdot\partial_{\tau_{j}}\partial_{\fF}\cL\cdot\Delta\fF
+ \sum_{j}\partial_{\tau_{j}}\partial_{\tau_{j}}\cL\cdot\Delta\tau_{j}
\\
 & \hspace{-8em}
= \cL
+ \partial_{\fF}\cL\cdot\Delta\fF
+ \sum_{j}\partial_{\tau_{j}}\cL\cdot\Zone_{j}
\\
 & \hspace{-4em}
+ \sum_{j}\partial_{\tau_{j}}\cL\cdot\Ztwo_{j}
+ \ctranspose{\Delta\fF}\cdot\partial_{\fF}\partial_{\fF}\cL\cdot\Delta\fF
+ \sum_{j}\ctranspose{\Delta\fF}\partial_{\fF}\partial_{\tau_{j}}\cL\cdot\Zone_{j}
\\
 & \hspace{-4em}
+ \sum_{j}\transpose{\Zone_{j}}\cdot\partial_{\tau_{j}}\partial_{\fF}\cL\cdot\Delta\fF
+ \sum_{j}\partial_{\tau_{j}}\partial_{\tau_{j}}\cL\cdot\Zone_{j}
\period 
\end{split}
\end{equation}
From this expression we can recover the corrections to the gradient and hessian resulting from the implicit relationship between $\fF$ and the $\{\Delta\tau_{j}\}$.

As an example of the kinds of manipulations that are required for these corrections, we'll discuss the simplest term: $\sum_{j}\partial_{\tau_{j}}\cL\cdot\Zone_{j}$, which serves as an addition to the original volumetric term $\partial_{\fF}\cL\cdot\Delta\fF$ from the gradient (see Eq. \ref{eq_dssnll_q3d_Gv}).

When unpacked, this additional term can be represented as:
\begin{multline}
\sum_{j}\partial_{\tau_{j}}\cL\cdot\Zone_{j} = \\
+ \sum_{j}\sum_{\iota,\iota'} \Re\langle \partial_{\tau_{j,\iota}}\fS(\fF,\tau_{j}),\fS(\fF,\tau_{j})-\fA_{j}\rangle \left[-{\tt X0}^{-1}\right]_{\iota,\iota'}\Re\langle\partial_{\tau_{j,\iota'}}\fS(\fF,\tau_{j}),\fSlice\circ\rotation(\tau_{j})\circ\Delta\fF\rangle 
\\
 + \sum_{j}\sum_{\iota,\iota'} \Re\langle \partial_{\tau_{j,\iota}}\fS(\fF,\tau_{j}),\fS(\fF,\tau_{j})-\fA_{j}\rangle \left[-{\tt X0}^{-1}\right]_{\iota,\iota'}\Re\langle \fS(\fF,\tau_{j})-\fA_{j},\partial_{\tau_{j,\iota'}}\fSlice\circ\rotation(\tau_{j})\circ\Delta\fF\rangle \period
\label{eq_implicit_delta_tau_term_1}
\end{multline}
In this expression the values of $\left[-{\tt X0}^{-1}\right]_{\iota,\iota'}$ are each just scalars that depend on $j$, $\iota$ and $\iota'$.
The same holds for other prefactors such as $\Re\langle \partial_{\tau_{j,\iota}}\fS(\fF,\tau_{j}),\fS(\fF,\tau_{j})-\fA_{j}\rangle$.
Thus, the first term on the right-hand-side of Eq. \ref{eq_implicit_delta_tau_term_1} can turned into a volumetric integral using monopole-impulses via Eq. \ref{eq_qbp_monopole_a} and \ref{eq_qbp_monopole_b}.
On the other hand, the second term on the right-hand-side of Eq. \ref{eq_implicit_delta_tau_term_1} involves a derivative, and can be turned into a volumetric integral using dipole-impulses via Eq. \ref{eq_qbp_dipole_a} and \ref{eq_qbp_dipole_b}.
Further correction terms for the hessian involve interactions between dipoles, quadrupoles, etc., producing a multipole expansion.

\section{Probing the spectrum of the hessian: case studies}
\label{sec_local_well_posedness_ansatz_0}

In this section we'll describe one approach that allows us to crudely bound $\sigma_{H}^{\critical}$ (see Eq. \ref{eq_sigma_H_crit}) by appealing to the image-aligned limit of the likelihood.
Specifically, we'll assume that the noise $\fsigma$ is sufficiently small that Eq. \ref{eq_P_AA_given_F_maximum_likelihood} is a good approximation to Eq. \ref{eq_P_AA_given_F}.
We'll then try and find perturbations $\{\Delta\fF^{\lowtemp},\Delta P^{\lowtemp}\}$ for which the leading-order term $\sigma_{H}^{\lowtemp}(\Delta\fF^{\lowtemp},\Delta P^{\lowtemp})$ is small (in this image-aligned limit).
These perturbations $\{\Delta\fF^{\lowtemp},\Delta P^{\lowtemp}\}$ will then serve as candidates for $\{\Delta\fF^{\critical},\Delta P^{\critical}\}$, from which we can bound $\sigma_{H}^{\critical}$

By using this approach, we'll be able to investigate certain relationships between the viewing-angle distribution and the stability/robustness of single-particle reconstruction.
By examining a few case-studies we'll see that, even if the data is sufficient for a single-particle reconstruction, non-uniformity in the viewing-angle distribution can compromise robustness.
Ultimately, we'll recapitulate a well-known phenomenon: images from a variety of viewing-angles are required for a robust reconstruction.

To simplify the discussion, we'll first make several additional assumptions which can be relaxed later on:
\begin{description}
\item[Assume a large pool of synthetic-images:] We'll assume that the pool of images is well modeled by the distribution of synthetic-images described in section \ref{sec_synthetic_images}. Furthermore, we'll neglect finite-size effects associated with the synthetic-image pool, and assume that $\nimage$ is sufficiently large that the distribution of synthetic-images is well-sampled.
\item[Ignore translations and CTF:] We'll assume that each synthetic-image is drawn with a displacement $\vd$ identically equal to 0 (i.e., $\sigma_{\vd}=0$) and a CTF identically equal to $1$.
\end{description}
We'll also suppress the notation for the coordinate $\vk$, noting that the synthetic-images are simply
\[ \fA_{j}^{\synthetic} = \fS(\tau_{j};\fF) \period \]

Let's now assume that we have a putative (local) minimum $\{\fF^{\opt},P(\tau\givenmed\fF^{\opt})\}$ of the negative-log-likelihood from Eq. \ref{eq_P_AA_given_F_maximum_likelihood}, which we'll denote by $-\log(P^{\opt})$.
This means that, for any particular image-index $j$, the corresponding synthetic-image $\fA^{\synthetic}_{j}=\fS(\tau^{\opt}_{j};\fF^{\opt})$ is associated with the maximum-likelihood alignment $\tau^{\opt}_{j}$ via Eq. \ref{eq_P_AA_given_F_maximum_likelihood_alignment}.
Furthermore, we expect that the volume $\fF^{\opt}$ is consistent with these alignments in the sense that $\fF^{\opt}$ can be reconstructed from the synthetic-images using Eq. \ref{eq_nll_qbp}.
Without loss of generality we can assume that the viewing-angles are described such that each $\egammaz_{j}\equiv 0$, and hence each $\tau_{j}=(0,\epolara_{j},\eazimub_{j})$.
With this notation the density of viewing-angles at a particular $(\epolara,\eazimub)$ is given by $P((0,\epolara,\eazimub)\givenmed\fF^{\opt})$, or just $P(\tau)$ for short.

Note that the assumptions above imply that the maximum-likelihood alignments $\{\tau^{\opt}_{j}\}$ are actually implicit functions of the $\{\fA^{\synthetic}_{j}\}$ as well as $\fF^{\opt}$; if the images or volume were to change, so too would the optimal alignments.
The alignments also technically depend on the prior distribution $P(\tau\givenmed\fF^{\opt})$, but only in the sense that each $\tau^{\opt}_{j}$ must lie in the support of $P(\tau\givenmed\fF^{\opt})$.
Thus, assuming the image-set is fixed and the prior distribution is allowed to be nonzero for all viewing-angles, any particular volumetric-perturbation will give rise to a perturbation in the maximum-likelihood alignments, and vice-versa.
We'll take advantage of this implicit relationship in a moment: instead of directly searching for infinitesimal volumetric-perturbations which best preserve the likelihood, we can instead search for infinitesimal perturbations in the maximum-likelihood alignments which best preserve the image-template similarities shown in Eq. \ref{eq_P_A_given_tau_vd_F}.

To be more specific, we'll denote by $\{\tau^{[0]}_{j}\}$ the `old' maximum-likelihood viewing-angles $\{\tau^{\opt}_{j}\}$ associated with $\fF^{[0]}:=\fF^{\opt}$, and define `temporary' perturbed viewing-angles via:
\begin{eqnarray}
\fF^{[0]} & := & \fF^{\opt} \comma \\
\tau^{[0]}_{j} & := & \tau^{\opt}_{j} \comma \\
\tau^{[1]}_{j} & := & \tau^{[0]}_{j} + \epsilon\Psi\left(\tau^{[0]}_{j}\right) \comma
\label{eq_viewing_angle_pos}
\end{eqnarray}
where $\epsilon\in\Real$ is small, and $\Psi:SO3\rightarrow {\mathfrak so}3$ is a flow-field on SO3.
Using the $\{\tau^{[1]}_{j}\}$, we then consider a new back-projection problem given by Eq. \ref{eq_nll_qbp} to form a new volume $\fF^{[1]}$.
To describe $\fF^{[1]}(\vkappa)$ for any point $\vkappa=\transpose{[\kappa_{1},\kappa_{2},0]}$, we can define $\perp(\vkappa)$ to be the set of viewing-angles that correspond to rotations for which $\vkappa$ is stationary:
\begin{equation}
\perp(\vkappa) = \{\tau \givenbig \rotation{\tau}\circ\vkappa = \vkappa\} \comma
\end{equation}
and then, noting that $\tau^{[0]}-\epsilon\Psi(\tau^{[0]})$ is the preimage of $\tau^{[0]}$ under the action of $\Psi$ (to lowest order), we can write:
\begin{eqnarray}
\fF^{[1]}(\vkappa) & = & \frac{1}{Z(\vkappa)}\int_{\tau\in \perp(\vkappa)}\fS\left(\vk;\fF^{[0]},\tau - \epsilon\Psi(\tau)\right) P\left(\tau - \epsilon\Psi(\tau)\right) d\tau \\
\ & \ & \quad \quad \text{\ where:\ } \vk=\rotation(\tau)\circ\vkappa=\transpose{[\kappa_{1},\kappa_{2}]} \comma \\
\ & = &  \frac{1}{Z(\vkappa)}\int_{\tau\in \perp(\vkappa)}\left[\fSlice\circ\rotation\left(\tau - \epsilon\Psi(\tau)\right)\circ\fF^{[0]}\right](\vk) P\left(\tau - \epsilon\Psi(\tau)\right) d\tau \\
\ & \ & \quad \quad \text{\ with:\ } Z(\vkappa) =  \int_{\tau\in \perp(\vkappa)} P\left(\tau - \epsilon\Psi(\tau)\right) d\tau \period
\label{eq_volume_pos}
\end{eqnarray}
An analogous equation holds (with a rotated frame of reference) for any general $\vkappa$.
In words: $\fF^{[1]}(\vkappa)$ at any point $\vkappa$ can be estimated by simply averaging over all the values of $\fF^{[0]}$ which are mapped onto $\vkappa$ by the viewing-angle perturbations.
In the case studies below, we'll choose $\Psi$ to be simple enough that we can estimate the back-projected $\fF^{[1]}$ `by hand'.
    
Obviously the volumetric-perturbation producing $\fF^{[1]}$ then induces an alignment-perturbation $\{\tau^{[2]}_{j}\}$:
\begin{eqnarray}
\tau^{[2]}_{j} & = & \argmax_{\tau} P\left( \fA^{\synthetic}_{j}\givenbig \tau ; \fF^{[1]} \right)  \\
\ & = & \argmax_{\tau} P\left( \fS(\tau^{[0]}_{j};\fF^{[0]}) \givenbig \tau ; \fF^{[1]} \right)  \period
\end{eqnarray}
when then induce another volumetric-perturbation $\fF^{[2]}$, and so forth.
This sequence (or iteration) may eventually converge, but we'll curtail it after the first few steps, typically using $\fF^{[1]}$ and $\{\tau^{[1]}_{j}\}$ as an estimate for the perturbed volume and alignments.
We'll define $\Delta\fF:=\fF^{[1]}-\fF^{[0]}$ and $\Delta\tau_{j} := \tau^{[1]}_{j} - \tau^{[0]}_{j}$.
Finally, we'll use these candidate $\Delta\fF$ and $\{\Delta\tau_{j}\}$ to measure the rayleigh-product $\sigma_{H}^{\lowtemp}(\Delta\fF,\{\Delta\tau_{j}\})$.

The relevant (image-aligned) likelihood is:
\begin{eqnarray}
-\log\left(P(\{\fA_{j}\}\givenbig \fF^{[1]},\{\tau^{[1]}_{j}\} \right) & \sim & \frac{1}{2\fsigma^{2}}\sum_{j=1}^{\nimage} \iint_{\vk\in\Omega_{\kmax}\in\Real^{2}} \left|\fA_{j}(\vk)-\fS\left(\vk;\tau^{[1]}_{j};\fF^{[1]}\right)\right|^{2} d\vk \\
 & = & \frac{1}{2\fsigma^{2}}\sum_{j=1}^{\nimage} \iint_{\vk\in\Omega_{\kmax}\in\Real^{2}} \left|\fS\left(\vk;\tau^{[0]}_{j};\fF^{[0]}\right)-\fS\left(\vk;\tau^{[1]}_{j};\fF^{[1]}\right)\right|^{2} d\vk \period
\label{eq_P_AA_given_F_maximum_likelihood_perturbation}
\end{eqnarray}

\subsection{Rotation Blur:\ }
One of the most straightforward perturbations to analyze is when $\Psi(\tau)$ corresponds to a rotation-blur.
To describe this scenario, let's first assume that $P(\tau)$ is uniform.
Now let's consider a particular $\tau^{[0]}$, for which the template $\fS(\vk;\tau^{0})$ has a pole $\vkappa_{0}:=\inverse{\rotation}{\tau^{[0]}}\circ\transpose{[0,0,1]}$.
Now let's define $\cN(\tau^{[0]},\epsilon^{2})$ to be an isotropic gaussian-distribution on the surface of the sphere with a mean at $\tau^{[0]}$ and a variance of $\epsilon^{2}$.
Finally, let's define $\Psi(\tau^{[0]})$ so that $\tau^{[1]}:=\tau^{[0]}+\epsilon\Psi(\tau^{[0]})$ is drawn independently from the distribution $\cN(\tau^{[0]},\epsilon^{2})$.

In this setting the back-projected solution $\fF^{[1]}$ on any $k$-shell can be constructed by simply convolving the values of $\fF^{[0]}$ on that shell with $\cN(\vzero,\epsilon^{2})$.
That is to say, the volume $\fF^{[1]}$ can be constructed by `rotationally blurring' the volume $\fF^{[0]}$: i.e.:
\begin{eqnarray}
\fF^{[1]}(k,\hkappa) & = & \int_{\hkappa'} \fF^{[0]}(k,\hkappa') \cN(\hkappa-\hkappa',\epsilon^{2}) d\hkappa' \period
\end{eqnarray}

This rotational-blur corresponds directly to an isotropic surface-diffusion of the volume $\fF^{[0]}$:
\begin{equation}
\left[\fF^{[1]}\right]_{l}^{m} = \exp(-l(l+1)\epsilon^{2}/2)\cdot\left[\fF^{[0]}\right]_{l}^{m} \period
\end{equation}

\subsection{Latitudinal perturbation:\ }
\label{sec_local_well_posedness_ansatz_latitudinal_perturbation_0}
Another relatively simple perturbation involves a fully latitudinal perturbation: $\Psi(\tau) = \left(0,\Theta\left(\epolara\right),0\right) = \left(0,\sin\left(2\epolara\right),0\right)$.
If $\epsilon>0$, this perturbation pushes each $\tau$ in the northern-hemisphere south (towards the equator), while pushing each $\tau$ in the southern-hemisphere north (again towards the equator).

If $P(\tau)$ is uniform, then in the limit as $\epsilon\rightarrow 0$ each value of $\fF^{[1]}(k,\hkappa)$ is determined by averaging the values of $\fF^{[0]}(k,\hkappa')$ on a small `circle' (on the surface of the sphere).
This circle can be described in terms of the polar-angle $a$ associated with $\hkappa$.
the circle is centered at a polar-angle of $a_{c}(a)$ with a radius (extending on the surface of the sphere of radius-$1$) of $r(a)$.
\begin{equation}
a_{c}(a) = a + 0.5\epsilon\Theta(a) \quad \text{\ and:\ } r(a) = 0.5\epsilon\Theta(a) \period
\label{eq_viewing_angle_perturbation_polar_cap}
\end{equation}
Note that the radius of this circle vanishes as $\vkappa$ approaches either the poles or the equator, and is largest in the temperate zones.

As a result of Eq. \ref{eq_viewing_angle_perturbation_polar_cap} the values of $\fF^{[1]}$ can each be obtained by integrating the nearby values of $\fF^{[0]}$ against a small local gaussian:
\begin{eqnarray}
\fF^{[1]}(k,\hkappa) & = & \int_{\hkappa'} \fF^{[0]}(k,\hkappa') \cN(\hkappa-\hkappa',\epsilon^{2}\sigma^{2}(\hkappa)) d\hkappa' \quad \text{\ with:\ } \sigma^{2}(\hkappa) = \frac{1}{4}\cdot\frac{1}{2}\Theta(a)^2 \period
\end{eqnarray}

By looking at this case-study we see that a latitudinal alignment-perturbation $\Theta(\epolara)$ results in a volumetric-perturbation $\Delta\fF$ which is well approximated by a $a$-dependent local-diffusion about each point $\vkappa$.
This local diffusion-coefficient drops to $0$ at the poles and equator (i.e., $a=0$, $\pi/2$ and $\pi$).
Furthermore, the latitudinal alignment-perturbation also shifts each value of $\fF^{[0]}$ in the latitudinal-direction, allowing the templates values to be unchanged to first-order if the viewing-angle is also perturbed latitudinally.
Thus, the template-perturbation
\[ \Delta\fS(\vk;\tau)= \fS\left(\vk;\fF^{[1]};\tau+0.5\epsilon\Theta(\epolara)\right)-\fS\left(\vk;\fF^{[0]};\tau\right) \]
will vanish as $\tau$ approaches the poles.
Consequently, any image with an initial viewing-angle near the poles will align almost as well to $\fF^{[1]}$ (at the latitudinally-perturbed $\tau^{[1]}$) as it did to $\fF^{[0]}$ (at its inital $\tau^{[0]}$).

Taking these considerations into account, we see that a latitudinal alignment-perturbation like $\Theta(\epolara)$ will introduce a likelihood-penalty to images that are oriented far from the poles, but will not introduce a likelihood-penalty to images that are oriented close to the poles.
As a result, any volume constructed from a distribution of viewing-angles concentrated near the poles will be especially susceptible to latitudinal-perturbations of this form.
To illustrate this phenomenon we can consider a non-uniform $P(\tau)$ which is concentrated around the polar-caps.
As $P(\tau)$ becomes more and more concentrated around the polar-caps, the likelihood of $\{\fF^{[1]},\{\tau^{[1]}_{j}\}\}$ becomes more and more similar to $\{\fF^{[0]},\{\tau^{[0]}_{j}\}\}$, even though $\fF^{[1]}$ may not be similar to $\fF^{[0]}$.
Equivalently, the average fisher-information per image with-respect to the perturbation $\{\Delta\fF,\{\Delta\tau_{j}\}\}$ vanishes as $P(\tau)$ becomes more concentrated near the poles.

Put another way, if the image-pool doesn't have many images that are oriented away from the poles, no amount of near-polar images will help improve the robustness of the reconstructed volume.
As mentioned above, this intuitive result aligns with many current methodologies: many images from a variety of viewing-angles are required for a robust reconstruction.


\begin{figure}[H]
\centering
\includegraphics[width=1.00\textwidth]{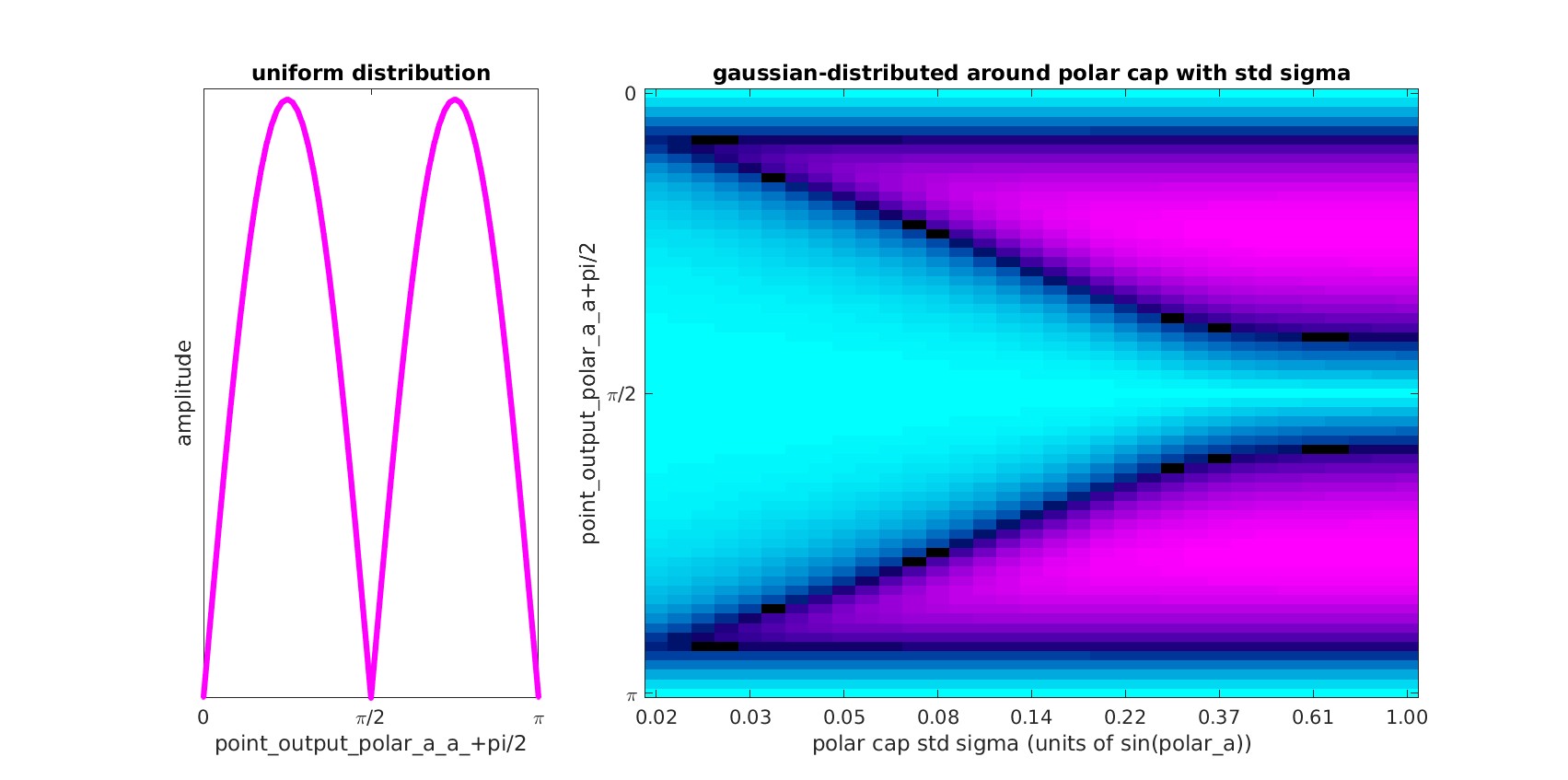}
\caption{
In this figure we illustrate some properties of a latitudinal-perturbation $\Theta(\epolara) = \sin\left(2\epolara\right)$.
On the right we illustrate the local diffusion-coefficient $\cD(a)$ (vertical) as a function of the polar-angle $a$ (horizontal) when the viewing-angle distribution is uniform.
Note that the local diffusion-coefficient $\cD(a)$ drops to $0$ when $a$ lies on either the equator or the poles.
This means that templates of the original volume $\xF$ which correspond to slices near the equatorial-zone (i.e., with viewing-angles close to the poles) will align almost as well to the perturbed volume $\xF+\Delta\xF$ as they did to the original volume $\xF$ (after latitudinally adjusting each template's viewing-angle $\tau$ by $\Delta\tau(\tau)$).
On the left we illustrate the same local-diffusion-coefficient $\cD(a;\sigma_{\epolara})$ as a function of the polar-angle $a$ (vertical) as the viewing-angle distribution becomes more concentrated around the poles (as parametrized by $\sigma_{\epolara}$).
The viewing-angle distribution considered for this figure is of the form $\rho(\epolara)\propto\exp\left( -\sin(\epolara)^{2} / (2\sigma_{\epolara}^{2}) \right)$, with $\sigma_{\epolara}$ shown along the horizontal.
The value of the diffusion-coefficient $\cD(a;\sigma_{\epolara})$ is shown as a color in this array, ranging from cyan (zero) to pink (the maximum shown on the left).
Note that, as $\sigma_{\epolara}$ shrinks the viewing-angle distribution $\rho(\epolara)$ becomes more and more concentrated towards the poles.
Such a `polar-cap' distribution introduces essentially no local-diffusion for values of $a$ in a large band around the equator.
This means that, as the viewing-angle distribution becomes concentrated near the poles, the range of templates of the original volume $\xF$ which can be well-aligned to the perturbed-volume $\xF + \Delta\xF$ increases to include templates with original-viewing-angles in the tropical- and temperate-zones.
}
\label{fig_M3d_shape_latitudinal_perturbation_std_vs_sigma_FIGA}
\end{figure}

\subsection{Longitudinal perturbation:\ }
\label{sec_local_well_posedness_ansatz_longitudinal_perturbation_0}
A slightly more elaborate perturbation involves $\Psi(\tau) := \Phi(\eazimub) = \sin(2\eazimub)$.
If $\epsilon>0$, this perturbation pushes each $\tau$ either east- or west-ward away from the great-circle corresponding to the prime-meridian.

If $P(\tau)$ is uniform, then in the limit as $\epsilon\rightarrow 0$ each value of $\fF^{[1]}(k,\hkappa)$ is determined by averaging the values of $\fF^{[0]}(k,\hkappa')$ on a small `collapsed circle' (on the surface of the sphere) projected onto (i.e., extending only in) the azimuthal-direction.
This collapsed circle can be described in terms of the polar-angle and azimuthal-angle $(a,b)$ associated with the point $\hkappa$.
The collapsed circle is centered at an azimuthal-angle of $b_{c}(a,b)$ and has an azimuthal-radius of $r(a)$.
\begin{eqnarray}
b_{c}(a,b) = b + \epsilon u(a)\Phi(b) \quad \text{\ and:\ } r^{2}(a) = \epsilon^{2}\left[1 + u(a)\right] \comma \\
u(a) = \frac{1-|\cos(a)|}{1+|\cos(a)|} \quad \text{\ and:\ } \Phi(b) = \sin(2b) \period 
\label{eq_viewing_angle_perturbation_equa_band}
\end{eqnarray}
Note that the azimuthal-radius of this collapsed-circle vanishes as $\vkappa$ approaches the equator, and is largest near the poles.

As a result of Eq. \ref{eq_viewing_angle_perturbation_equa_band} the values of $\fF^{[1]}$ can each be obtained by integrating the nearby values of $\fF^{[0]}$ against a small local gaussian extending in the azimuthal-direction:
\begin{eqnarray}
\fF^{[1]}(k,\hkappa) & = & \int_{b} \fF^{[0]}(k,\hkappa') \cN(b-b',\epsilon^{2}\sigma^{2}(a)) d\hkappa' \quad \text{\ with:\ } \sigma^{2}(a) = \frac{1}{2}r^{2}(a) \period
\end{eqnarray}

By looking at this case-study we see that a longitudinal alignment-perturbation $\Phi(\eazimub)$ results in a volumetric-perturbation $\Delta\fF$ which is well approximated by a $b$-dependent local azimuthal-diffusion about each point $\vkappa$.
This local azimuthal-diffusion-coefficient drops to $0$ at the equator (i.e., $a=\pi/2$).
However, the longitudinal-perturbation induces an azimuthal-shift in each value of $\fF^{[0]}$ that is both $a$- and $b$-dependent.
Thus, the original synthetic-images (i.e., templates from $\fF^{[0]}$) won't typically align well to the volume $\fF^{[1]}$.
Taking these considerations into account, we see that a longitudinal alignment-perturbation like $\Phi(\eazimub)$ will introduce likelihood-penalties for all the images.

However, the results above are strongly dependent on the (near) uniformity of the viewing-angle distrubution $P(\tau)$.
If $P(\tau)$ were to be concentrated near the equatorial-belt, then the results change substantially.
In fact, as $P(\tau)$ becomes more concentrated around the equatorial-belt there are two major changes:
\begin{enumerate}
\item The azimuthal-displacement $b_{c}(a,b)$ becomes dependent mainly on the azimuthal-angle $b$ of $\vkappa$ (becoming proportional to $\Phi(b)$) and becomes more and more independent of the polar-angle $a$.
\item the azimuthal-variance $r^{2}(a)/2$ drops to zero away from the poles.
\end{enumerate}
In this scenario (i.e., an equatorially-concentrated $P(\tau)$), the situation is quite similar to the poorly-posed scenario involving {\em only} equatorial viewing-angles.
That is to say, any volume constructed from a distribution of viewing-angles concentrated near the equator will be especially susceptible to longitudinal-perturbations.

To illustrate this phenomenon we can consider a non-uniform $P(\tau)$ which is concentrated around the equator.
Similar to the latitudinal-perturbation described above, as $P(\tau)$ becomes more and more concentrated around the equator, the likelihood of $\{\fF^{[1]},\{\tau^{[1]}_{j}\}\}$ becomes more and more similar to $\{\fF^{[0]},\{\tau^{[0]}_{j}\}\}$, and the average fisher-information per image with-respect to the perturbation $\{\Delta\fF,\{\Delta\tau_{j}\}\}$ vanishes.
Put another way, if the image-pool only has images that have viewing-angles near the equator, their relative orientation cannot be pinned down.
Once again, this result echoes conventional-wisdom: many images from a variety of viewing-angles are required for a robust reconstruction.


\begin{figure}[H]
\centering
\includegraphics[width=1.00\textwidth]{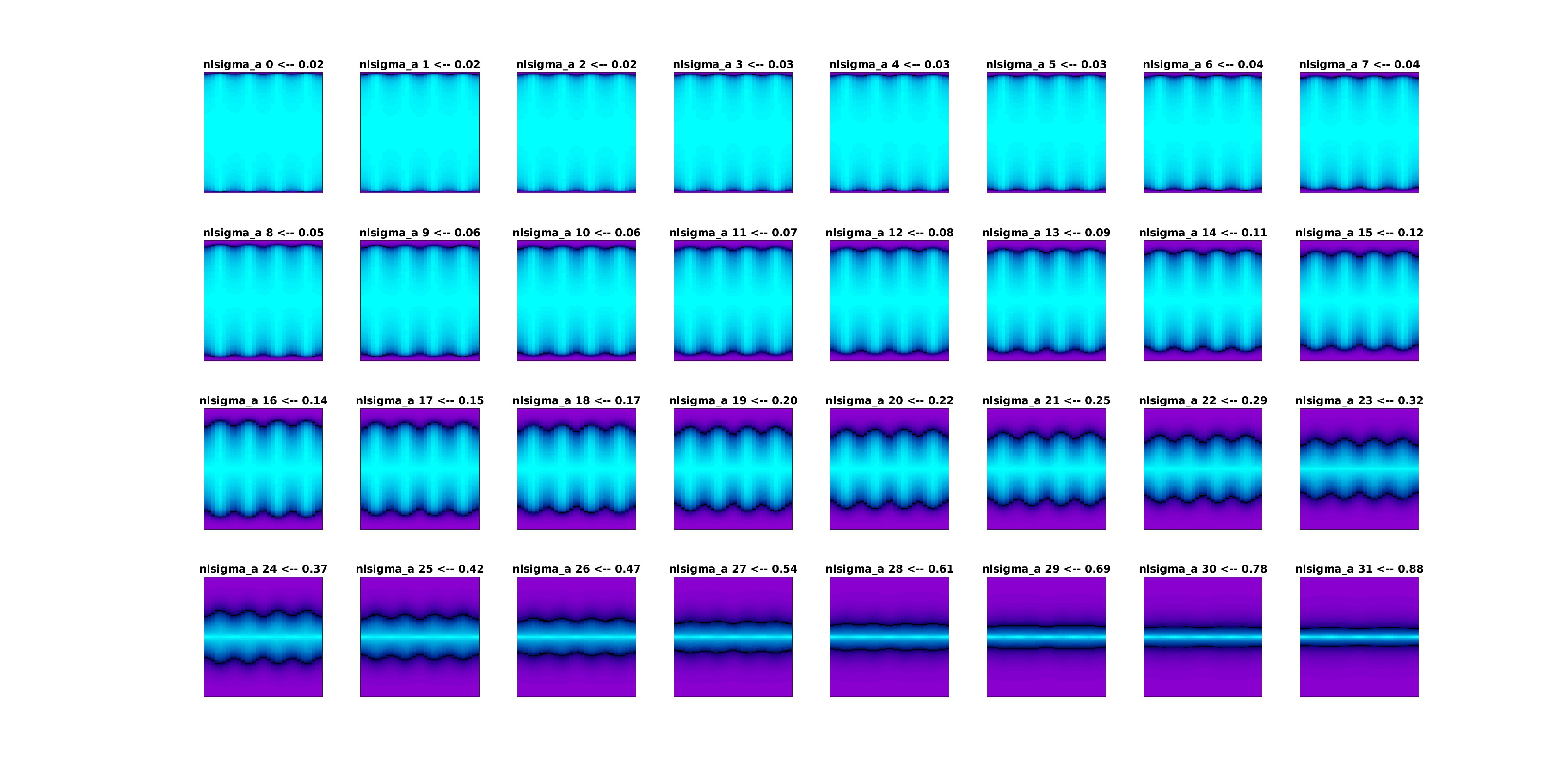}
\caption{ 
In this figure we illustrate some properties of a longitudinal-perturbation $\Phi(\eazimub) = \sin\left(2\eazimub\right)$.
In this figure we show multiple subplots.
Within each subplot (indexed by $\sigma_{\epolara}$) we show a heatmap of the local-azimuthal-diffusion $\cE(a,b;\sigma_{\epolara})$ as a function of the polar- and azimuthal angles $(a,b)$, corresponding to the vertical- and horizontal-axes respectively.
Within each subplot the value of $\cE(a,b;\sigma_{\epolara})$ ranges from cyan (zero) to pink (maximum).
The different subplots correspond to different viewing-angle distributions (parametrized by $\sigma_{\epolara}$).
The viewing-angle distribution considered for this figure is of the form $\rho(\epolara)\propto\exp\left( -\cos(\epolara)^{2} / (2\sigma_{\epolara}^{2}) \right)$, with $\sigma_{\epolara}$ shown in the title of each subplot.
In this case, when $\sigma_{\epolara}$ is large the viewing-angle distribution is essentially uniform (lower-right subplot).
On the other hand, when $\sigma_{\epolara}$ becomes small the viewing-angle distribution concentrates around the equator (upper-left subplot).
Note that, as $\sigma_{\epolara}$ shrinks and the viewing-angle distribution concentrates around the equator, the azimuthal-diffusion $\cE(a,b;\sigma_{\epolara})$ drops close to zero for large regions on the sphere.
Thus, for these equatorially-concentrated viewing-angle-distributions, templates of the original volume $\xF$ will align almost as well to the perturbed volume $\xF+\Delta\xF$ as they did to the original volume $\xF$ after longitudinally adjusting each template's viewing-angle $\tau$ by $\Delta\tau(\tau)$.
}
\label{fig_M3d_shape_longitudinal_perturbation_std_vs_sigma_FIGA}
\end{figure}

\subsection{Relating the laplacian to a local average:\ }
In the above calculations we've used the following fact.
Given a twice-differentiable function $u(\vx):\Real^{2}\rightarrow\Complex$, we can use the laplacian $\Delta u$ to approximate the average of that function along the perimeter $\partial B_{r}$ of a small disk $B_{r}$ of radius $r$.
\[ I(r) = \frac{1}{|\partial B_{r}|} \int_{B_{r}} u(\vx) d\Gamma. \]
\[ \partial_{r} I(r) = \frac{1}{|\partial B_{r}|} \int_{B_{r}} \partial_{x_{1}}^{2}u(\vx) + \partial_{x_{2}}^{2}u(\vx) d\vx = \frac{1}{|\partial B_{r}|} \int_{B_{r}} \Delta u(\vx) d\vx. \]
\[ \partial_{r} I(r) \approx \frac{1}{|\partial B_{r}|} \pi r^{2} \Delta u(0) \approx \frac{1}{2}\ r\ \Delta u(0). \]
\[ I(r) = I(0) + \int_{0}^{r} \partial_{r} I(s)ds \approx u(0) + \frac{1}{4}\ r^{2}\ \Delta u(0). \]
We can also describe a similar relationship: if we were to define $D_{r}$ to be the projection of $B_{r}$ onto the $x_{1}$ axis, then:
\[ I(r) = \frac{1}{|\partial D_{r}|} \int_{D_{r}} u(x_{1}) d\Gamma \approx u(0) + \frac{1}{4}\ r^{2}\ \partial^{2}_{x_{1}} u(0) \comma \]
which is what one would expect if $u$ did not vary (or only varied linearly) orthogonal to the $x_{1}$ direction (and hence $\partial_{x_{2}}^{2}u\equiv 0$).

\subsection{Notes for the longitudinal-perturbation:\ }
For the longitudinal-perturbation we set $\Theta(\eazimub) = 2\sin(2\eazimub)$.

Let's pick a particular point on the sphere, and describe it via the polar- and azimuthal-angles $(a,b)$.
The value of the third euler-angle $c$ can be used to parametrize the poles associated with that point.
Now all the pole-locations for $(a,b)$ are given by the same formula used to determine the templates:
\begin{equation}
(x,y,z) = [ \cos(b) \cos(a) \cos(c) - \sin(b) \sin(c) , \sin(b) \cos(a) \cos(c) + \cos(b) \sin(c) , -\sin(a) \cos(c) ]
\end{equation}
The norm $(x^2 + y^2) = w^2$ is given by:
\begin{eqnarray}
w^2 & = & ( (\cos(b) \cos(a) \cos(c) - \sin(b) \sin(c))^2 + (\sin(b) \cos(a) \cos(c) + \cos(b) \sin(c))^2 ) \\
\ & = & (\cos(b) \cos(a) \cos(c))^2 + (\sin(b) \sin(c))^2 - 2 (\cos(b) \cos(a) \cos(c) \sin(b) \sin(c)) \\
\ & \ & + (\sin(b) \cos(a) \cos(c))^2 + (\cos(b) \sin(c))^2 + 2 (\sin(b) \cos(a) \cos(c) \cos(b) \sin(c))  \\
\ & = & ( (\cos(b) \cos(a) \cos(c))^2 + (\sin(b) \sin(c))^2 + (\sin(b) \cos(a) \cos(c))^2 + (\cos(b) \sin(c))^2 ) \\
\ & = & (\cos(a) \cos(c))^2 + (\sin(c))^2 \period
\end{eqnarray}
The polar-angle at $(x,y,z)$ is:
\begin{equation}
\arctan(y/x) = \arctan\left( \frac{\sin(b) \cos(a) \cos(c)+\cos(b) \sin(c)}{\cos(b) \cos(a) \cos(c)-\sin(b) \sin(c)}\right) =: \phi(a,b,c) \period
\end{equation}
Note that, obviously, as any particular pole is perturbed azimuthally, the associated point on the sphere (associated with that pole) is perturbed azimuthally by the same amount.
Thus, for any pole $\phi$, the preimage $\phi^{\pre} = \phi^{\pre} - \epsilon \Theta(\phi^{\pre})$, or, more simply: $\phi^{\pos} - \phi^{\pre} = -\epsilon \sin(2 \phi^{\pre})$, which is the azimuthal change in the point induced by the azimuthal change in the pole.

Now we can estimate the average shift $\Delta b$ for any point on the sphere by averaging over $c$ (parametrizing the poles associated with that point).
\begin{multline}
\int_{0}^{2\pi} -\epsilon \sin(2 \phi) dc = -\epsilon \int_{0}^{2\pi} 2 \sin(\phi) \cos(\phi) dc = -\epsilon \int_{0}^{2\pi} 2 \frac{xy}{w^{2}} dc \\
= -\epsilon \int_{0}^{2\pi} 2 \frac{ (\sin(b) \cos(a) \cos(c) + \cos(b) \sin(c)) (\cos(b) \cos(a) \cos(c) - \sin(b) \sin(c)) }{ (\cos(a) \cos(c))^2 + (\sin(c))^2 } dc \period
\end{multline}

Now seeking a separable solution we fix the $b$ at, say, $\pi/4$ to get:
\begin{eqnarray}
 \ & = & -2 \epsilon \int_{0}^{2\pi} \frac{ (\sqrt{0.5} \cos(a) \cos(c) + \sqrt{0.5} \sin(c)) (\sqrt{0.5} \cos(a) \cos(c) - \sqrt{0.5} \sin(c)) }{ (\cos(a) \cos(c))^2 + (\sin(c))^2 } dc \\
 \ & = & -\epsilon \int_{0}^{2\pi} \frac{ (\cos(a) \cos(c))^2 - (\sin(c))^2 }{ (\cos(a) \cos(c))^2 + (\sin(c))^2 } dc \\
 \ & = & -\epsilon \int_{0}^{2\pi} \frac{ (\cos(a))^2 (\cos(c))^2 - (\sin(c))^2 }{ (\cos(a))^2 (\cos(c))^2 + (\sin(c))^2 } dc \period
\end{eqnarray}

Now using Gradshteyn \& Ryzhik: 3.647 (p402) BI (47)(20) we see: ;
\[ \int_{0}^{\pi/2} \frac{\cos(x)^p   \cos(px)}{a^2 \sin(x)^2 + b^2 \cos(x)^2}dx = \frac{\pi}{2 b} \frac{a^{p-1}}{(a+b)^p} \]
\[ \frac{1}{2 \pi}   \int_{0}^{2 \pi} \frac{\cos(x)^2}{\sin(x)^2 + b^2 \cos(x)^2}dx = \frac{1}{b (1+b)} \]
\[ +\int_{0}^{2\pi} \frac{ \alpha^2   \cos(c)^2 }{ \alpha^2 \cos(c)^2 + \sin(c)^2 }dc = \frac{\alpha^2}{\alpha (1+\alpha)} = -\frac{\alpha}{1+\alpha} \]
\[ +\int_{0}^{2\pi} \frac{ \sin(c)^2 }{ \alpha^2 \cos(c)^2 + \sin(c)^2 }dc = +\frac{1}{\alpha (1+\alpha)} - \frac{1}{\alpha} \]
\begin{eqnarray}
 +\int_{0}^{2\pi} \frac{ \alpha^2 \cos(c)^2 - \sin(c)^2 }{ \alpha^2 \cos(c)^2 + \sin(c)^2 }dc & = & +\frac{\alpha}{1+\alpha} + \frac{1}{\alpha (1+\alpha)} - \frac{1}{\alpha} \\
\ & = & \frac{\alpha^2 - \alpha }{ \alpha^2 + \alpha } = \frac{\alpha-1}{\alpha+1} \period
\end{eqnarray}
Note that,  the equations above hold for $\alpha>0$. Thus, in practice we'll need absolute-values in the equations above: i.e., $(|\alpha|-1)/(|\alpha|+1)$.

Aside: as we perturb the point-polar-angle $a$, the pole-azimuthal-angle $\phi$ changes as $\partial_{a} \phi$. ;
Note that:
\[ \partial\arctan(y/x) = \frac{1}{1+\frac{y^2}{x^2}} \cdot \left(\frac{y'}{x} - \frac{yx'}{x^2}\right) = \frac{x^2}{x^2+y^2}\cdot\frac{y'x - yx}{x^2} = \frac{y'x-yx'}{x^2+y^2} \period \]
Thus, after some arithmetic:
\begin{eqnarray}
\partial_{a} \phi & = & \frac{ (\cos(c))^2 (-\sin(b) \sin(a) \cos(b) \cos(a)) + (\sin(b))^2 (\sin(a) \cos(c) \sin(c)) }{ (\cos(a) \cos(c))^2 + (\sin(c))^2 } \\
\ & \ & + \frac{ (\cos(c))^2 (+(\cos(c))^2 (+\sin(b) \cos(a) \cos(b) \sin(a)) + (\cos(b))^2 (\sin(a) \cos(c) \sin(c)) }{ (\cos(a) \cos(c))^2 + (\sin(c))^2 } \\
\ & = & \frac{ \sin(a) \cos(c) \sin(c) }{ (\cos(a) \cos(c))^2 + (\sin(c))^2 } \period
\end{eqnarray}

\section{The Noise-marginalized limit}
\label{sec_noise_marginalized_limit_0}

In section \ref{sec_Volume_likelihood_mle} we discussed one version of the image-aligned limit involving the `low-noise' limit $\fsigma\rightarrow 0$, for which the likelihood in Eq. \ref{eq_P_AA_given_F} can be greatly simplified, producing Eq. \ref{eq_P_AA_given_F_maximum_likelihood}.
We leverage this simplified expression to produce Eq. \ref{eq_nll_qbp}, which affords certain computational and conceptual advantages.

A similar simplification can occur when the likelihood is marginalized over the noise $\fsigma$.
To demonstrate this, we can return to Eq. \ref{eq_P_A_given_B} and rewrite the likelihood of observing an image $\fA$, given a particular template $\fS$ and noise-level $\fsigma$:
\begin{eqnarray}
P(\fA \givenbig \fS,\fsigma) & := & \frac{1}{(\sqrt{2\pi}\ \fsigma)^{\dof}} \exp\left( - \frac{1}{\fsigma^{2}} \ell(\fA\givenmed \fS) \right) \comma
\label{eq_P_A_given_S_noise_marginal}
\end{eqnarray}
where $\ell(\fA\givenmed \fS)$ takes the form:
\begin{eqnarray}
\ell(\fA \givenmed \fS) & := & \frac{1}{2} \iint_{\vk\in\Omega_{\kmax}} |\fA(\vk) - \fS(\vk)|^{2} d\vk \comma
\label{eq_ell_A_given_S_noise_marginal}
\end{eqnarray}
and $\dof$ refers to the number of independent degrees-of-freedom in the image.

Now we can use Gradshteyn and Ryzhik 3.326.2:
\[ \int_{0}^{+\infty} \frac{1}{\xlambda^{\npixel}}\exp\left(-\frac{\ell}{\xlambda^{2}}\right)d\xlambda = \frac{1}{2}\Gamma\left(\frac{\npixel-1}{2}\right) \frac{1}{\ell^{(\npixel-1)/2}} = \frac{1}{2}\Gamma\left(\frac{\npixel-1}{2}\right) \cdot \exp\left(-\frac{\npixel-1}{2}\log\ell\right) \comma \]
and marginalize over the noise $\fsigma$ to produce:
\begin{eqnarray}
P\left(\fA\givenbig\fS,\strikefsigma\right) & = & \int_{0}^{+\infty} P\left(\fA\givenbig\fS,\fsigma\right)\ d\fsigma \\
\ & = & \frac{1}{(2\pi)^{\dof/2}}\cdot\frac{1}{2}\Gamma\left(\dof'\right) \cdot \exp\left(-\dof'\log\ell(\fA\givenmed\fS)\right) \comma
\label{eq_marginal1_likelihood_freq_null_noise_marginal}
\end{eqnarray}
where we have denoted $\fsigma$-marginalization via the strikethrough $\strikefsigma$, and set $\dof'=(\dof-1)/2$.

Note that $\dof'$ is often quite large in practice, and thus this marginalized likelihood is typically very well approximated using standard asymptotics (e.g., laplace's approximation).
In this regime we can marginalize Eq. \ref{eq_P_A_given_F} over $\fsigma$ and write:
\begin{eqnarray}
P\left(\fA \givenbig \fF ,\strikefsigma \right) & = & \int_{\vd\in \Real^{2}} \int_{\tau\in SO3} P\left(\fA \givenbig \tau ; \vd ; \fF ; \strikefsigma \right)\cdot P\left(\vd \givenbig \tau ; \fF \right) \cdot P\left(\tau \givenbig \fF \right) d\tau d\vd  \comma
\label{eq_P_A_given_F_noise_marginal} 
\end{eqnarray}
where $P\left(\fA \givenbig \tau ; \vd ; \fF ; \strikefsigma \right)$ now represents a marginalized likelihood similar to Eq. \ref{eq_marginal1_likelihood_freq_null_noise_marginal}.
In the large-$\dof$ limit we now have a statement very similar to Eq. \ref{eq_P_AA_given_F_maximum_likelihood_alignment}.
Namely, as $\dof\rightarrow\infty$, the integral in Eq. \ref{eq_P_A_given_F_noise_marginal} becomes dominated by the contribution associated with the `maximum-likelihood' alignment-angle for the image $\fA$.
Just as before, this maximum-likelihood alignment occurs for the viewing-angle, displacement and volume which maximize the summand $P(\fA \givenmed \tau ; \vd ; \fF ; \strikefsigma)\cdot P( \tau , \vd  \givenmed \fF ) $ in the integrand of Eq. \ref{eq_P_A_given_F_noise_marginal}.

In the context of single-particle reconstruction, the negative-log-likelihood for a collection of images in the large-$\dof$ limit is now dominated by the following leading-order expansion:
\begin{eqnarray}
-\log\left(P\left(\{\fA_{j}\}\givenbig \fF ,\strikefsigma \right)\right) \ \sim \ +\dof'\cdot\sum_{j=1}^{\nimage} \log\left(\ell\left(\fA_{j}\givenmed\fS_{j}^{\opt}\right)\right) \comma
\label{eq_P_AA_given_F_maximum_likelihood_noise_marginal_0}
\end{eqnarray}
with:
\[ \fS_{j}^{\opt}(\vk) := \fS\left(\vk;\tau^{\opt}_{j};\vd^{\opt}_{j};CTF_{j};\fF\right) \period \]
While the term $\log\left(\ell\right)$ in Eq. \ref{eq_P_AA_given_F_maximum_likelihood_noise_marginal_0} is nonlinear in $\ell$ (as opposed to previous equations, such as Eq. \ref{eq_P_AA_given_F_maximum_likelihood}), it is possible for the numerical-range of $\ell$ to allow for a good linear approximation, which we'll discuss at the end of the next section.

\subsection{marginalizing with respect to offset $\mu$:}\
In practice it is often convenient to take into consideration an image-specific real-space offset $\mu$:
\begin{eqnarray}
P(\fA \givenbig \fS,\fsigma,\mu) & := & \frac{1}{(\sqrt{2\pi}\ \fsigma)^{\dof}} \exp\left( - \frac{1}{\fsigma^{2}} \ell(\fA\givenmed \fS,\mu) \right) \comma
\label{eq_P_A_given_S_mu_noise_marginal}
\end{eqnarray}
where $\ell(\fA\givenmed \fS,\mu)$ takes the form:
\begin{eqnarray}
\ell(\fA \givenmed \fS,\mu) & := & \frac{1}{2} \iint_{\vx\in\Omega_{\xmax}} |\xA(\vx) + \mu - \xS(\vx)|^{2} d\vx \period
\label{eq_ell_A_given_S_mu_noise_marginal}
\end{eqnarray}

To marginalize this likelihood with respect to the image-offset $\mu$ we can use Gradshteyn and Ryzhik 2.33:
\[ \int_{-\infty}^{+\infty} \exp\left(-\frac{a x^{2}\pm 2b x+c}{\lambda}\right) dx = \sqrt{\frac{\lambda\pi}{a}}\exp\left(-\frac{ac-b^{2}}{\lambda a}\right) \comma \]
which holds for any normalizing factor $\lambda$.

Using this relationship, we can write:
\begin{eqnarray}
P\left(\fA\givenbig\fS,\fsigma,\strikemu\right) & = & \int_{-\infty}^{+\infty} P\left(\fA\givenbig\fS,\fsigma,\mu\right) d\mu  \\
\ & = & \frac{1}{(2\pi)^{\dof/2}}\cdot\frac{1}{\fsigma^{\dof}}\cdot (2\pi\fsigma^{2})\cdot \exp\left(-\frac{1}{\fsigma^{2}}\ell\left(\fA\givenbig\fS,\strikemu\right)\right) \comma
\label{eq_marginal0_likelihood_freq}
\end{eqnarray}
\begin{equation}
\text{with:}\ \ell\left(\fA\givenbig\fS,\strikemu\right) := \frac{1}{2} \iint_{\vx\in\Omega_{\xmax}} \left|\left(\xA(\vx)-\bar{A}\right) - \left(\xS(\vx)-\bar{S}\right)\right|^{2} d\vx \comma
\label{eq_marginal0_ell_freq_null}
\end{equation}
where we have used $\bar{A}$ and $\bar{S}$ to denote the average image- and template-value.

Given the structure of $\ell\left(\fA\givenbig\fS,\strikemu\right)$, we can assume without loss of generality that the images are all `centered' (i.e., with mean $0$).
Similarly, we can assume that the volume $\xF$ is also mean $0$, and hence every template $\xS$ is centered as well.

After performing this offset-marginalization we can once again marginalize over the noise to obtain the negative-log-likelihood:
\begin{eqnarray}
-\log\left(P\left(\{\fA_{j}\}\givenbig \fF ,\strikefsigma,\strikemu \right)\right) \ \sim \ +\dof'\cdot\sum_{j=1}^{\nimage} \log\left(\ell\left(\fA_{j}\givenmed\fS_{j}^{\opt},\strikemu\right)\right) \period
\label{eq_P_AA_given_F_maximum_likelihood_mu_noise_marginal_0}
\end{eqnarray}

\subsection{approximation to $\log(\ell)$:}
\begin{figure}[H]
\centering
\includegraphics[width=1.0\textwidth]{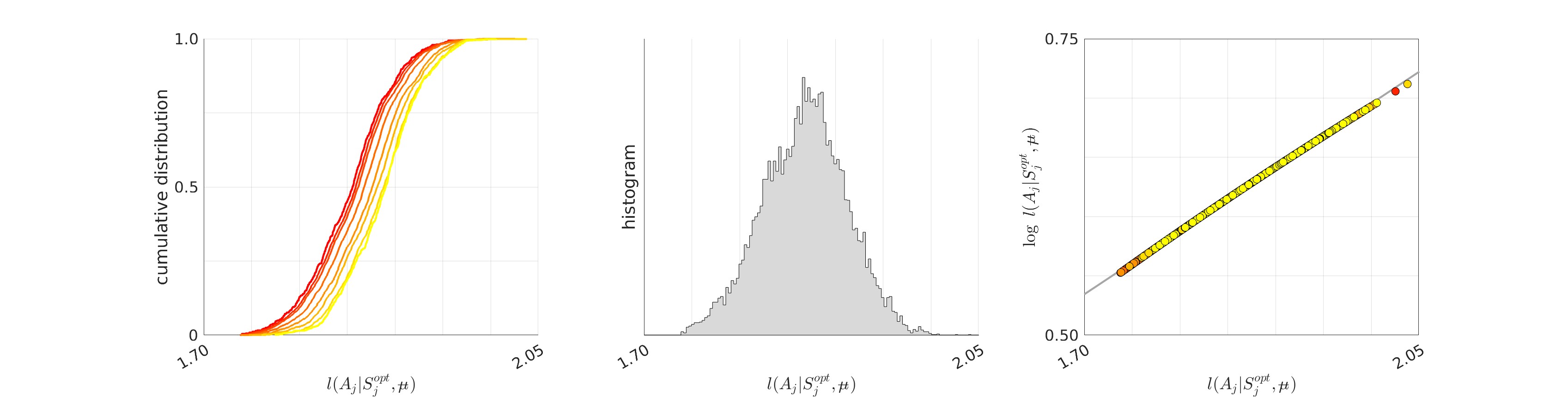}
\caption{
  In this figure we illustrate values of $\ell(\fA_{j}\givenmed\fS_{j}^{\opt},\strikemu)$ taken from micrographs of FIB-milled yeast lamellae as described in \protect\cite{ZCRLG2024} (see Fig 1 and Fig 15 of \protect\cite{ZCRLG2024}).
  These micrographs have been scanned for picked particle images that resemble mature 60S ribosomes.
  For each of these picked particle images $\xA_{j}$, we measure the optimal pose-parameters for the 60S ribosome, and calculate the corresponding projection $\fS_{j}^{\opt}$, as well as the function $\ell_{j}:=\ell(\fA_{j}\givenmed\fS_{j}^{\opt},\strikemu)$.
  On the far left we show the cumulative-distribution function of $\{\ell_{j}\}$ for the original set of picked-particles (bright red).
  We also show the corresponding cumulative-distribution-functions for the particles that were detected after adding various amounts of gaussian-noise to the original micrograph.
  The noise added to the original micrograph is quantified by the additional variance per-pixel (relative to the original noise-level), which ranges across the values $\{0.0,0.1,0.2,0.5,1.0,1.5,2.0,2.5\}$ (ranging from red to yellow, respectively).
  In the middle we show the histogram of values for $\{\ell_{j}\}$ accumulated across all $16014$ picked-particles.
  On the right we show a scatterplot of $\ell_{j}$ along the horizontal, and $\log(\ell_{j})$ along the vertical.
  The color indicates the amount of added noise, and uses the same scale as the right subplot.
  The affine-fit is shown in grey in the background.
  Note that, for this experimental data, the values of $\log(\ell_{j})$ can be very well approximated by an affine function $\log(\ell)\approx p_{1}\ell + p_{0}$ for values of $p_{1}\approx 0.537$ and $p_{0}\approx -0.378$ that are fixed and independent of the image-index $j$.
  Indeed, for this data-set the average relative error between the the values of $\log(\ell_{j})$ and the linear-fit is $\approx 6\times 10^{-4}$.
}
\label{fig_test_normalized_cc_yeast_ribo_20241115_FIGC}
\end{figure}

\
In many practical scenarios the values of $\ell(\fA_{j}\givenmed\fS_{j}^{\opt},\strikemu)$ are rather tightly distributed across the image-pool $j$, and it is not unreasonable to approximate the function $\log(\ell)$ as an affine function $\log(\ell)\approx p_{1}\ell+p_{0}$ for some fixed $p_{1}$ and $p_{0}$.
These $p_{1}$ and $p_{0}$ can be taken as regression-coefficients calculated across the image-pool.
One such example is shown in Fig \ref{fig_test_normalized_cc_yeast_ribo_20241115_FIGC}.

When this approximation can be made, the negative-log-likelihood can be approximated by:
\begin{eqnarray}
-\log\left(P\left(\{\fA_{j}\}\givenbig \fF , \strikefsigma ,\strikemu \right)\right) \ \approx \ +\dof'\cdot p_{0}\cdot \nimage + \dof'\cdot p_{1}\cdot \sum_{j=1}^{\nimage} \ell\left(\fA_{j}\givenmed\fS_{j}^{\opt},\strikemu\right) \comma
\label{eq_P_AA_given_F_maximum_likelihood_mu_noise_marginal_1}
\end{eqnarray}
Which is once again amenable to a volumetric-representation very similar to Eq. \ref{eq_nll_qbp}.

\subsection{consistency of image-aligned limits:}\
We remark that the high-degree-of-freedom limit obtained after noise-marginalization is compatible with the low-noise limit discussed in section \ref{sec_Volume_likelihood_mle}.
To elaborate, consider an idealized scenario involving synthetic-images, as described in section \ref{sec_synthetic_images}.
In this scenario, each synthetic-image is obtained by taking one of the templates $\fS(\vk;\tau;\fF)$ and then adding additional noise with variance $\fsigma^{2}$ per degree-of-freedom.
Thus, for each synthetic-image, the term $\ell\left(\fA_{j}\givenmed\fS_{j}^{\opt}\right)$ is simply $\frac{1}{2}\|\fA_{j}^{\noise}\|$, which itself is drawn from $\cN\left(\frac{1}{2}\ \dof\ \fsigma^{2},\frac{1}{2}\ \dof\ \fsigma^{4}\right)$.
The corresponding linear-regression coefficient (calculated across all the images) will then satisfy:
\[ p_{1} \sim \frac{2}{\dof} \cdot \frac{1}{\fsigma^{2}} \comma \]
in the limit as the number of images and number of degrees-of-freedom become large.
Plugging this expression for $p_{1}$ into Eq \ref{eq_P_AA_given_F_maximum_likelihood_noise_marginal_0} or \ref{eq_P_AA_given_F_maximum_likelihood_mu_noise_marginal_1} recovers the same structure as Eq. \ref{eq_nll_q2d} in section \ref{sec_Volume_likelihood_3d}.

\section{Ill-posedness: Case studies}
\label{sec_ill_posedness_case_studies}
\subsection{Equatorially-distributed viewing-angles}

As described in section \ref{sec_local_well_posedness_ansatz_longitudinal_perturbation_0}, the single-particle reconstruction problem is generically poorly posed when the viewing angles are concentrated around the equator.
More specifically, assume we are given a volume $\fF$ and a collection of templates taken from a viewing-angle-distribution $\mu(\tau)$ supported on the equator.
Now any equatorial-perturbation $\Delta\tau(\tau)$ of these viewing-angles can be applied to the original viewing-angles.
These perturbed viewing-angles can then be used to reconstruct a volume $\fG$.

As long as the map $\tau \leftarrow \tau+\Delta\tau(\tau)$ is one-to-one, the reconstructed $\fG$ will then produce a collection of templates that perfectly match those templates used to reconstruct it.
In other words, the collection of templates of $\fG$ will include the original collection of templates from $\fF$.

\begin{figure}[H]
\centering
\includegraphics[width=1.0\textwidth]{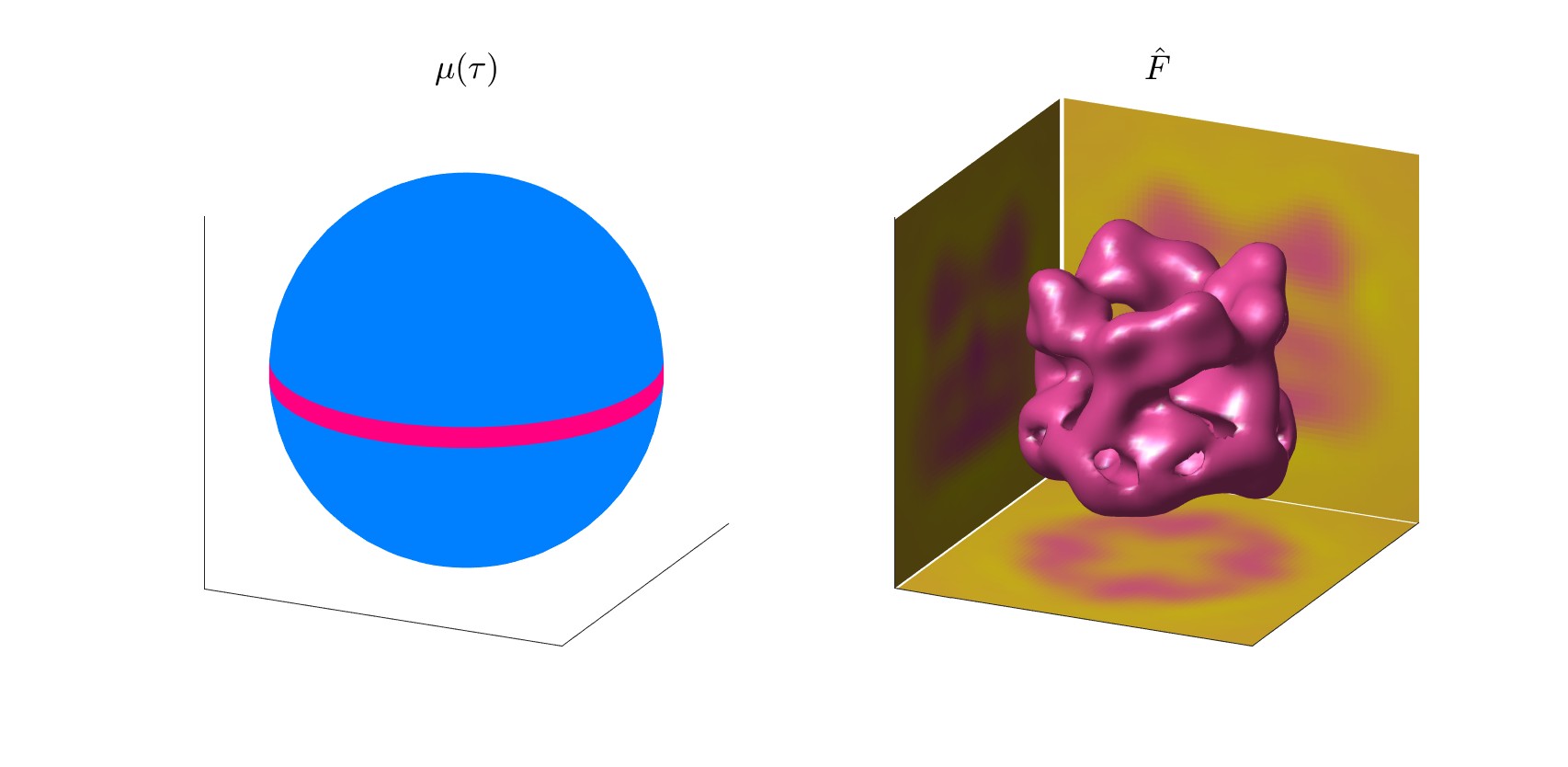}
\caption{
On the left we show an idealized viewing-angle-distribution $\mu(\tau)$ which is uniformly distributed along the equator.
On the right we show a low-res ($2\pi\kmax=48$) reconstruction $\xF$ of the TRPV1 molecule obtained by appropriately aligning many synthetic-images (i.e., noisy projections) taken from these equatorial viewing-angles.
This low-res reconstruction $\xF$ agrees with the reconstruction referenced in {\tt EMPIAR-10005}.
}
\label{fig_trpv1_k48_eig_i1_from_synth_nltInfpm49_p_equa_band_FIGL}
\end{figure}

\begin{figure}[H]
\centering
\includegraphics[width=1.0\textwidth]{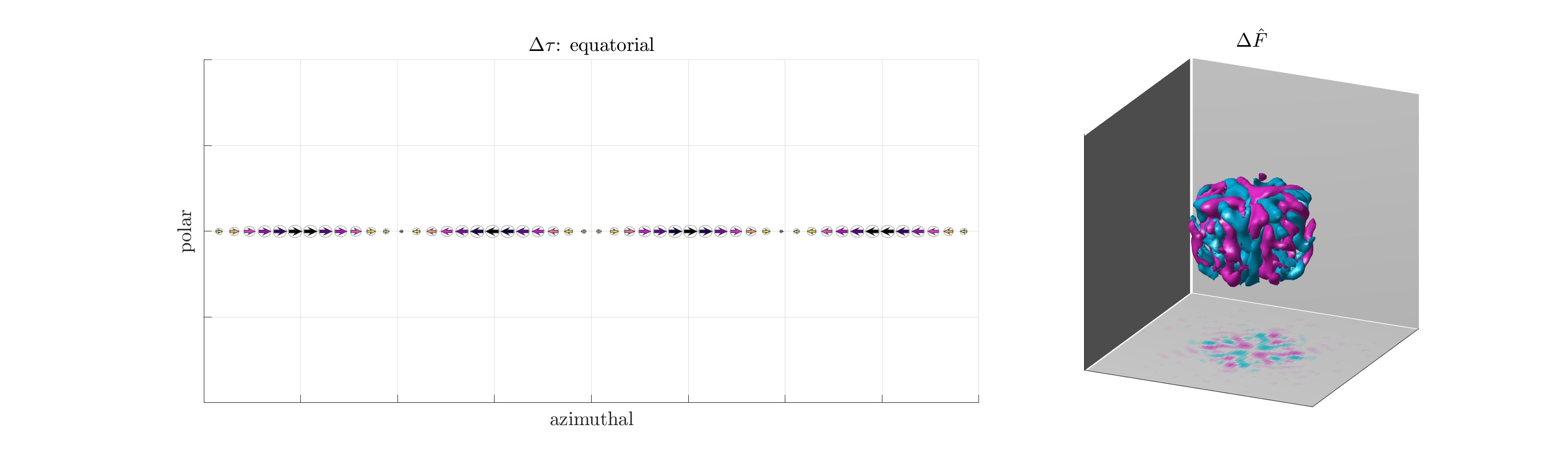}
\caption{
In this figure we show a collection of $\tau$-specific alignment-perturbations $\{\Delta\tau(\tau)\}$ (left) along with a corresponding volumetric-perturbation $\delta\xF$ (right).
The arrows on the left indicate the direction of the alignment-perturbations for each $(\epolara,\eazimub)$, with $\Delta\egammaz(\tau)=0$.
Larger/darker arrows indicate larger values of $\Delta\tau(\tau)$, with the arrow-area proportional to $|\Delta\tau|$ (and dark purple arrows corresponding to the largest values of $|\Delta\tau|$.
On the right we show an isosurface of $|\Delta\xF|$, colored by the sign of $\Delta\xF$ (with blue and pink corresponding to negative and positive, respectively).
The projections of $\Delta\xF$ are shown along each axis (background and bottom).
These perturbations have been selected so that, when applied to the reconstruction $\xF$ shown in Fig \ref{fig_trpv1_k48_eig_i1_from_synth_nltInfpm49_p_equa_band_FIGL}, the likelihood remains the same.
Put another way, each equatorial-template of the original $\xF$ at angle $\tau$ corresponds to a template of the perturbed volume $\xF+\Delta\xF$ at angle $\tau+\Delta\tau(\tau)$.
Indeed, when inspecting the right-hand subplot, one can see that the equatorial projections of $\Delta\xF$ shown along the background are identically $0$. Only the polar-projection shown on the bottom is nonzero (however, for this example there are no viewing-angles that have a polar-component).
}
\label{fig_trpv1_k48_eig_i1_from_synth_nltInfpm49_p_equa_band_FIGN}
\end{figure}

\begin{figure}[H]
\centering
\includegraphics[width=1.0\textwidth,trim=000 150 000 175,clip]{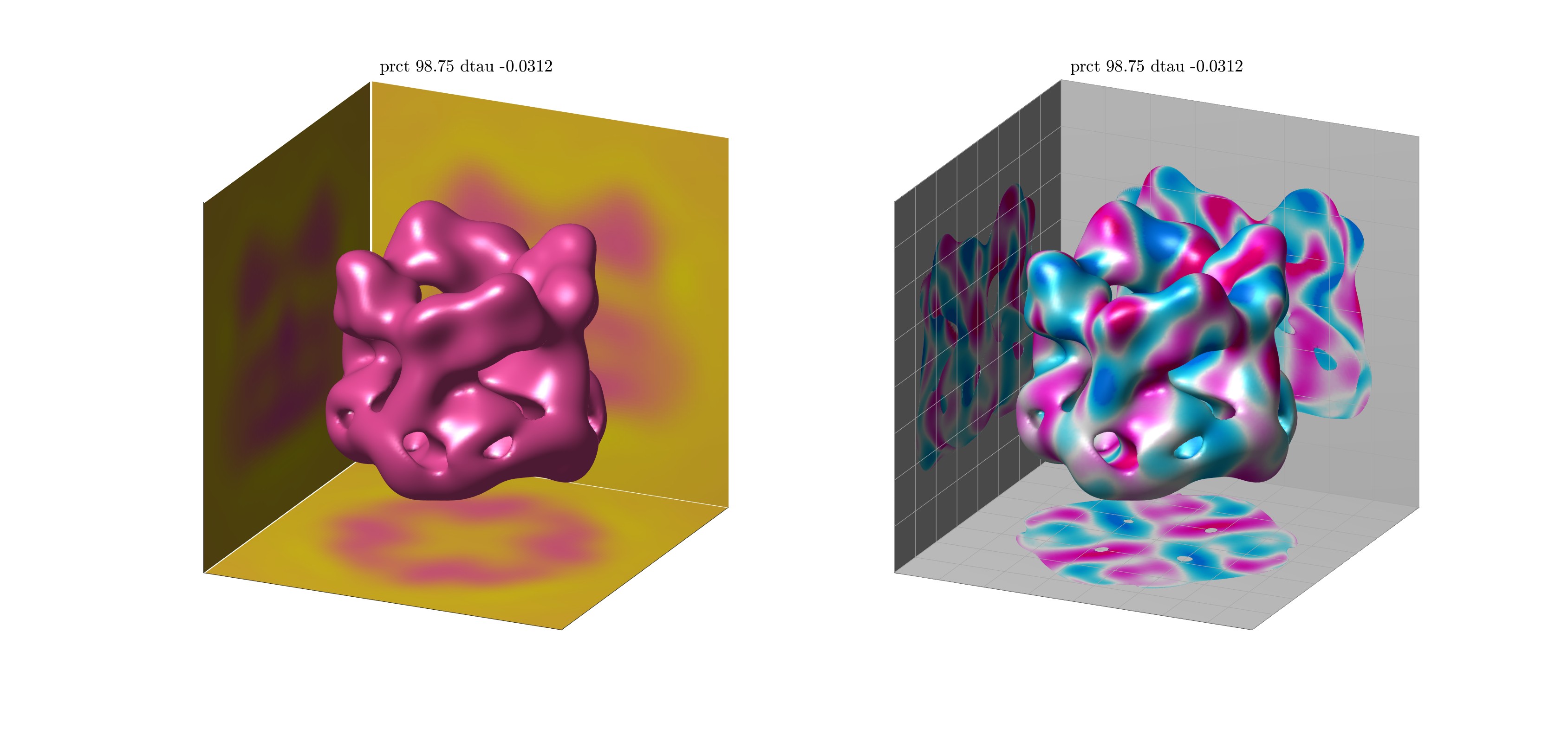}
\caption{
In this figure we show one of the perturbed volumes $\xF-\Delta\xF$ of the TRPV1 molecule (at $2\pi\kmax=48$).
On the left subplot we show an isosurface (center), along with projections along each axis (background and bottom).
On the right subplot we show the same isosurface, re-colored to indicating the value of the volumetric-deformation $\Delta\xF$ at each point.
The color on the right ranges from blue (negative) to pink (positive).
}
\label{fig_trpv1_k48_eig_i1_from_synth_nltInfpm49_p_equa_band_FIGK_p9875_dtau8}
\end{figure}

\begin{figure}[H]
\centering
\includegraphics[width=1.0\textwidth,trim=000 150 000 175,clip]{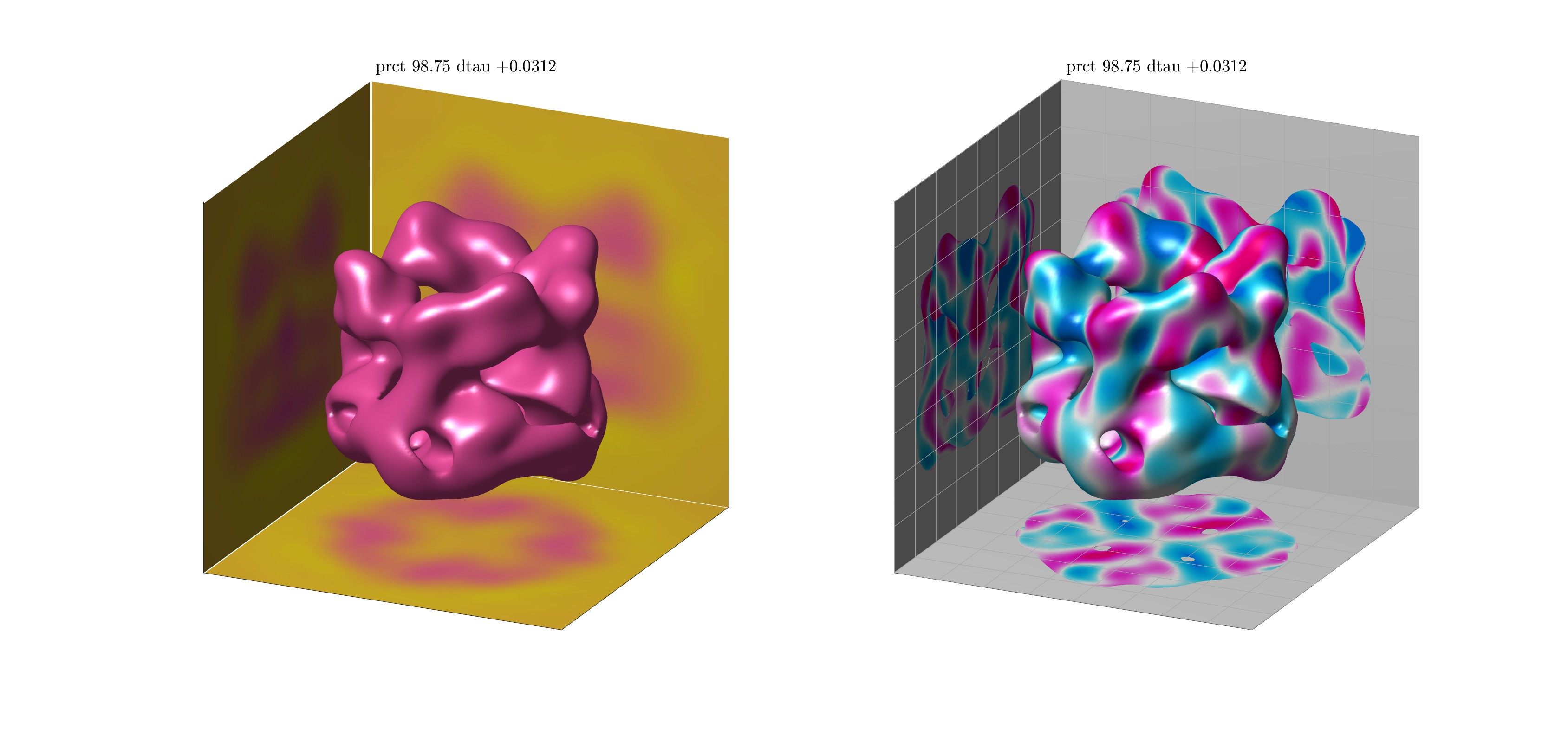}
\caption{
This figure has the same format as Fig \ref{fig_trpv1_k48_eig_i1_from_synth_nltInfpm49_p_equa_band_FIGK_p9875_dtau8}, except that we illustrate $\xF+\Delta\xF$.
}
\label{fig_trpv1_k48_eig_i1_from_synth_nltInfpm49_p_equa_band_FIGK_p9875_dtau9}
\end{figure}

\begin{figure}[H]
\centering
\includegraphics[width=1.0\textwidth,trim=000 150 000 175,clip]{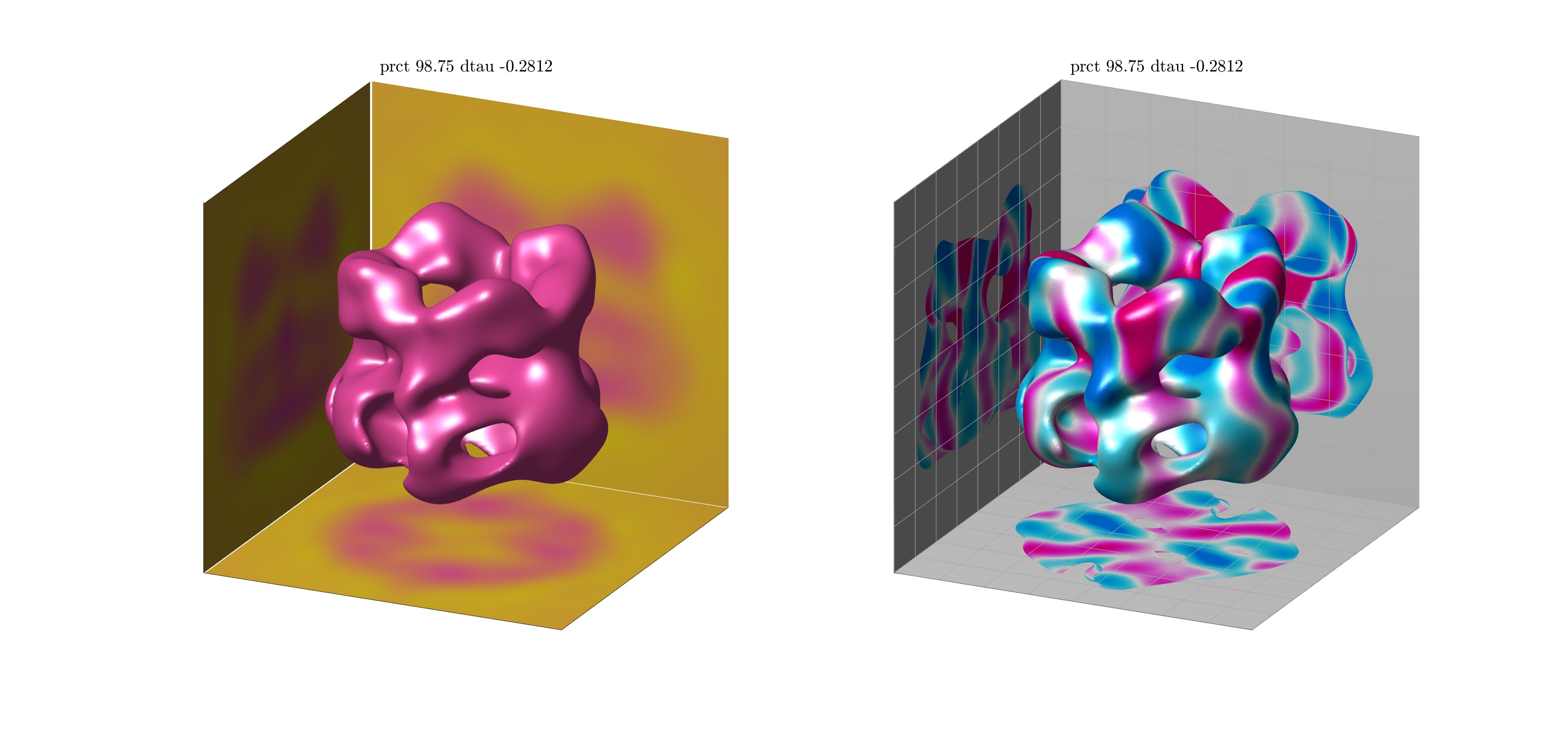}
\caption{
This figure illustrates an even more extreme volumetric perturbation that, again, has the same collection of equatorial-templates as the original $\xF$.
}
\label{fig_trpv1_k48_eig_i1_from_synth_nltInfpm49_p_equa_band_FIGK_p9875_dtau4}
\end{figure}

\begin{figure}[H]
\centering
\includegraphics[width=1.0\textwidth,trim=000 150 000 175,clip]{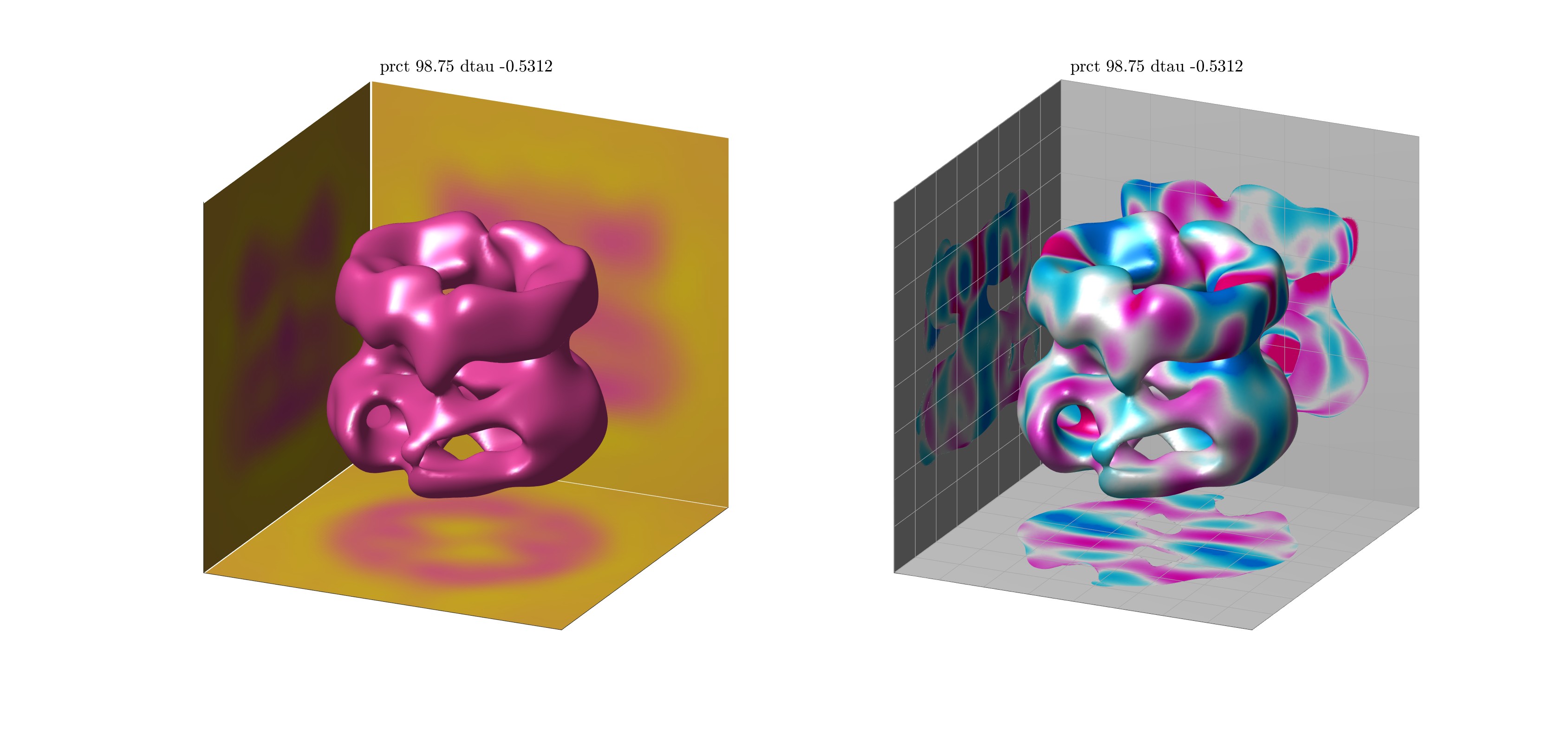}
\caption{
This figure illustrates an even more extreme volumetric perturbation that, again, has the same collection of equatorial-templates as the original $\xF$.
}
\label{fig_trpv1_k48_eig_i1_from_synth_nltInfpm49_p_equa_band_FIGK_p9875_dtau0}
\end{figure}

\begin{figure}[H]
\centering
\includegraphics[width=1.0\textwidth,trim=000 150 000 175,clip]{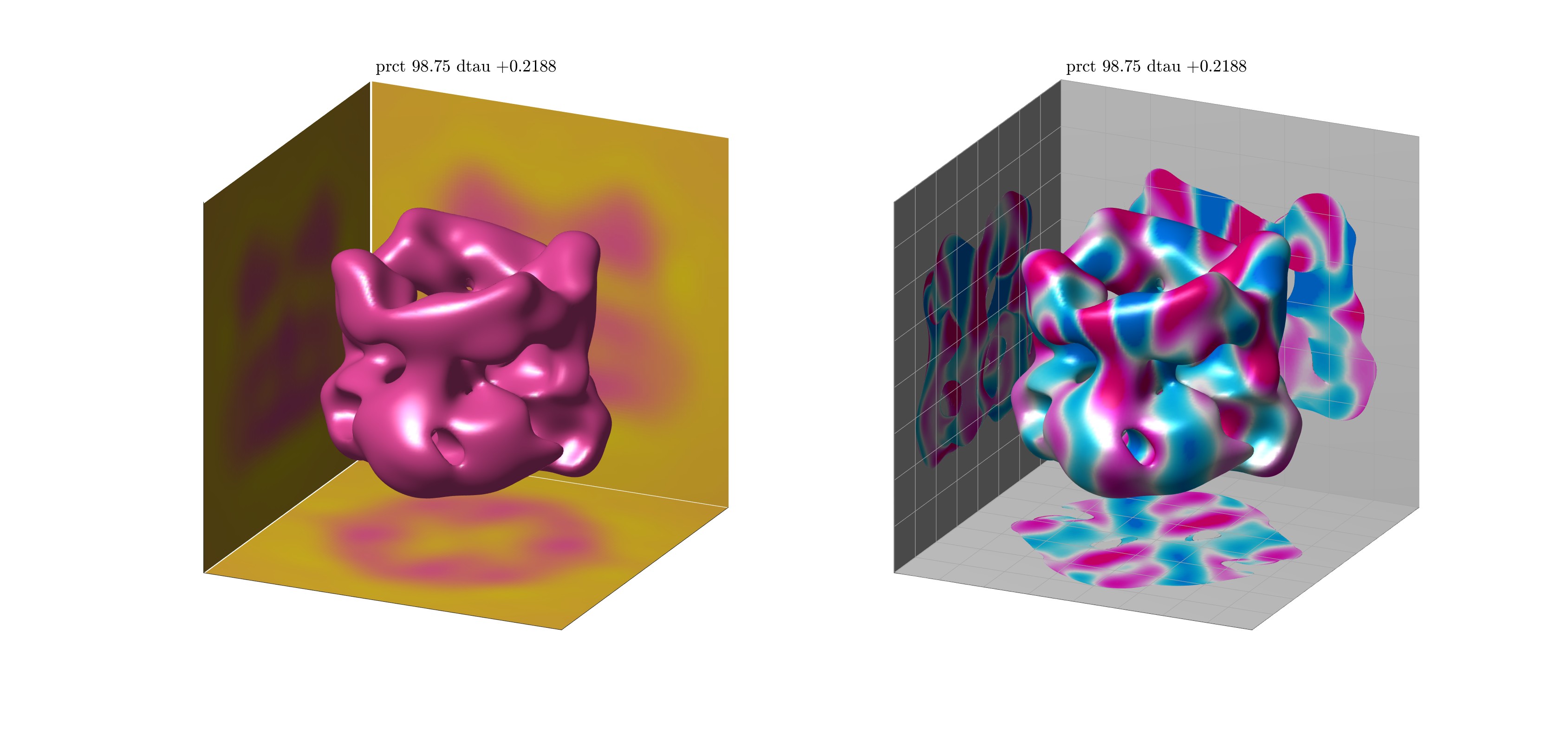}
\caption{
This figure illustrates an even more extreme volumetric perturbation that, again, has the same collection of equatorial-templates as the original $\xF$.
}
\label{fig_trpv1_k48_eig_i1_from_synth_nltInfpm49_p_equa_band_FIGK_p9875_dtau12}
\end{figure}

\begin{figure}[H]
\centering
\includegraphics[width=1.0\textwidth,trim=000 150 000 175,clip]{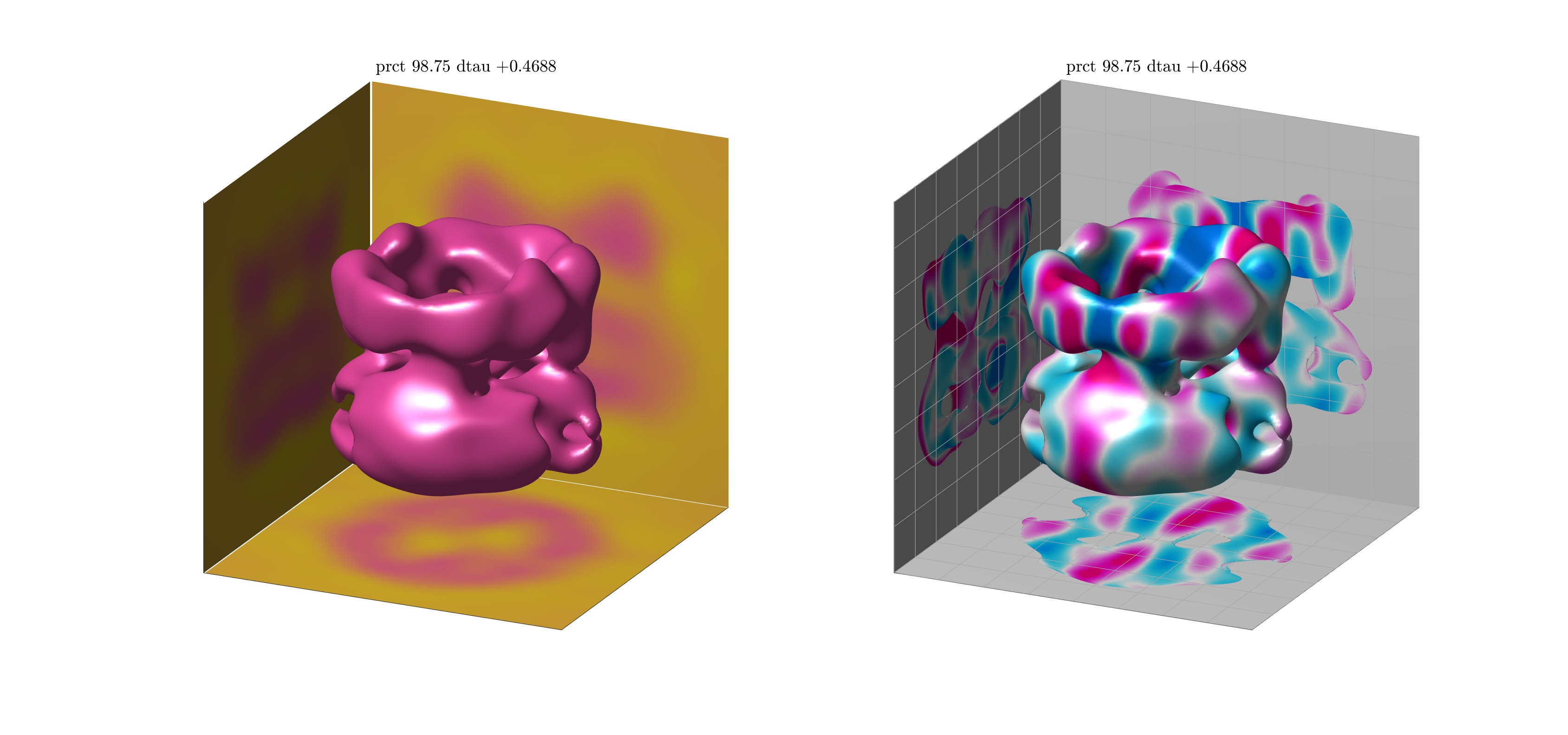}
\caption{
This figure illustrates an even more extreme volumetric perturbation that, again, has the same collection of equatorial-templates as the original $\xF$.
To summarize, because the viewing-angle distribution is supported along the equator, any one-to-one equatorial perturbation of the viewing-angles gives rise to a reconstructed volume with the same collection of equatorial-projections (i.e., templates).
In other words, the single-particle reconstruction problem is ill-posed for this example.
}
\label{fig_trpv1_k48_eig_i1_from_synth_nltInfpm49_p_equa_band_FIGK_p9875_dtau16}
\end{figure}

This issue is a result of the fact that, when $\mu(\tau)$ is restricted to the equator, different templates only overlap at the poles.
Any invertible equatorial-perturbation simply reorganizes the azimuthal-angle associated with each template, without disturbing this polar overlap.
Examples of this phenomenon are shown in Figs \ref{fig_trpv1_k48_eig_i1_from_synth_nltInfpm49_p_equa_band_FIGL}-\ref{fig_trpv1_k48_eig_i1_from_synth_nltInfpm49_p_equa_band_FIGK_p9875_dtau16}.


\subsection{Uniformly-distributed viewing-angles}
In the section above we discussed a somewhat degenerate scenario: one where the viewing-angles $\tau_{j}$ for each observed image were all drawn from the equator.
In a more general scenario we would expect the distribution $\mu(\tau)$ of viewing-angles to be more uniformly distributed.
In this more general situation we typically don't find the same kinds of degeneracies.
Indeed, in the zero-noise limit it is typically the case that there will be a unique volume that satisfies the single-particle reconstruction problem \cite{ZS2014,BBMZS2018}.

\begin{figure}[H]
\centering
\includegraphics[width=1.0\textwidth]{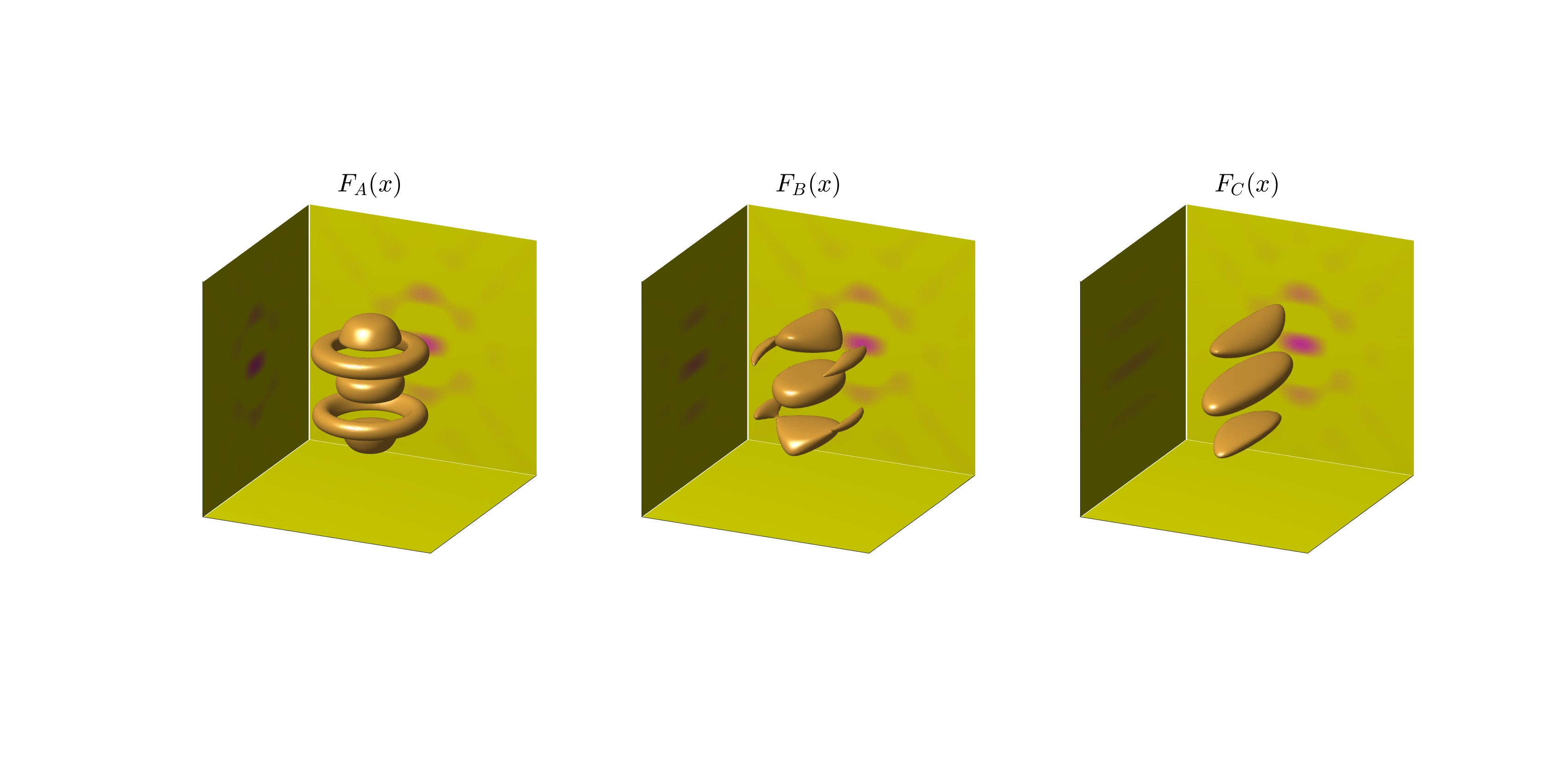}
\caption{
In this figure we illustrate the frequency-shells $\fF_{A}(k,\hk)$, $\fF_{B}(k,\hk)$ and $\fF_{C}(k,\hk)$.
The real- and imaginary-parts of these frequency-shells are shown from both a three-quarters view and a polar view.
Each frequency shell adopts the background values of $b$ and $b^{\dagger}$ in the northern- an southern-hemispheres, respectively.
in the regions surrounding the north- and south-poles, these frequency-shells adopt the values of $d$ and $d^{\dagger}$, respectively.
Note that any great-circle cutting through $\fF_{A}(k,\hk)$ will correspond to at least one great-circle cutting through $\fF_{B}(k,\hk)$ and at least one great-circle cutting through $\fF_{C}(k,\hk)$.
}
\label{fig_test_heterogeneity_spurious_s1sg125sm650bv8i8sr6sv16i12cr5cv12i10_FIGA}
\end{figure}
\begin{figure}[H]
\centering
\includegraphics[width=1.0\textwidth]{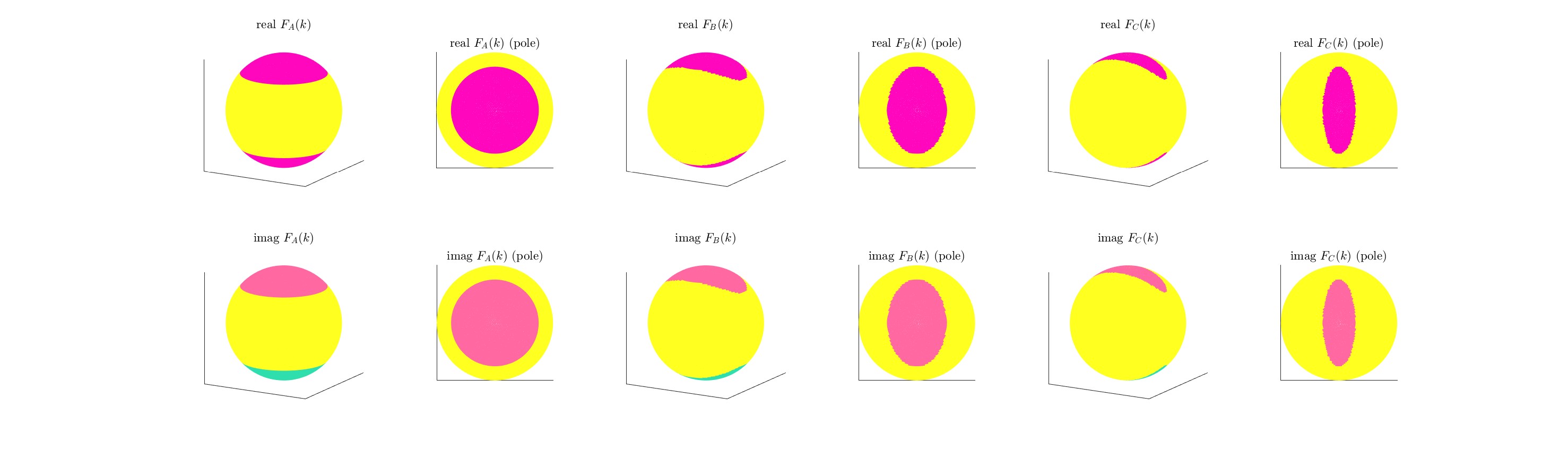}
\caption{
In this figure we illustrate level-sets of the volumes produced by convolving the frequency-shells shown in Fig \ref{fig_test_heterogeneity_spurious_s1sg125sm650bv8i8sr6sv16i12cr5cv12i10_FIGA} with a radial-kernel.
Note that, by construction, each projection of the volume $\xF_{A}$ will correspond to some projection of $\xF_{B}$ and also to some projection of $\xF_{C}$.
In other words, given the full collection of templates produced by $\xF_{A}$, the single-particle reconstruction problem is poorly posed.
}
\label{fig_test_heterogeneity_spurious_s1sg125sm650bv8i8sr6sv16i12cr5cv12i10_FIGB}
\end{figure}

Nevertheless, even though they are not generic, degeneracies can still exist.
One of many such examples is shown in Figs \ref{fig_test_heterogeneity_spurious_s1sg125sm650bv8i8sr6sv16i12cr5cv12i10_FIGA} and \ref{fig_test_heterogeneity_spurious_s1sg125sm650bv8i8sr6sv16i12cr5cv12i10_FIGB}.
To explain this example, let's first consider a single frequency-shell $\fF_{A}(k,\hk)$ (with fixed frequency-amplitude $k$).
On this frequency-shell let's imagine that there is a baseline value of $b$ in the northern-hemisphere and -- due to the conjugacy constraint -- the corresponding value of $b^{\dagger}$ in the southern-hemisphere.
Let's also assume there exists a localized feature, in this case a circular disc `$D$' around the north-pole where the value of $\fF_{A}(k,\hk)$ is elevated from the baseline value of `$b$' to some different value '$d$'.
Because of the conjugacy-constraint applied to $\fF_{A}$, the presence of $D$ mandates an antipodal region $D'$ on which $\fF_{A}$ takes the value $d^{\dagger}$.
On this frequency-shell each projection corresponds to a great-circle.
Most such great-circles do not intersect $D$ or $D'$, and thus correspond to projections $\fS$ which cut through the background-values of $b$ and $b^{\dagger}$ (i.e., for which $\fS(\psi)\equiv b$ or $b^{\dagger}$ for all $\psi$).
However, great-circles that intersect $D$ and $D'$ will produce projections $\fS$ that have intervals of $\psi$ for which $\fS(\psi)=d$ and $\fS(\psi+\pi)=d^{\dagger}$.

Now consider another frequency-shell $\fF_{B}(k,\hk)$ which is similar to $\fF_{A}(k,\hk)$, except that the localized feature now lies on an elliptical zone `$E$'.
The zone $E$ has a major-axis that matches that of $D$, but a minor-axis which is smaller.
Let's assume that $\fF_{B}$ takes on the values $b$ on most of the sphere, and the values $d$ and $d^{\dagger}$ on the elliptical-regions $E$ and $E'$ (where $E'$ is antipodal to $E$).

Now each great-circle from $\fF_{A}(k,\hk)$ corresponds to at least one great-circle from $\fF_{B}(k,\hk)$, and vice-versa.
A collection of templates corresponding to a uniform distribution of viewing-angles from $\fF_{A}$ can now be associated with a collection of templates corresponding to a non-uniform distribution of viewing-angles from $\fF_{B}$.
Of course the same story holds if we reduce the minor-axis of the elliptical-region $E$ to produce a shell $\fF_{C}(k,\hk)$.
Thus, in this case the single-particle reconstruction problem is poorly posed.
Give only the collection of templates from $\xF_{A}$, but without knowledge of the viewing-angles nor of the viewing-angle-distribution $\mu(\tau)$, one cannot determine if the true underlying volume is $\xF_{A}$, $\xF_{B}$ or any other admissible function (e.g., $\xF_{C}$).

\begin{figure}[H]
\centering
\includegraphics[width=1.0\textwidth]{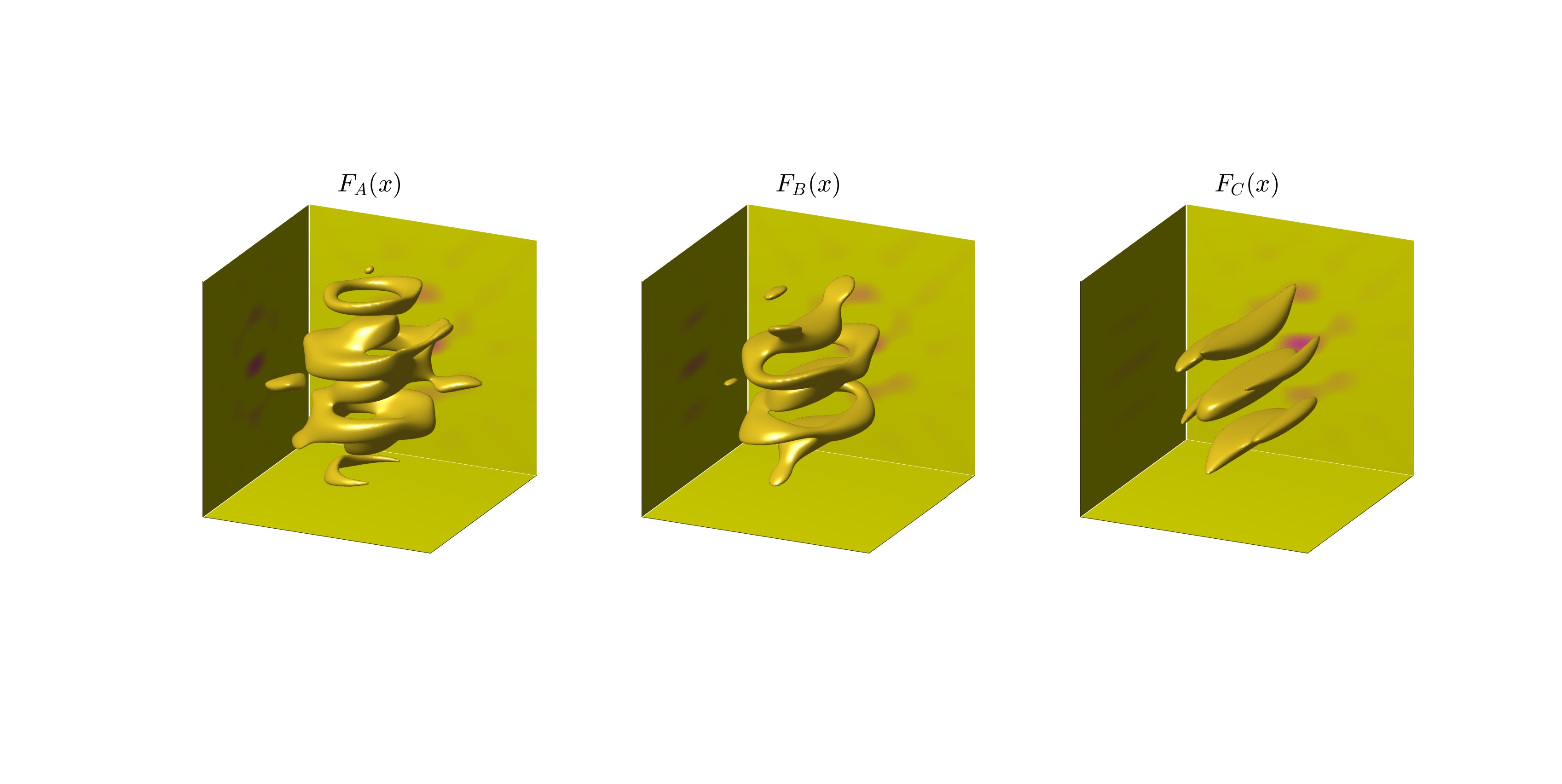}
\caption{
This figure is similar to Fig \ref{fig_test_heterogeneity_spurious_s1sg125sm650bv8i8sr6sv16i12cr5cv12i10_FIGA}.
This time the frequency-shells adopt more than one value in the polar-regions.  
Note that any great-circle cutting through $\fF_{A}(k,\hk)$ will correspond to at least one great-circle cutting through either $\fF_{B}(k,\hk)$ and/or $\fF_{C}(k,\hk)$.
}
\label{fig_test_heterogeneity_spurious_s2sg125sm650bv8i8sr6sv16i12cr5cv12i10_FIGA}
\end{figure}

\begin{figure}[H]
\centering
\includegraphics[width=1.0\textwidth]{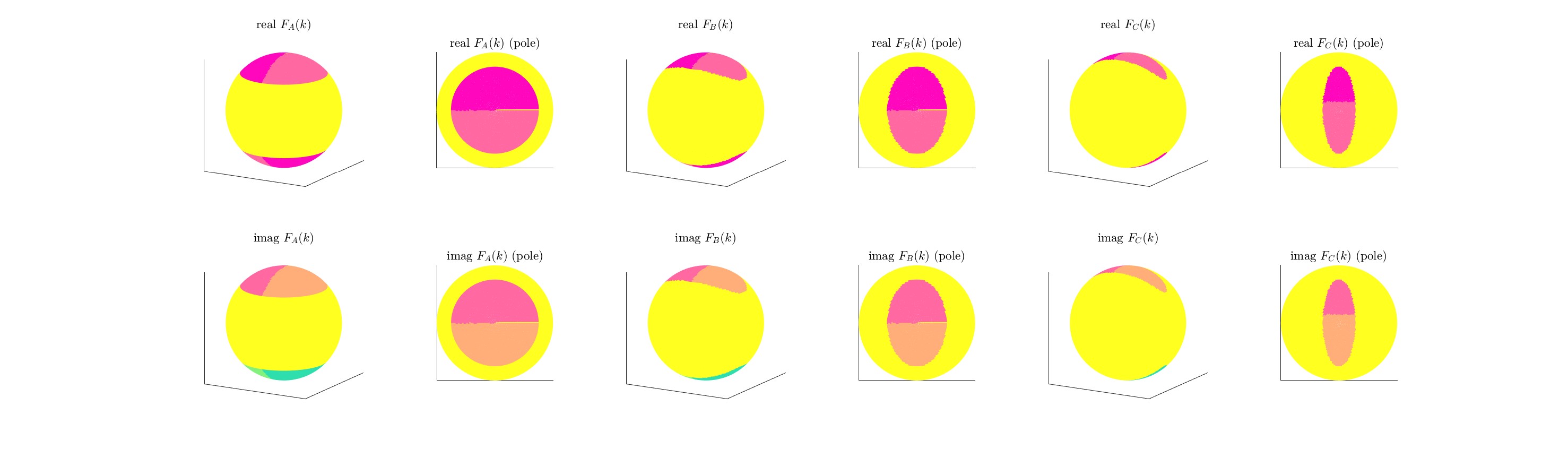}
\caption{
In this figure we illustrate level-sets of the volumes produced by convolving the frequency-shells shown in Fig \ref{fig_test_heterogeneity_spurious_s2sg125sm650bv8i8sr6sv16i12cr5cv12i10_FIGA} with a radial-kernel.
Note that, by construction, each projection of the volume $\xF_{A}$ will correspond to some projection of $\xF_{B}$ and/or to some projection of $\xF_{C}$.
In other words, given the full collection of templates produced by $\xF_{A}$, the multi-particle reconstruction problem is poorly posed.
}
\label{fig_test_heterogeneity_spurious_s2sg125sm650bv8i8sr6sv16i12cr5cv12i10_FIGB}
\end{figure}

The scenario described above is just one example of the kinds of degeneracies that can occur.
Figs \ref{fig_test_heterogeneity_spurious_s2sg125sm650bv8i8sr6sv16i12cr5cv12i10_FIGA} and \ref{fig_test_heterogeneity_spurious_s2sg125sm650bv8i8sr6sv16i12cr5cv12i10_FIGB} illustrate another similar degenerate scenario.
In this case each projection of the volume $\xF_{A}$ corresponds to at least one projection of $\xF_{B}$ or of $\xF_{C}$.
Thus, given only an image-pool comprising the collection of projections from $\xF_{A}$, one cannot determine if the image-pool was produced by the single volume $\xF_{A}$, or a combination of other volumes (e.g., $\xF_{B}$ and $\xF_{C}$).


\subsection{Noisy images}
The previous section discussed degenerate scenarios exhibiting spurious heterogeneity: a collection of projections from one volume can be indistinguishable from a subset of projections collected from a pair of other volumes.

\begin{figure}[H]
\centering
\includegraphics[width=0.65\textwidth,trim=250 100 150 100,clip]{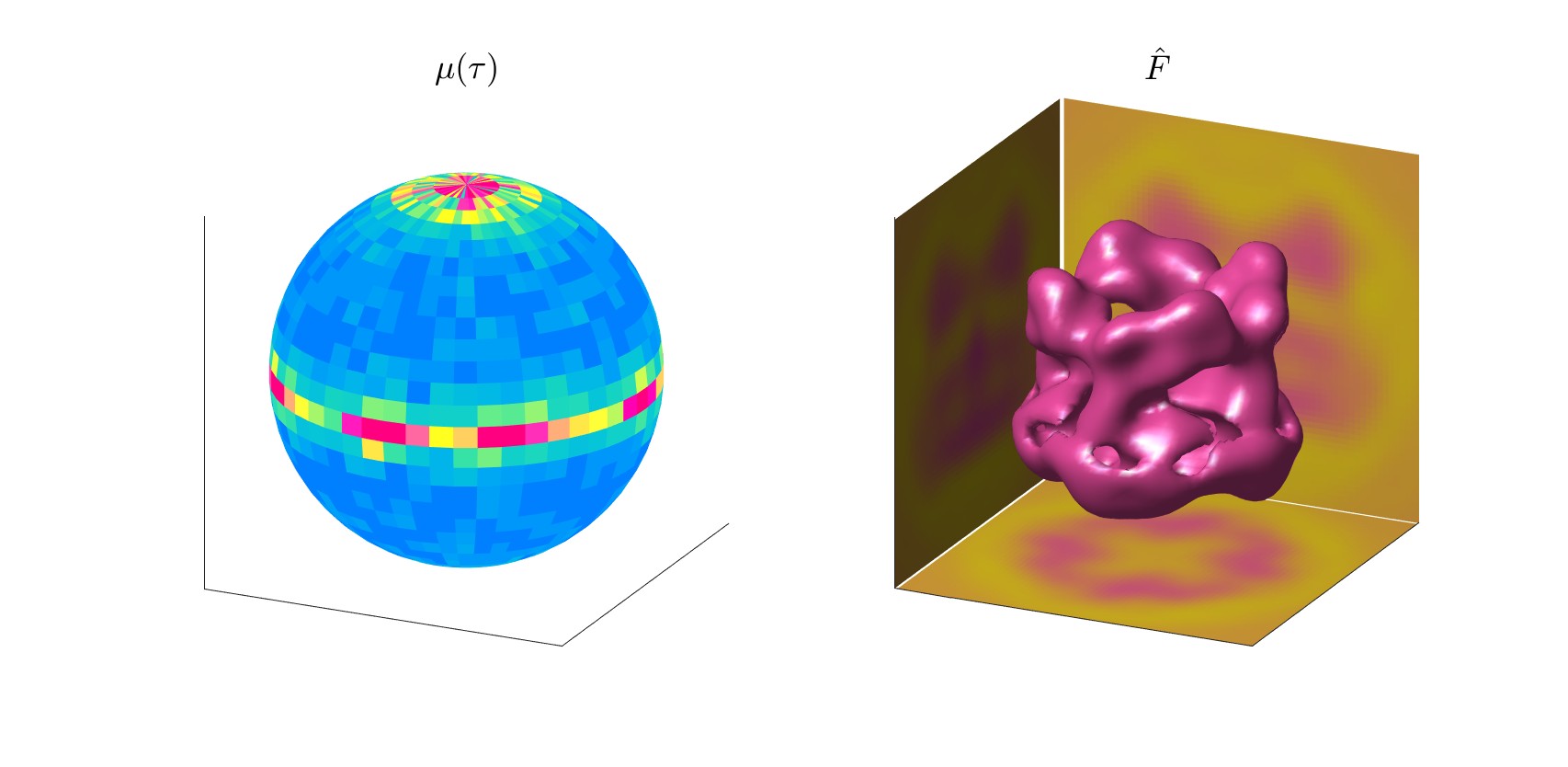}
\caption{
This figure illustrates the empirical viewing-angle distribution (left) and a level-surface of a $13\angstrom$ resolution reconstruction from {\tt emd-5778}.
}
\label{fig_trpv1_k48_eig_i1_from_synth_nlt30pm7_p_empirical_FIGL}
\end{figure}

A similar scenario can manifest in real data, particularly when the number of images $\nimage$ is relatively small and the viewing-angle distribution $\mutauemp$ of the data is non-uniform.
As an example, consider the trpv1-molecule and empirical viewing-angle distribution shown in Fig \ref{fig_trpv1_k48_eig_i1_from_synth_nlt30pm7_p_empirical_FIGL}.

Now let's consider the multi-particle reconstruction problem, given only the first $\nimage=1024$ images from the image-stack at {\tt EMPIAR\_10005}.
For simplicity, let's ignore translations for now (assuming that each image is appropriately centered).
As described in section \ref{sec_Volume_likelihood_mle_multi_particle}, one might approximately solve this problem by constructing a collection of volumes $\{\xF_{i}\}$ and, for each image $\xA_{j}$, determining the maximum-likelihood alignments $\{\tau^{\opt}_{j}\}$ and associated volume-label $i^{\opt}_{j}$.

There is a trivial (and useless) solution to this problem involving a number of volumes equal to the number of images, where volume $j$ is fit to image $\xA_{j}$.
Clearly, this trivial solution is overfit to the data, and one might hope to prevent such overfitting by restricting the total number of volumes.
However, as we'll see in a moment, such a strategy is not guaranteed to succeed, and in this scenario we still observe spurious heterogeneity even if the number of volumes is restricted to only $2$. 

\paragraph{Spurious heterogeneity with synthetic images:}

To see why this might be the case, recall that the empirical viewing-angle distribution from {\tt EMPIAR\_10005} is partially concentrated around the equator, with the remainder concentrated near the poles.
Let's now consider performing `surgery' on $\mutauemp$, and dividing this into two disjoint distributions, one $\mutauequa$ supported in a band around the equator, and the second $\mutaupole$ supported on a pair of antipodal polar caps.

As discussed in section \ref{sec_local_well_posedness_ansatz_longitudinal_perturbation_0}, the equatorial viewing-angles in $\mutauequa$ can, by themselves, give rise to a single-particle reconstruction which is not very robust.
That is to say, there are volumes that are not close to the true volume, but which can produce a collection of equatorial-templates which are a very close match to the equatorial-templates of the true volume.
Generally speaking, the quality of this match, as measured by the likelihood in Eq. \ref{eq_P_AA_given_F_maximum_likelihood} will increase as the viewing-angle distribution $\mutauequa$ becomes concentrated around the equator (see Fig \ref{fig_M3d_shape_longitudinal_perturbation_std_vs_sigma_FIGA}).

Similarly, as discussed in section \ref{sec_local_well_posedness_ansatz_latitudinal_perturbation_0}, the polar viewing-angles in $\mutaupole$ can, by themselves, also produce a reconstruction which is not very robust.
Once again, there are volumes that are not close to the true volume, but which can produce nearly identical polar-templates.
Generally speaking, the quality of this match will increase as the viewing-angle distribution $\mutaupole$ becomes concentrated around the poles (see Fig \ref{fig_M3d_shape_latitudinal_perturbation_std_vs_sigma_FIGA}).

\begin{figure}[H]
\centering
\includegraphics[width=1.0\textwidth]{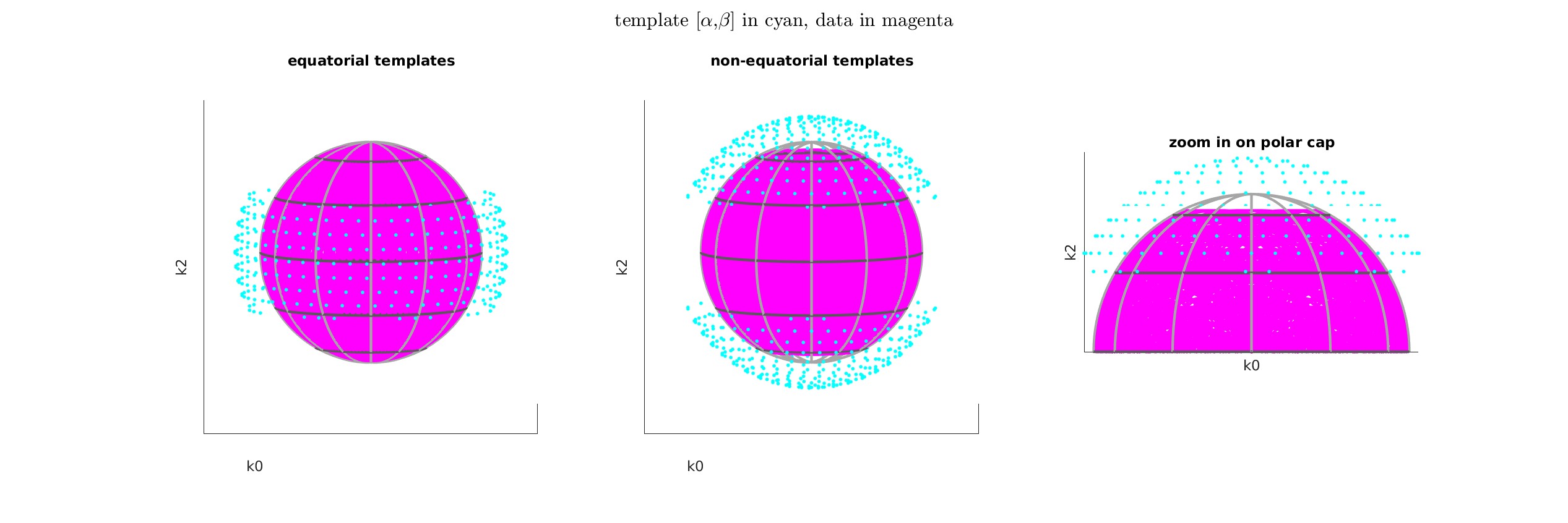}
\caption{
In this figure we illustrate the viewing-angles corresponding to $\mutauequa$ (cyan dots, left) and $\mutaupole$ (cyan dots, middle and right).
Each viewing-angle is indicated by a cyan dot hovering above the surface of the sphere.
The fourier-slice corresponding to each viewing-angle is supported on a great-circle orthogonal to the viewing-angle itself.
This data (accumulated across all viewing-angles) is illustrated by pink dots covering the surface of the sphere.
Note that, while the data from $\mutauequa$ technically covers the surface of the sphere, the data from $\mutaupole$ does not.
As show on the right, there is a `blind spot' near each pole where the data does not constrain the values of $\fF$ near the poles.
}
\label{fig_test_heterogeneity_spurious_trpv1c_equ_vs_pol_FIGA}
\end{figure}

For this particular case-study we'll simply chop $\mutauemp$ into two parts, as indicated in Fig \ref{fig_test_heterogeneity_spurious_trpv1c_equ_vs_pol_FIGA}.
Note that, while the data from $\mutauequa$ technically covers the surface of the sphere, the data from $\mutaupole$ does not.
As show on the right of Fig \ref{fig_test_heterogeneity_spurious_trpv1c_equ_vs_pol_FIGA}, there is a `blind spot' at each pole where the data from $\mutaupole$ does not constrain the values of $\fF$ near the poles.

\begin{figure}[H]
\centering
\includegraphics[width=1.0\textwidth]{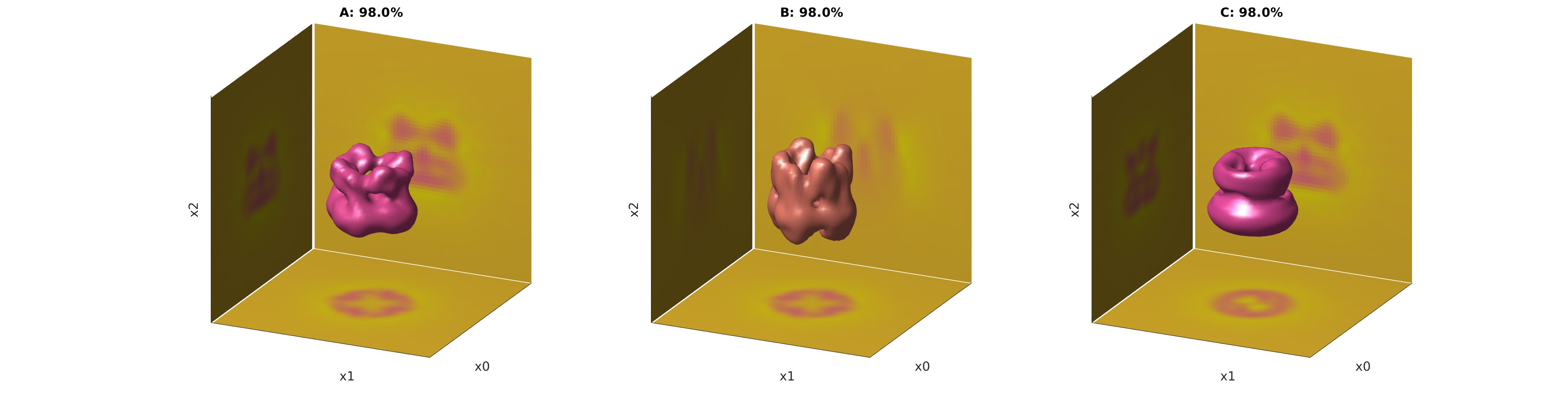}
\caption{
In this figure we illustrate three level-sets of volumes $\xF_{A}$, $\xF_{B}$ and $\xF_{C}$.
The volume $\xF_{A}$ on the left is the reconstruction {\tt EMD-5778} of trpv1 referenced in {\tt EMPIAR-10005}.
The empirical-distribution of viewing-angles $\mutauemp$ for this volume in the data-set {\tt EMPIAR-10005} is concentrated around the equator and the poles (see Fig \ref{fig_trpv1_k48_eig_i1_from_synth_nlt30pm7_p_empirical_FIGL}).
Templates of $\xF_{A}$ with viewing-angles $\tau\sim\mutaupole$ are very likely to have a close match to one of the templates of $\xF_{B}$.
Similarly, templates of $\xF_{A}$ with viewing-angles $\tau\sim\mutauequa$ are very likely to have a close match to one of the templates of $\xF_{C}$.
Thus, by construction, each projection of the volume $\xF_{A}$ will correspond to some projection of $\xF_{B}$ and/or to some projection of $\xF_{C}$.
In other words, given a collection of synthetic-images produced by $\xF_{A}$, the multi-particle reconstruction problem can be poorly posed.
}
\label{fig_test_heterogeneity_spurious_trpv1c_p9800_FIGD}
\end{figure}

\begin{figure}[H]
\centering
\includegraphics[width=1.0\textwidth]{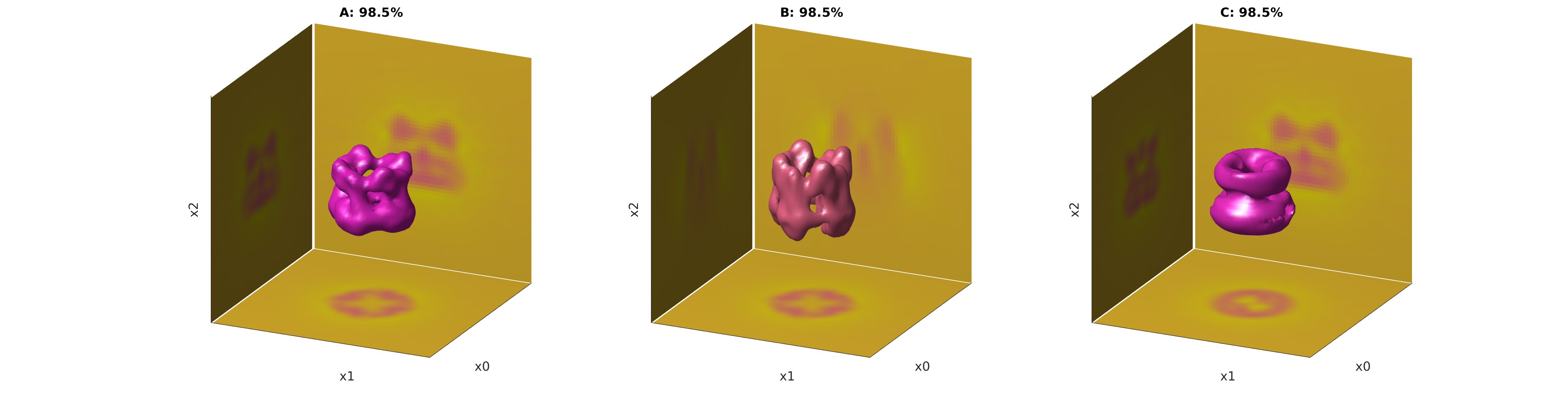}
\caption{
This is similar to Fig \ref{fig_test_heterogeneity_spurious_trpv1c_p9800_FIGD}, except for a different level-surface.
}
\label{fig_test_heterogeneity_spurious_trpv1c_p9850_FIGD}
\end{figure}

\begin{figure}[H]
\centering
\includegraphics[width=1.0\textwidth]{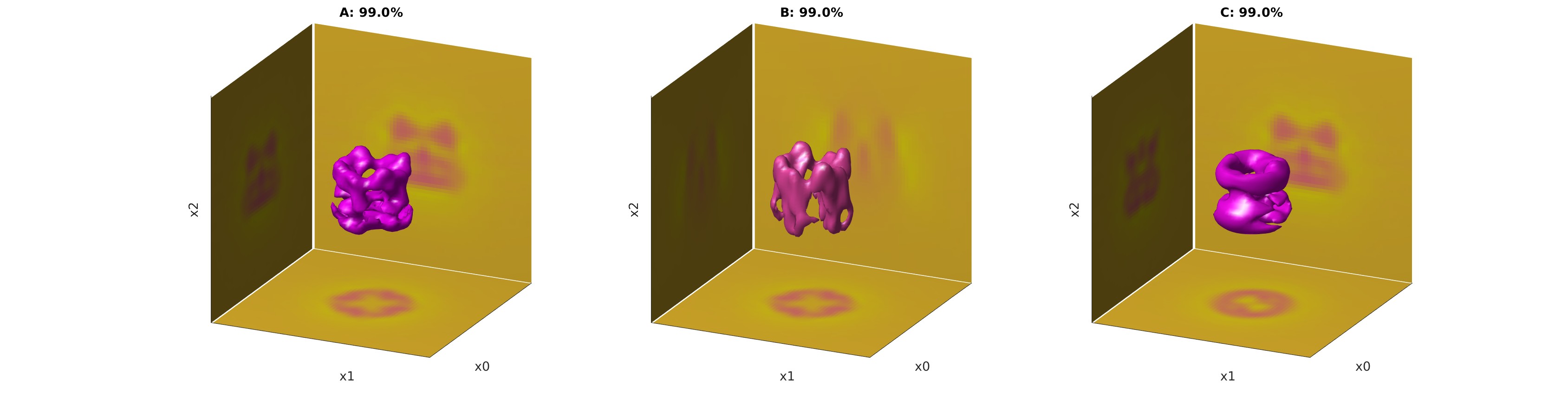}
\caption{
This is similar to Fig \ref{fig_test_heterogeneity_spurious_trpv1c_p9800_FIGD}, except for a different level-surface.
}
\label{fig_test_heterogeneity_spurious_trpv1c_p9900_FIGD}
\end{figure}

\begin{figure}[H]
\centering
\includegraphics[width=1.0\textwidth]{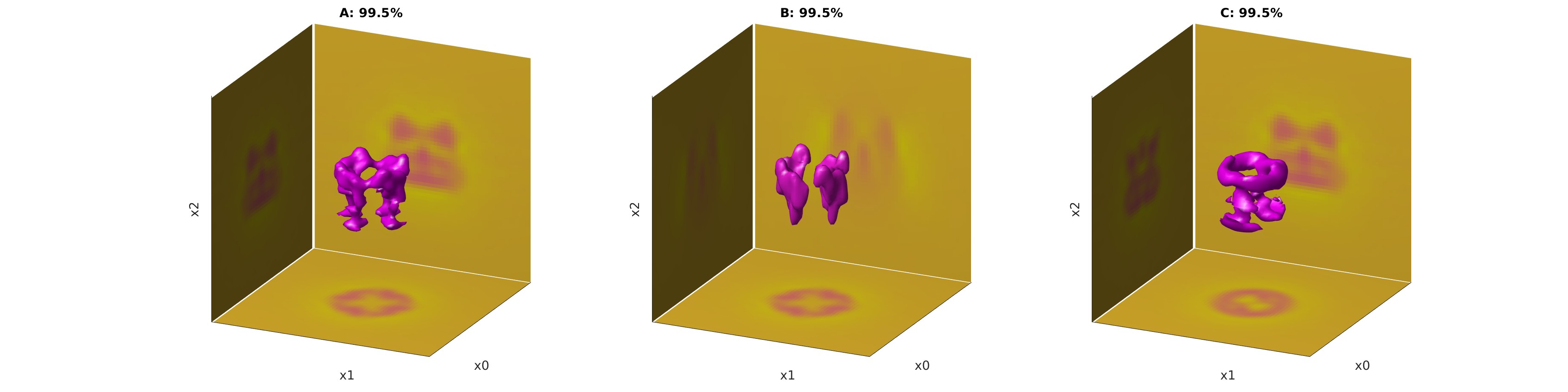}
\caption{
This is similar to Fig \ref{fig_test_heterogeneity_spurious_trpv1c_p9800_FIGD}, except for a different level-surface.
}
\label{fig_test_heterogeneity_spurious_trpv1c_p9950_FIGD}
\end{figure}

We'll return to these blind-spots in a moment, but for now we'll illustrate in Figs \ref{fig_test_heterogeneity_spurious_trpv1c_p9800_FIGD}-\ref{fig_test_heterogeneity_spurious_trpv1c_p9950_FIGD} two volumes $\xF_{B}$ and $\xF_{C}$.
These two volumes correspond to reconstructions using the distributions $\mutaupole$ and $\mutauequa$, respectively.
These volumes are illustrated alongside the original volume $\xF_{A}$, which corresponds to the true volume {\tt EMD-5778}.

The volume $\xF_{B}$ has been crafted so that, for each template $\xS(\cdot;\tau^{A},\xF_{A})$ with a $\tau^{A}$ drawn from $\mutaupole$, there exists a template $\xS(\cdot;\tau^{B},\xF_{B})$ which is very close to $\xS(\cdot;\tau^{A},\xF_{A})$.
Similarly, the volume $\xF_{C}$ has been crafted so that, for each template $\xS(\cdot;\tau^{A},\xF_{A})$ with a $\tau^{A}$ drawn from $\mutauequa$, there exists a template $\xS(\cdot;\tau^{C},\xF_{C})$ which is very close to $\xS(\cdot;\tau^{C},\xF_{C})$.

To be more explicit, consider taking a sample of $\nimage$ synthetic-images $\{\fA_{j}^{\synthetic}\}$, each produced using the true volume $\fF_{A}$ and a viewing-angle $\tau^{A}_{j}$ drawn from $\mutauemp$.
If we were to align $\fA_{j}^{\synthetic}$ to $\fF_{A}$, then the optimal viewing-angle would be $\tau_{j}^{\opt,A}\equiv \tau^{A}_{j}$, by construction.
This collection of synthetic-images forms our `sythetic-image-pool', and is the data we will use immediately below.

Let's first align each $\fA_{j}^{\synthetic}$ to $\fF_{B}$, producing an optimal viewing-angle $\tau_{j}^{\opt,B}$.
This optimal viewing-angle will only correspond to a close match if $\fA_{j}^{\synthetic}$ corresponds to an polar-template (i.e., if $\tau^{A}_{j}$ lies in the support of $\mutaupole$).
To indicate this, let's define $\tau_{j}^{\opt,B+C}:=\tau_{j}^{\opt,B}$ and set the volume-assignment $i_{j}^{\opt,B+C}$ to be `$B$' for these images.

In a similar fashion, we can align each $\fA_{j}^{\synthetic}$ to $\fF_{C}$, producing an optimal viewing-angle $\tau_{j}^{\opt,C}$.
This optimal viewing-angle will only correspond to a close match if $\fA_{j}^{\synthetic}$ corresponds to a equatorial-template (i.e., if $\tau^{A}_{j}$ lies in the support of $\mutauequa$).
To indicate this, let's define define $\tau_{j}^{\opt,B+C}:=\tau_{j}^{\opt,C}$  and set the volume-assignment $i_{j}^{\opt,B+C}$ to be `$C$' for these images.

Now we can use the optimal-alignments $\{\tau_{j}^{\opt,A}\}$ to calculate the log-likelihood $\cL_{A}:=\log\left(P\left(\{\fA_{j}^{\synthetic}\}\givenbig \fF_{A} \right)\right)$, as described in Eq. \ref{eq_P_AA_given_F_maximum_likelihood}.
Similarly, we can can use the optimal-alignments $\{\tau_{j}^{\opt,B+C}\}$ and optimal-assignments $\{i_{j}^{\opt,B+C}\}$ to calculate the log-likelihood $\cL_{B+C}:=\log\left(P\left(\{\fA_{j}\}\givenbig\{\fF_{B},\fF_{C}\}\right)\right)$ as described in Eq. \ref{eq_P_AA_given_FF_maximum_likelihood}.

Now we can compare two different hypothesis.
The first hypothesis, denoted by `$\hypothesis_{A}$', is that the synthetic-image-pool was generated by the single volume $xF_{A}$ and the alignments $\{\tau_{j}^{\opt,A}\}$.
The second hypothesis, denoted by `$\hypothesis_{B+C}$' is that the synthetic-image-pool was generated by the pair of volumes $\{\xF_{B},\xF_{C}\}$ along with the alignments $\{\tau_{j}^{\opt,B+C}\}$ and the volume-assignments $\{i_{j}^{\opt,B+C}\}$.
Because we are using synthetic-images, we are guaranteed that the `true' hypothesis $\hypothesis_{A}$ is more likely than the `spurious' hypothesis $\hypothesis_{B+C}$ (see section \ref{sec_synthetic_images}).
However, for this case-study we have prepared the volumes $\xF_{B}$ and $\xF_{C}$ so that, given the empirical noise-level (estimated from {\tt EMPIAR\_10005}), the log-likelihood ratio $\cL_{A}-\cL_{B+C}$ is actually slightly less than $\log(20)$.
This means that, under standard-practice, the spurious hypothesis $\hypothesis_{B+C}$ would not be rejected in favor of the true hypothesis $\hypothesis_{A}$.

\paragraph{Spurious heterogeneity with experimental-images:}

Above we considered a somewhat ideal setting where the image-pool was constructed using only sythetic images drawn from the true volume $\xF_{A}$.
If the size $\nimage$ of the image-pool is not too large and the empirical viewing-angle distribution $\mutauemp$ is nonuniform, then we can encounter spurious heterogeneity.
Notably, if we revisit the same setting and use actual experimental-images, we see that spurious heterogeneity can be even more pernicious.

Recall that the distribution $\mutaupole$ was quite concentrated near the poles, yielding a `blind-spot' near each pole shown in Fig \ref{fig_test_heterogeneity_spurious_trpv1c_equ_vs_pol_FIGA}.
That is to say, the data from the synthetic-images drawn from $\mutaupole$ do not constrain the values of $\fF_{B}$ in these blind-spots.
Thus, these blind-spots represents a collection of degrees-of-freedom that are unconstrained by the true projections of the volume.

Above, to produce the polar-aligned volume $\xF_{B}$ shown in Figs \ref{fig_test_heterogeneity_spurious_trpv1c_p9800_FIGD}-\ref{fig_test_heterogeneity_spurious_trpv1c_p9950_FIGD}, we merely set the values of $\fF_{B}$ to be constant in each blind-spot (subject to the conjugacy conditions).
However, in practice these blind-spots are not likely to be treated so simply.

\begin{figure}[H]
\centering
\includegraphics[width=1.0\textwidth]{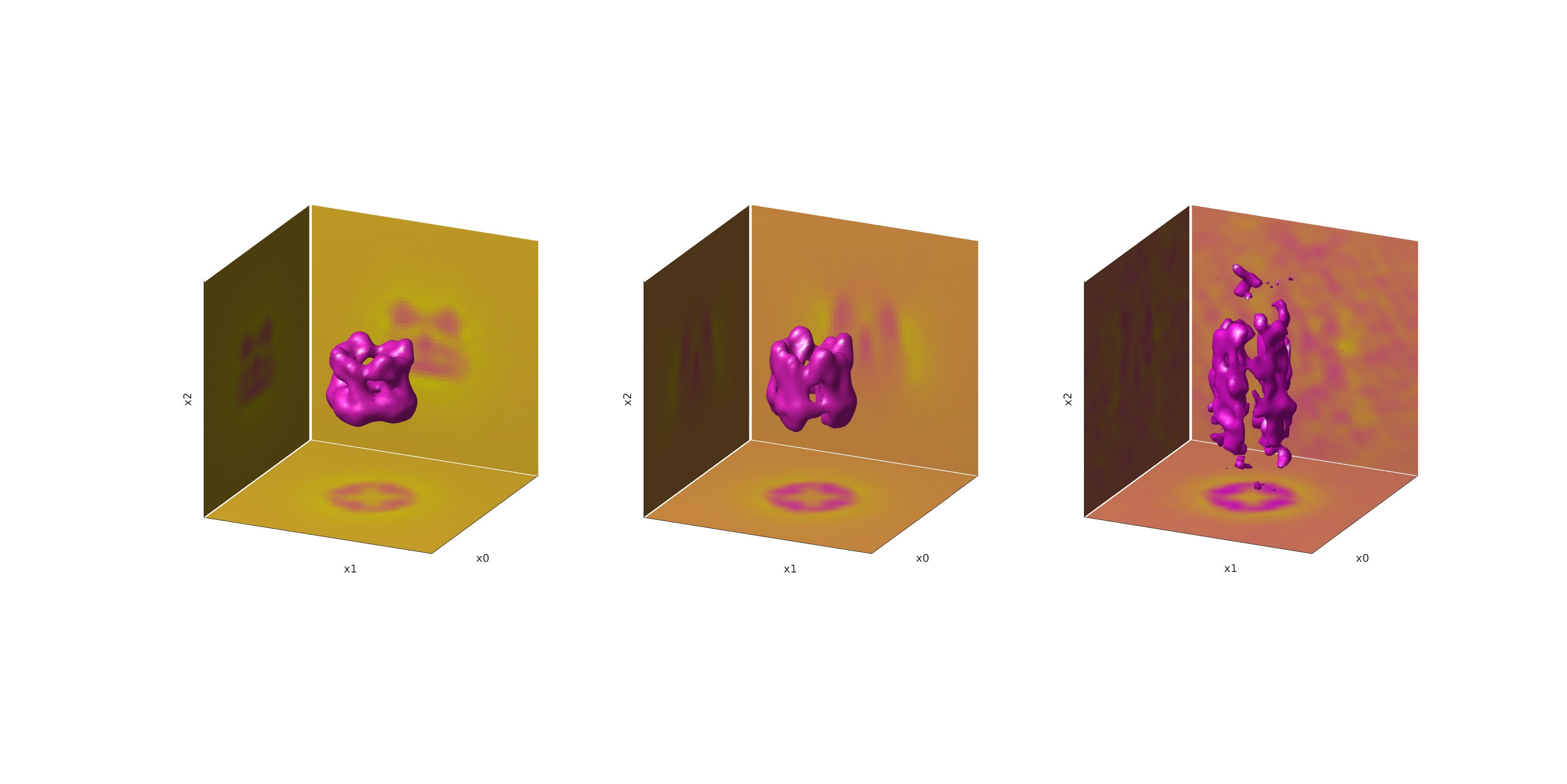}
\caption{
This is similar to Fig \ref{fig_test_heterogeneity_spurious_trpv1c_p9850_FIGD}.
We show the original $\xF_{A}$ on the left, and the polar-matched volume $\xF_{B}$ in the middle.
On the far right we now show an alternative volume $\xF_{B'}$.
This alternative volume again matches the polar-templates from $\xF_{A}$, but has its `blind-spots' fit to a small subset of $32$ noisy images.
}
\label{fig_test_heterogeneity_spurious_trpv1c_reco_rem2_rem3_R0032_FIGB}
\end{figure}

One such example is shown on the right of Fig \ref{fig_test_heterogeneity_spurious_trpv1c_reco_rem2_rem3_R0032_FIGB}.
This shows a volume $\xF_{B'}$ which has the same polar-templates as $\xF_{B}$ from Fig \ref{fig_test_heterogeneity_spurious_trpv1c_p9850_FIGD}, but has different data in its blind spot, corresponding to `overfitting' to noise.

To explain this scenario more carefully, first recall the original pool of $\nimage=1024$ synthetic-images described above.
As described above, these $\nimage$ synthetic-images were constructed using projections of the true molecule $\xF_{A}$.
When aligning these synthetic-images to the spurious-volumes $\xF_{B}$ and $\xF_{C}$, those synthetic-images corresponding to polar-templates fit best to $\xF_{B}$, while those corresponding to equatorial-templates fit best to $\xF_{C}$.

Now let's add $32$ additional noisy experimental-images to the image-pool.
These additional noisy images have viewing-angles that are not too far from the poles, and naturally align best to $\xF_{B}$.
However, due to the noise, their optimal alignment-angle $\tau_{j}^{\opt,B}$ is somewhat far from the poles, so that the images themselves correspond to values of $\hk$ that pass through the polar `blind-spots' associated with $\mutaupole$.

Now, by using these optimally-aligned noisy images (along with the optimally-aligned sythetic-images corresponding to polar-templates), we can reconstruct $\xF_{B'}$.
Unlike the volume $\xF_{B}$ shown in the middle of Fig \ref{fig_test_heterogeneity_spurious_trpv1c_reco_rem2_rem3_R0032_FIGB}, the volume $\xF_{B'}$ shown on the right of Fig \ref{fig_test_heterogeneity_spurious_trpv1c_reco_rem2_rem3_R0032_FIGB} has its polar-data fit to the additional noisy experimental-images.
This additional polar-information distorts the volume $\xF_{B'}$, but does not appreciably change the polar-projections.
Thus while the volumes $\xF_{B}$ and $\xF_{B'}$ provides equally good matches to the polar-templates, the volume $\xF_{B'}$ provides a better match than the `cleaner' volume $\xF_{B}$ to the noisy experimental-images.

\begin{figure}[H]
\centering
\includegraphics[width=1.0\textwidth]{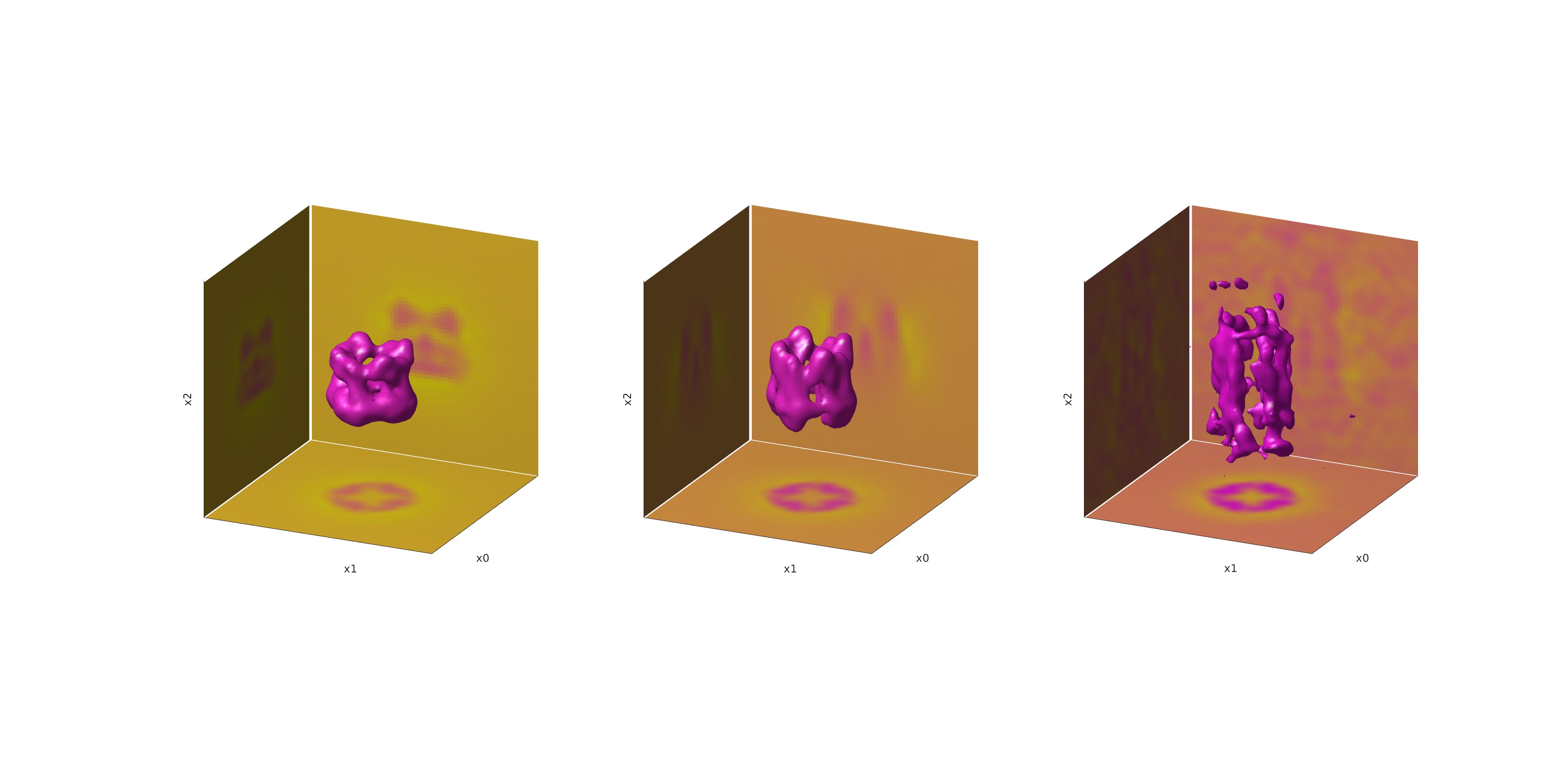}
\caption{
This is similar to Fig \ref{fig_test_heterogeneity_spurious_trpv1c_reco_rem2_rem3_R0032_FIGB}.
This time the alternative volume on the right is fit to a different subset of $64$ noisy images.
}
\label{fig_test_heterogeneity_spurious_trpv1c_reco_rem2_rem3_R0064_FIGB}
\end{figure}

This results in a classical instance of `overfitting': When considering the expanded image-pool comprising both the idealized synthetic-images and the additional $32$ noisy experimental-images, the hypothesis $\hypothesis_{B'+C}$ is actually \textit{more} likely than $\hypothesis_{A}$.
A similar scenario is shown in Fig \ref{fig_test_heterogeneity_spurious_trpv1c_reco_rem2_rem3_R0064_FIGB}, using a different collection of $64$ noisy experimental-images.

Finally, we remark that spurious heterogeneity of this kind can be difficult to correct for a-posteriori.
In many cases it might not be clear which images are `good' images of particles, and which are too strongly polluted by noise.
It also might be unclear which features of the reconstructed volume (or volumes) are a result of overfitting to noise, or if such a distinction can even be made.
Moreover, some commonly used techniques for validation, such as aligning the images to an `averaged' volume of the form $0.5\times\left(\xF_{B}+\xF_{C}\right)$, could erroneously provide support for the spurious hypothesis.

In these situations we would recommend calculating the softest modes of the Hessian associated with each reconstructed volume, as described in the main text.
While such a calculation won't immediately identify a heterogeneous reconstruction as spurious, it can certainly provide a warning.
For example, in the case-studies shown above the spurious volume $\xF_{B}$ is susceptible to latitudinal perturbations, while the spurious volume $\xF_{C}$ is susceptible to longitudinal perturbations.
Both of these spurious volumes are less robust to perturbations than the volume $\xF_{A}$ produced via single-particle reconstruction.
This kind of analysis can help indicate possible errors associated with a heterogeneous reconstruction.


\bibliographystyle{unsrt}
\bibliography{Acta_D_bib}

\end{document}